\documentclass{llncs}

\usepackage{amsmath,amssymb}
\usepackage{QED}
\usepackage[all]{xy}

\usepackage{pstricks,pst-node}
 \psset{rowsep=.1cm,colsep=.1cm}
 \psset{arrowsize=3.5pt 3.8,arrowlength=0.7}
 \psset{labelsep=.1cm,nodesep=2pt}
\def\_{\kern.08em\vbox{\hrule width.35em height.6pt}\kern.08em}
\newcommand{\coqdockw}[1]{\textsf{#1}}
\newcommand{\coqdocid}[1]{\textit{#1}}
\newcommand{\coqdoceol}{\setlength\parskip{0pt}\par}
\newcommand{\coqdocindent}[1]{\noindent\kern#1}

\newcommand{\choiceline}{\ncline[doubleline=true]}

\author{St\'ephane Le Roux\thanks{This research was partly supported by Coll\`ege Doctoral Franco-Japonais}\thanks{Also available as LIP research report RR2007-18}}
 
\institute{\'Ecole normale sup\'erieure de Lyon, LIP, CNRS, INRIA, UCBL}

\title{Acyclicity of Preferences, Nash Equilibria, and Subgame Perfect Equilibria: a Formal and Constructive Equivalence}

\begin{document}

\maketitle

\begin{abstract}
Sequential game and Nash equilibrium are basic key concepts in game theory. In 1953, Kuhn showed that every sequential game has a Nash equilibrium. The two main steps of the proof are as follows: First, a procedure expecting a sequential game as an input is defined as ``backward induction'' in game theory. Second, it is proved that the procedure yields a Nash equilibrium. ``Backward induction'' actually yields Nash equilibria that define a proper subclass of Nash equilibria. In 1965, Selten named this proper subclass subgame perfect equilibria. In game theory, payoffs are rewards usually granted at the end of a game. Although traditional game theory mainly focuses on real-valued payoffs that are implicitly ordered by the usual total order over the reals, there is a demand for results dealing with non totally ordered payoffs. In the mid 1950's, works of Simon or Blackwell already involved partially ordered payoffs. This paper further explores the matter: it generalises the notion of sequential game by replacing real-valued payoff functions with abstract atomic objects, called outcomes, and by replacing the usual total order over the reals with arbitrary binary relations over outcomes, called preferences. This introduces a general abstract formalism where Nash equilibrium, subgame perfect equilibrium, and ``backward induction'' can still be defined. Using a lemma on topological sorting, this paper proves that the following three propositions are equivalent: 1) Preferences over the outcomes are acyclic. 2) Every sequential game has a Nash equilibrium. 3) Every sequential game has a subgame perfect equilibrium. The result is fully computer-certified using the (highly reliable) constructive proof assistant called Coq. Beside the additional guarantee of correctness provided by the proof in Coq, the activity of formalisation also helps clearly identify the useful definitions and the main articulations of the proof.
\end{abstract}

\emph{Abstract sequential game, Nash equilibrium, subgame perfect equilibrium, constructivism, induction, proof assistant}

\section{Introduction}

\subsection{Game Theory in Short}

Game theory embraces the theoretical study of processes involving (more or less conscious) possibly interdependent decision makers. Game theory originates in economics, politics, law, and also games dedicated to entertainment. Instances of game theoretic issues may be traced back to Babylonian times when the Talmud would prescribe marriage contracts that seem to be solutions of some relevant games described in~\cite{Aumann85}. In 1713, a simple card game raised questions and solutions involving probabilities, as discussed in~\cite{Kuhn68}. During the XVIIth and XVIIIth centuries, philosophers such as Hobbes~\cite{Hobbes51} adopted an early game theoretical approach to study political systems. In 1838, Cournot~\cite{Cournot38} introduced the notion of equilibrium for pricing in duopoly. It is said that game theory became a major field in 1944, when von Neumann and Morgenstern~\cite{Neumann44} published a summary of prior works and a systematic study of a few classes of games. In 1950, the notion of equilibrium and the corresponding solution concept discussed by Cournot were generalised by Nash~\cite{Nash50} for a class of games called \emph{strategic games}. In 2006, the notion of Nash equilibrium was abstracted~\cite{LLV06} as a situation that no stakeholder can convert into another situation that he prefers. In addition to economics, politics and law, modern game theory is consciously involved in many other fields such as biology, computer science, and sociology.

\subsection{Sequential Game Theory}

Another class of games is that of \emph{sequential games}, also called \emph{games in extensive form}. It traditionally refers to games where players play in turn till the play ends and \emph{payoffs} are granted. For instance, the game of chess is naturally modelled by a sequential game where payoffs are ``win'', ``lose'' and ``draw''. Sequential games are often represented by finite rooted trees each of whose internal nodes is owned by a player, and each of whose external nodes encloses one payoff per player. In 1912, Zermelo~\cite{Zermelo12} proved about the game of chess that either white can win, or black can win, or both sides can force a draw. This is sometimes considered as the first non-trivial theoretical results in game theory. Although the concept of Nash equilibrium referred to strategic games in the first place, it is natural and relevant to extend that concept to sequential games. In 1953, Kuhn~\cite{Kuhn53} showed the existence of Nash equilibrium for sequential games. For this, he built a specific Nash equilibrium through what is called ``backward induction'' in game theory. In 1965, Selten (\cite{Selten65} and~\cite{Selten75}) introduced the concept of subgame perfect equilibrium in sequential games. This is a refinement of Nash equilibrium that seems to be even more meaningful than Nash equilibrium for sequential games. The two concepts of ``backward induction'' and subgame perfect equilibrium happen to coincide, so Kuhn's result also guarantees existence of subgame perfect equilibrium in sequential games. In 2006, Vestergaard~\cite{Vestergaard06} formalised part of Kuhn's result with the proof assistant Coq, for the subclass of games represented by binary trees and whose payoffs range over the natural numbers. For this, he defined sequential games and corresponding strategy profiles inductively.

\subsection{Ordering Payoffs\label{subsect:ou}}

Game theory has mostly studied games with real-valued payoffs, perhaps for the following reason: In 1944, von Neumann and Morgernstern~\cite{Neumann44} suggested that the notion of payoff in economics could be reduced to real numbers. They argued that more and more physical phenomena were measurable; therefore, one could reasonably expect that payoffs in economics, although not yet measurable, would become reducible to real numbers some day. However, game theory became popular soon thereafter, and its scope grew larger. As a result, several scientists and philosophers questioned the reducibility of payoffs to real numbers. In 1955, Simon~\cite{Simon55} discussed games where agents are awarded (only partially ordered) vectors of real-valued payoffs instead of single real-valued payoffs. In 1956, Blackwell~\cite{Blackwell56} proved a result involving vectors of payoffs. Those vectors model agents that take several non-commensurable dimensions into consideration; such games are sometimes called  multi-criteria games. In 1994, Osborne and Rubinstein~\cite{OR94} mentioned arbitrary preferences, without any further results. In 2003, Krieger~\cite{Krieger03} noticed that ``backward induction'' on sequential multi-criteria games may not yield Nash equilibria, and yet showed that sequential multi-criteria games have Nash equilibria. The proof seems to invoke probabilities and Nash's theorem for strategic games. In 2006, a paper \cite{SLR06} discussed sequential games with arbitrary partially ordered payoffs and proved in Coq the existence of non-deterministic Nash equilibria {\it via} ``backward induction''.

\subsection{Contribution}

This paper contributes at both the technical and the presentation level.

There are five main technical contributions: First, an inductive formalism is designed to represent sequential games in the constructive proof assistant Coq (\cite{coq} and~\cite{BC04}), and all the results in this paper are proved in Coq. Second, the new formalism allows the paper to introduce an abstraction of traditional sequential games and of a few related concepts. The abstraction preserves the tree structure of the game but replaces the real-valued payoff functions, enclosed in the leaves of the trees, by arbitrary outcomes. Each agent has a preference over outcomes, which is given {\it via} an explicit and arbitrary binary relation. This preference replaces the implicit and traditional ``usual total order'' over the real numbers. Nash equilibria and subgame perfect equilibria are defined accordingly. Third, Kuhn's result~\cite{Kuhn53} is translated into the new formalism when agents' preferences are totally ordered. Fourth, the notion of ``backward induction'' is naturally generalised for arbitrary preferences, but a simple example shows that total ordering of preferences is needed for ``backward induction'' to guarantee subgame perfect equilibrium: the two notions of ``backward induction'' and subgame perfect equilibrium coincide for total orders but not in general. Fifth, Kuhn's result is substantially generalised as follows. On the one hand, an intermediate result proves that smaller preferences, {\it i.e.}, binary relations with less arcs, yield more equilibria than bigger preferences. On the other hand, a topological sorting result was formally proved in~\cite{SLR2007-14}. By both results mentioned above, acyclicity of the preferences proves to be a necessary and sufficient condition for every game to have a Nash equilibrium/subgame perfect equilibrium.

This paper deals with basic notions of game theory that are all exemplified and defined before they are used. Most of the time, these notions are explained in three different ways, with the second one helping make the connection between the two others: First, the notions are presented in a graphical formalism close to traditional game theory. Second, they are presented in a graphical formalism suitable for induction. Third, they are presented in a light Coq formalism close to traditional mathematics, so that only a basic understanding of Coq is needed. (A quick look at the first ten pages of~\cite{SLR2007-14} will introduce the reader to the required notions.) The proofs are structured along the corresponding Coq proofs but are written in plain English.

\subsection{Contents}

Section~\ref{sect:tsgt} gives an intuitive and informal presentation of traditional sequential games through graphical examples. Section~\ref{sect:wto} explores further the relevance of non-totally ordered payoffs. Section~\ref{sect:pre} discusses general concepts that are not specially related to game theory but required in the remainder of the paper. Section~\ref{sect:sg} presents the above-mentioned abstraction of sequential games and their new formalism. Section~\ref{sect:sp} presents the notion of strategy profile at the same level of abstraction as for sequential games. Section~\ref{sect:c} defines convertibility between strategy profiles. Section~\ref{sect:ce} discusses the notion of preference, happiness, Nash equilibrium, and subgame perfect equilibrium. Section~\ref{sect:ee} generalises ``backward induction'', translates Kuhn's result into the new formalism, and proves the triple equivalence between acyclicity of preferences and existence of Nash equilibrium/subgame perfect equilibrium.

\section{Traditional Sequential Game Theory\label{sect:tsgt}}

Through graphical examples and explanations in plain English, this section gives an intuitive and informal presentation of traditional sequential games and of a few related concepts such as Nash equilibrium. 

A traditional sequential game involves finitely many agents and payoffs. Payoffs usually are real numbers. A payoff function is a function from the agents to the payoffs. A sequential game is a rooted finite tree with internal nodes labelled with agents and external nodes, \textit{i.e.}, leaves, labelled with payoff functions. Consider the following graphical example of such games. It involves the two agents $a$ and $b$. At every leaf, a payoff function is represented by two numbers separated by a comma: the payoff function maps agent $a$ to the first number and agent $b$ to the second number.

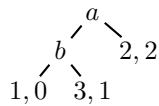
\[\psmatrix
      &&&[name=n1] a\\
      &&[name=n2] b &&[name=l3]  2,2\\
      &[name=l1] 1,0 &&[name=l2] 3,1 
      \ncline{n1}{n2}
      \ncline{n1}{l3}
      \ncline{n2}{l1}
      \ncline{n2}{l2}
    \endpsmatrix\]

Such game trees are interpreted as follows: A play of a game starts at the root of the game tree. If the tree is a leaf then agents are rewarded according to the enclosed payoff function. Otherwise, the agent owning the root chooses the next node among the children of the root. The subtree rooted at the chosen child is considered and the play continues from there. In the game above, if $a$ chooses to go right then both agents get $2$. If $a$ chooses left then $b$ has to choose too, etc. Now, a specific play of a game is described and drawn. Double lines represent choices made during the play. Agent $a$ first chooses to go right, then left, then $b$ chooses to go right. Eventually, $a$ gets $1$, $b$ gets $0$, and $c$ gets $2$. The stars $*$ represent arbitrary payoff functions irrelevant to the discussion. 

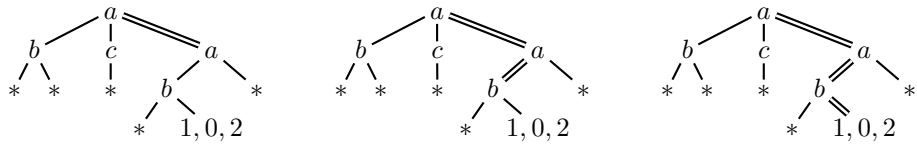
\[\psmatrix
   &&&&&[name=n] a &&&&& &&&&&[name=n'] a &&&&& &&&&&[name=n''] a\\
   &&[name=n1] b &&&[name=n2] c &&&[name=n3] a &&  &&[name=n1'] b &&&[name=n2'] c &&&[name=n3'] a &&
   &&[name=n1''] b &&&[name=n2''] c &&&[name=n3''] a \\
   &[name=l11] * &&[name=l12] * &\phantom{aa}&[name=l21] * &\phantom{aa}&[name=n31] b &&[name=l32] * &\phantom{aaaa}
   &[name=l11'] * &&[name=l12'] * &\phantom{aa}&[name=l21'] * &\phantom{aa}&[name=n31'] b &&[name=l32'] * &\phantom{aaaa}
   &[name=l11''] * &&[name=l12''] * &\phantom{aa}&[name=l21''] * &\phantom{aa}&[name=n31''] b &&[name=l32''] * \\
   &&&&&&[name=l311] *&&[name=l312] 1,0,2 && &&&&&&[name=l311'] *&&[name=l312'] 1,0,2 && &&&&&&[name=l311''] *&&[name=l312''] 1,0,2
   \ncline{n}{n1}
   \ncline{n}{n2}
   \choiceline{n}{n3}
   \ncline{n1}{l11}
   \ncline{n1}{l12}
   \ncline{n2}{l21}
   \ncline{n3}{n31}
   \ncline{n3}{l32}
   \ncline{n31}{l311}
   \ncline{n31}{l312}
   \ncline{n'}{n1'}
   \ncline{n'}{n2'}
   \choiceline{n'}{n3'}
   \ncline{n1'}{l11'}
   \ncline{n1'}{l12'}
   \ncline{n2'}{l21'}
   \choiceline{n3'}{n31'}
   \ncline{n3'}{l32'}
   \ncline{n31'}{l311'}
   \ncline{n31'}{l312'}
   \ncline{n''}{n1''}
   \ncline{n''}{n2''}
   \choiceline{n''}{n3''}
   \ncline{n1''}{l11''}
   \ncline{n1''}{l12''}
   \ncline{n2''}{l21''}
   \choiceline{n3''}{n31''}
   \ncline{n3''}{l32''}
   \ncline{n31''}{l311''}
   \choiceline{n31''}{l312''}
\endpsmatrix\]

In game theory, the strategy of an agent is an object that accounts for the decisions of the agent in all situations that the agent might encounter. A strategy profile is a tuple combining one strategy per agent. So, for sequential games, a strategy profile amounts to choices made at all internal nodes of the tree. Below is an example of a strategy profile. Double lines between nodes represent choices and the stars $*$ represent arbitrary payoff functions irrelevant to the discussion. The choice of $b$ at the leftmost internal node may seem rather ineffective, but it can be interpreted as $b$'s choice if a play ever reach this very node.

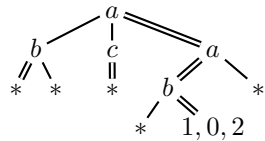
\[\psmatrix
   &&&&&[name=n] a\\
   &&[name=n1] b &&&[name=n2] c &&&[name=n3] a\\
   &[name=l11] * &&[name=l12] * &\phantom{aa}&[name=l21] * &\phantom{aa}&[name=n31] b &&[name=l32] *\\
   &&&&&&[name=l311] *&&[name=l312] 1,0,2 
   \ncline{n}{n1}
   \ncline{n}{n2}
   \choiceline{n}{n3}
   \choiceline{n1}{l11}
   \ncline{n1}{l12}
   \choiceline{n2}{l21}
   \choiceline{n3}{n31}
   \ncline{n3}{l32}
   \ncline{n31}{l311}
   \choiceline{n31}{l312}
\endpsmatrix\]

Given a strategy profile, starting from the root and following the agents' consecutive choices leads to one leaf. The payoff function enclosed in that specific leaf is called the induced payoff function. The induced payoff function of the strategy profile above is: $a$ gets $1$, $b$ gets $0$, and $c$ gets $2$.

The (usually implicit) preference of agents for strictly greater payoffs induces a (usually implicit) preference of an agent for payoff functions that grants him strictly greater payoffs. This, in turn, yields a (usually implicit) preference of an agent for strategy profiles inducing preferred payoff functions. Below, agent $a$ prefers the left-hand strategy profile to the right-hand one since $3$ is greater than $2$, but it is the opposite for agent $b$ since $1$ is less than $2$.

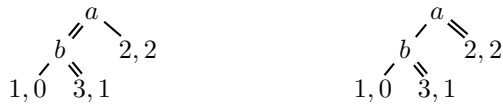
\[\psmatrix
      &&&[name=n1] a&& &&&[name=n1'] a \\
      &&[name=n2] b &&[name=l3]  2,2& &&[name=n2'] b &&[name=l3']  2,2 \\
      &[name=l1] 1,0 &&[name=l2] 3,1 && \phantom{aaaaaaaaaaaaa}&[name=l1'] 1,0 &&[name=l2'] 3,1
      \choiceline{n1}{n2}
      \ncline{n1}{l3}
      \ncline{n2}{l1}
      \choiceline{n2}{l2}
      \ncline{n1'}{n2'}
      \choiceline{n1'}{l3'}
      \ncline{n2'}{l1'}
      \choiceline{n2'}{l2'}
    \endpsmatrix\]

An agent is (usually implicitly) granted the ability to change his choices at all nodes he owns. For instance, below, the agent $b$ can convert the strategy profile on the left to the one on the right by changing his choices exactly at the nodes where $\bold{b}$ is displayed in bold font. The stars $*$ represent arbitrary payoff functions irrelevant to the discussion.

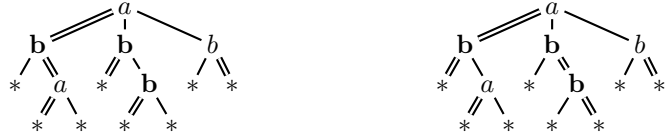
\[\psmatrix
      &&&&&&[name=n0] a&&&&&&   &&&&&&[name=n0'] a \\
      &&[name=n1] \bold{b} &&&&[name=n2] \bold{b}&&&&[name=n3] b &&&&[name=n1'] \bold{b} &&&&[name=n2'] \bold{b} &&&&[name=n3'] b \\
      &[name=l11] * &&[name=n12] a &&[name=l21] * &&[name=n22] \bold{b}  &&[name=l31] * &&[name=l32] *
      &\phantom{aaaaaaaaaaaaa}& [name=l11'] * &&[name=n12'] a &&[name=l21'] * &&[name=n22'] \bold{b} &&[name=l31'] * &&[name=l32'] *\\
&& [name=l121] * && [name=l122] * && [name=l221] * && [name=l222] * &&&&&& [name=l121'] * && [name=l122'] * && [name=l221'] * && [name=l222'] * 
      \choiceline{n0}{n1}
      \ncline{n0}{n2}
      \ncline{n0}{n3}
      \ncline{n1}{l11}
      \choiceline{n1}{n12}
      \choiceline{n2}{l21}
      \ncline{n2}{n22}
      \ncline{n3}{l31}
      \choiceline{n3}{l32}
      \choiceline{n12}{l121}
      \ncline{n12}{l122}
      \choiceline{n22}{l221}
      \ncline{n22}{l222}
      \choiceline{n0'}{n1'}
      \ncline{n0'}{n2'}
      \ncline{n0'}{n3'}
      \choiceline{n1'}{l11'}
      \ncline{n1'}{n12'}
      \ncline{n2'}{l21'}
      \choiceline{n2'}{n22'}
      \ncline{n3'}{l31'}
      \choiceline{n3'}{l32'}
      \choiceline{n12'}{l121'}
      \ncline{n12'}{l122'}
      \ncline{n22'}{l221'}
      \choiceline{n22'}{l222'}
      \endpsmatrix\]

An agent is said to be happy with a strategy profile if he cannot convert it to another strategy profile that he prefers. A Nash equilibrium is a strategy profile that makes all agents happy. Below, the strategy profile to the left is not a Nash equilibrium since its sole player gets $0$ but can convert it to the right-hand strategy profile and get $1$. However, the right-hand strategy profile is a Nash equilibrium since $a$ cannot convert it and get a payoff strictly greater than $1$. 

 \[\psmatrix
      &&&[name=n] a&&& &&&[name=n'] a \\
      &[name=l1] 0 &\phantom{a}&[name=l2] 1&\phantom{a}&[name=l3] 1&\phantom{aaaaaaaaaaaaa} &[name=l1'] 0 &\phantom{a}&[name=l2'] 1 &\phantom{a}&[name=l3'] 1
      \choiceline{n}{l1}
      \ncline{n}{l2}
      \ncline{n}{l3}
      \ncline{n'}{l1'}
      \choiceline{n'}{l2'}
      \ncline{n'}{l3'}
    \endpsmatrix\]

Here is another example of Nash equilibrium.

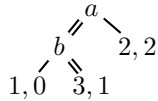
\[\psmatrix
      &&&[name=n1] a \\
      &&[name=n2] b &&[name=l3]  2,2 \\
      &[name=l1] 1,0 &&[name=l2] 3,1
      \choiceline{n1}{n2}
      \ncline{n1}{l3}
      \ncline{n2}{l1}
      \choiceline{n2}{l2}
    \endpsmatrix\]

Indeed, agent $a$ could only convert the strategy profile above to the left-hand strategy profile below, and would get $2$ instead of $3$. Therefore $a$ is happy. Agent $b$ could only convert the strategy profile above to the right-hand strategy profile below, and would get $0$ instead of $1$. Therefore $b$ is happy too. The strategy profile above makes all players happy; it is a Nash equilibrium. 

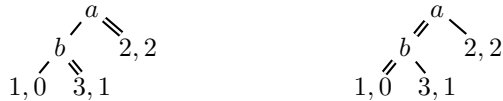
\[\psmatrix
      &&&[name=n1] a&& &&&[name=n1'] a \\
      &&[name=n2] b &&[name=l3]  2,2& &&[name=n2'] b &&[name=l3']  2,2 \\
      &[name=l1] 1,0 &&[name=l2] 3,1 && \phantom{aaaaaaaaaaaaa}&[name=l1'] 1,0 &&[name=l2'] 3,1
      \choiceline{n1}{l3}
      \ncline{n1}{n2}
      \ncline{n2}{l1}
      \choiceline{n2}{l2}
      \ncline{n1'}{l3'}
      \choiceline{n1'}{n2'}
      \choiceline{n2'}{l1'}
      \ncline{n2'}{l2'}
    \endpsmatrix\]

The underlying game of a strategy profile is computed by forgetting all the choices made at the internal nodes of the strategy profile. The next picture displays a strategy profile, to the left, and its underlying game, to the right.

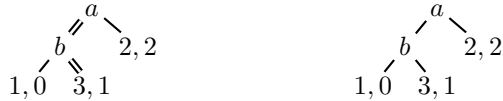
\[\psmatrix
      &&&[name=n1] a&& &&&[name=n1'] a \\
      &&[name=n2] b &&[name=l3]  2,2& &&[name=n2'] b &&[name=l3']  2,2 \\
      &[name=l1] 1,0 &&[name=l2] 3,1 && \phantom{aaaaaaaaaaaaa}&[name=l1'] 1,0 &&[name=l2'] 3,1
      \choiceline{n1}{n2}
      \ncline{n1}{l3}
      \ncline{n2}{l1}
      \choiceline{n2}{l2}
      \ncline{n1'}{n2'}
      \ncline{n1'}{l3'}
      \ncline{n2'}{l1'}
      \ncline{n2'}{l2'}
    \endpsmatrix\]

Two Nash equilibria inducing different payoff functions may have the same underlying game: indeed, consider the previous Nash equilibrium and the following one, where $b$'s choice is ineffective in terms of induced payoff function. 

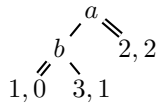
\[\psmatrix
      &&&[name=n1'] a \\
      &&[name=n2'] b &&[name=l3']  2,2 \\
      &[name=l1'] 1,0 &&[name=l2'] 3,1
      \ncline{n1'}{n2'}
      \choiceline{n1'}{l3'}
      \choiceline{n2'}{l1'}
      \ncline{n2'}{l2'}
    \endpsmatrix\]

A given game is said to have a Nash equilibrium if there exists a Nash equilibrium whose underlying game is the given game. In order to prove that every sequential game has a Nash equilibrium, one can use a construction called ``backward induction'' in game theory. It consists in building a strategy profile from a sequential game. Performing ``backward induction'' on the following example will help describe and interpret the idea of the construction.

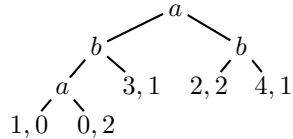
\[\psmatrix
      & &&&&[name=n] a\\
      & &&[name=n1] b &&&&[name=n2] b \\
      & &[name=n11] a &&[name=l12] 3,1 &&[name=l21] 2,2 &&[name=l22] 4,1\\
      & [name=l111] 1,0 &&[name=l112] 0,2\\
      \ncline{n}{n1}
      \ncline{n}{n2}
      \ncline{n1}{n11}
      \ncline{n1}{l12}
      \ncline{n2}{l21}
      \ncline{n2}{l22}
      \ncline{n11}{l111}
      \ncline{n11}{l112}
    \endpsmatrix\]

 If a play starts at the leftmost and lowest node of the game above, then agent $a$ faces the following game:

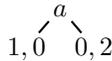
\[\psmatrix
      &&[name=n] a \\
      &[name=l1] 1,0 &&[name=l2] 0,2
      \ncline{n}{l1}
      \ncline{n}{l2}
    \endpsmatrix\]

So, provided that $a$ is ``rational'' in some informal sense, he chooses left and get 1 instead of 0. In the same way, if the play starts at the rightmost node, $b$ chooses left and get 2 instead of 1. These two remarks correspond to the leftmost picture below. Provided that agent $b$ is aware of $a$'s ``rational'' behaviour, if a play starts at the left node owned by $b$, then $b$ chooses right and get 1 instead of 0, as shown on the second picture below. The last picture shows that when a play starts at the root, as in the usual interpretation of a sequential game, $a$ chooses left and gets 3 instead of 2. In such a process, an agent facing several options equivalent in terms of payoffs may choose either of them.  

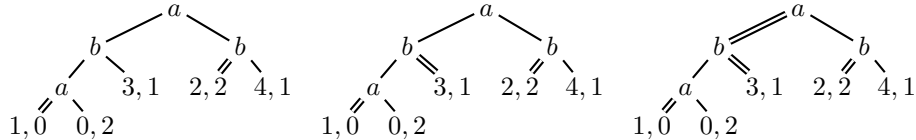
\[\psmatrix
      & &&&&[name=n] a &&&& & &&&& [name=n'] a &&&& & &&&& [name=n''] a \\
      & &&[name=n1] b &&&&[name=n2] b && & &&[name=n1'] b &&&&[name=n2'] b
      && & &&[name=n1''] b &&&&[name=n2''] b  \\
      & &[name=n11] a &&[name=l12] 3,1 &&[name=l21] 2,2 &&[name=l22] 4,1 &\phantom{a}
      & &[name=n11'] a &&[name=l12'] 3,1 &&[name=l21'] 2,2 &&[name=l22'] 4,1&\phantom{a}
      & &[name=n11''] a &&[name=l12''] 3,1 &&[name=l21''] 2,2 &&[name=l22''] 4,1\\
      & [name=l111] 1,0 &&[name=l112] 0,2 &&&&&& & [name=l111'] 1,0 &&[name=l112'] 0,2
      &&&&&& & [name=l111''] 1,0 &&[name=l112''] 0,2\\
      \ncline{n}{n1}
      \ncline{n}{n2}
      \ncline{n1}{n11}
      \ncline{n1}{l12}
      \choiceline{n2}{l21}
      \ncline{n2}{l22}
      \choiceline{n11}{l111}
      \ncline{n11}{l112}
      \ncline{n'}{n1'}
      \ncline{n'}{n2'}
      \ncline{n1'}{n11'}
      \choiceline{n1'}{l12'}
      \choiceline{n2'}{l21'}
      \ncline{n2'}{l22'}
      \choiceline{n11'}{l111'}
      \ncline{n11'}{l112'}
      \choiceline{n''}{n1''}
      \ncline{n''}{n2''}
      \ncline{n1''}{n11''}
      \choiceline{n1''}{l12''}
      \choiceline{n2''}{l21''}
      \ncline{n2''}{l22''}
      \choiceline{n11''}{l111''}
      \ncline{n11''}{l112''}
    \endpsmatrix\]

A strategy profile built by ``backward induction'' is a Nash equilibrium whose underlying game is the original game. (A formal proof relying on formal definitions is presented in a later section.) This way, it is proved that all sequential games have Nash equilibria. However, the next example shows that not all Nash equilibria are obtained by ``backward induction''. Even stronger, a Nash equilibrium may induce a payoff function induced by no ``backward induction'' strategy profile. Indeed, the left-hand strategy profile below is a Nash equilibrium that is not a ``backward induction'', and the only ``backward induction'' on the same underlying game is shown on the right-hand side.

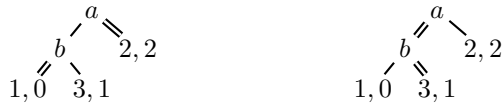
\[\psmatrix
      &&&[name=n1] a&& &&&[name=n1'] a \\
      &&[name=n2] b &&[name=l3]  2,2& &&[name=n2'] b &&[name=l3']  2,2 \\
      &[name=l1] 1,0 &&[name=l2] 3,1 && \phantom{aaaaaaaaaaaaa}&[name=l1'] 1,0 &&[name=l2'] 3,1
      \ncline{n1}{n2}
      \choiceline{n1}{l3}
      \choiceline{n2}{l1}
      \ncline{n2}{l2}
      \choiceline{n1'}{n2'}
      \ncline{n1'}{l3'}
      \ncline{n2'}{l1'}
      \choiceline{n2'}{l2'}
    \endpsmatrix\]

Traditional game theory uses the following definition of subtree: all the descendants of a node of a tree define a subtree. For instance consider the following game.

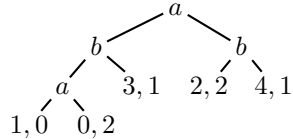
\[\psmatrix
      & &&&&[name=n] a\\
      & &&[name=n1] b &&&&[name=n2] b \\
      & &[name=n11] a &&[name=l12] 3,1 &&[name=l21] 2,2 &&[name=l22] 4,1\\
      & [name=l111] 1,0 &&[name=l112] 0,2\\
      \ncline{n}{n1}
      \ncline{n}{n2}
      \ncline{n1}{n11}
      \ncline{n1}{l12}
      \ncline{n2}{l21}
      \ncline{n2}{l22}
      \ncline{n11}{l111}
      \ncline{n11}{l112}
    \endpsmatrix\]

In addition to the leaves, the proper subtrees of the game above are listed below.

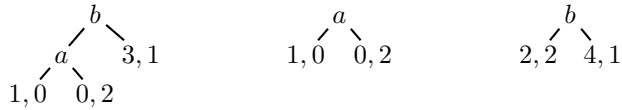
\[\psmatrix  
      & &&[name=n] b &&&&[name=n'] a  &&&&[name=n''] b\\
      & &[name=n1] a &&[name=l3] 3,1 &\phantom{aaaaaaaa}& [name=l1'] 1,0 &&[name=l2'] 0,2 &\phantom{aaaaaaaa}& [name=l1''] 2,2 &&[name=l2''] 4,1\\
      & [name=l1] 1,0 &&[name=l2] 0,2\\
      \ncline{n}{n1}
      \ncline{n}{l3}
      \ncline{n1}{l1}
      \ncline{n1}{l2}
      \ncline{n'}{l1'}
      \ncline{n'}{l2'}
      \ncline{n''}{l1''}
      \ncline{n''}{l2''}
    \endpsmatrix\]

With this definition of subtree, a subgame perfect equilibrium is defined as a Nash equilibrium all of whose substrategy profiles are also Nash equilibria. In traditional game theory, the notions of subgame perfect equilibrium and ``backward induction'' strategy profile coincide.

\section{Why Total Order?\label{sect:wto}}

Section \ref{sect:tsgt} only deals with real-valued, totally ordered payoffs. However, as mentioned in subsection \ref{subsect:ou}, there is a need for game theoretic results involving non totally ordered payoffs. This section adds simple and informal arguments to the discussion. In particular, it shows that the class of traditional sequential games naturally induces two classes of games slightly more general than itself. For these classes of games, the question whether Nash equilibria/subgame perfect equilibria exist or not is still relevant, and has yet not been addressed by Kuhn's result.

\subsection{Selfishness Refinements}\label{subsect:self}

An agent that gives priority to his own payoffs without taking other agents into consideration is called selfish. It is the case in traditional game theory. Now consider a benevolent agent that takes all agents, including him, into account, \textit{e.g.}, in a Pareto style. More specifically, consider two payoff functions $p$ and $p'$. If for each agent, $p$ grants a payoff greater than or equal to the one granted by $p'$, and if there exists one agent to whom $p$ grants strictly greater payoff than $p'$, then the benevolent agent prefers $p$ to $p'$. For instance, consider three agents $a$, $b$ and $c$, and three payoff functions $(1,3,0)$, $(1,2,1)$, and $(1,2,0)$. (The first component corresponds to $a$, the second to $b$, and the third to $c$.) A benevolent agent prefers the first two to the last one, but has no preference among the first two.

An agent is selfish-benevolent if his preference is the union of selfish and benevolent preferences. Put otherwise, an agent prefers a payoff function to another one if he prefers it either selfishly or benevolently. For instance, consider the previous example. Assume that all agents are selfish-benevolent. So, $b$ prefers $(1,3,0)$ to $(1,2,1)$ since $3$ is greater than $2$, prefers $(1,2,1)$ to $(1,2,0)$ by benevolence, and prefers $(1,3,0)$ to $(1,2,0)$ by both selfishness and benevolence. Selfishness-benevolence induces a partial order over real-valued payoff functions.

In the same way, an agent may be selfish-malevolent. More specifically, consider two payoff functions $p$ and $p'$. If $p$ grants a selfish-malevolent agent a payoff greater than the one granted by $p'$ then the agent prefers $p$ to $p'$. If $p$ and $p'$ grant the same payoff to the selfish-malevolent agent, and if $p$ grants every other agent a payoff lesser than or equal to the one granted by $p'$, and if there exists one agent to whom $p$ grants strictly lesser payoff than $p'$, then the selfish-malevolent agent prefers $p$ to $p'$. Selfishness-malevolence induces a partial order over real-valued payoff functions too.

\subsection{Lack of Information}

Consider an agent that prefers greater payoffs and that is ``rational'', \textit{i.e.}, acts according to his preferences. Imagine the following 1-player game played by the above-mentioned agent: when a play starts, the agent has two options, say left and right. If he chooses left then he gets either $0$ or $5$, and if he chooses right then he gets either $1$ or $2$ or $3$. After payoffs are granted, the play ends. This game is represented below.

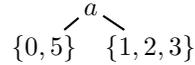
\[\psmatrix
      &&[name=a] a\\
      &[name=leafOne] \{0,5\} && [name=leafTwo] \{1,2,3\} \\
      \ncline{a}{leafOne}
      \ncline{a}{leafTwo}
    \endpsmatrix\]

The wording ``either... or...'', in the phrase ``either $0$ or $5$'' does not refer to any procedure whatsoever. Therefore, in each case the agent has no clue how payoffs are going to be granted. It is worth stressing that, in particular, ``either $0$ or $5$'' does not refer to probability half for $0$ and probability half for $5$. It does not refer to probabilities at all. As a result, the agent cannot dismiss for sure any of his options in the game above. The two options are not comparable; the payoffs are not totally ordered. This type of game can even help understand traditional sequential games better. Indeed, consider the traditional sequential game below, where $a$ and $b$ are ``rational'', and therefore only care about maximising their own payoffs. Also assume that $a$ knows $b$'s being ``rational''.

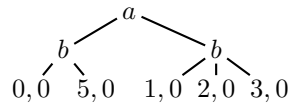
\[\psmatrix
      &&&&[name=n0] a \\
      &&[name=n1] b &&&&&[name=n2] b  \\
      &[name=l11] 0,0 &&[name=l12] 5,0 &&[name=l21] 1,0 &&[name=l22] 2,0&&[name=l23] 3,0\\     
      \ncline{n0}{n1}
      \ncline{n0}{n2}
      \ncline{n1}{l11}
      \ncline{n1}{l12}
      \ncline{n2}{l21}
      \ncline{n2}{l22}
      \ncline{n2}{l23}
  \endpsmatrix\]

Whatever agent $a$ may choose, $b$'s options are equivalent since they yield the same payoff. Therefore $a$ has no clue how $b$ is going to choose: Go left when left and right are equivalent? Toss a coin? Phone a friend? So, from agent $a$'s viewpoint, the game above reduces to the 1-player game discussed before. In this subsection, non totally ordered payoffs represents lack of information without invoking a formalism dedicated to knowledge representation such as epistemic logic.

When payoffs are non-empty sets of real numbers instead of single real numbers, as in the first example above, there are several ways to define relevant preferences. For instance, agents can focus either on the lower bound of the set, which amounts to guaranteeing a minimum, or on the upper bound of the set, which amounts to letting hope for a maximum, or both at once. Most of these yield partial orders over non-empty sets of reals.

\section{Preliminaries\label{sect:pre}}

Prior to the game theoretic development presented in Coq in later sections, a few general concepts and related results are needed. Part of them are mentioned in~\cite{SLR2007-14}: list inclusion, subrelation, restriction of a relation to the elements of a list, etc. This section completes the inventory of the required notions, in the Coq formalism.

A first useful result reads as follows: given a binary relation and two lists, one included in the other, the restriction of the relation to the smaller list is a subrelation of the restriction of the relation to the bigger list. The proof is a straightforward unfolding of the definitions, and the result is formally written below.

\medskip
\noindent
\coqdockw{Lemma} \coqdocid{sub\_rel\_restriction\_incl} : \ensuremath{\forall} (\coqdocid{B} : \coqdocid{Set})(\coqdocid{l} \coqdocid{l'} : \coqdocid{list} \coqdocid{B}) \coqdocid{R}, \coqdoceol
\noindent
\coqdocid{incl} \coqdocid{l} \coqdocid{l'} \ensuremath{\rightarrow} \coqdocid{sub\_rel} (\coqdocid{restriction} \coqdocid{R} \coqdocid{l}) (\coqdocid{restriction} \coqdocid{R} \coqdocid{l'}).\coqdoceol
\medskip

The remainder of this section presents the extension to lists of four concepts usually pertaining to one or two objects only.

\subsection{Extension of Predicates to Lists}

Let \coqdocid{A} be a \coqdocid{Set}. The function \coqdocid{listforall} expects a predicate on \coqdocid{A}, \textit{i.e.}, an object of type \coqdocid{A} \ensuremath{\rightarrow} \coqdocid{Prop}, and returns a predicate on lists, \textit{i.e.}, an object of type \coqdocid{list} \coqdocid{A} \ensuremath{\rightarrow} \coqdocid{Prop}, stating that all the elements in the list comply with the original predicate. It is recursively defined along the inductive structure of the list argument.

\medskip
\noindent
\coqdockw{Fixpoint} \coqdocid{listforall} (\coqdocid{Q} : \coqdocid{A} \ensuremath{\rightarrow} \coqdocid{Prop})(\coqdocid{l} : \coqdocid{list} \coqdocid{A})\{\coqdocid{struct} \coqdocid{l}\} : \coqdocid{Prop} :=\coqdoceol
\noindent
\coqdocid{match} \coqdocid{l} \coqdocid{with}\coqdoceol
\noindent
$\mid$ \coqdocid{nil} \ensuremath{\Rightarrow} \coqdocid{True}\coqdoceol
\noindent
$\mid$ \coqdocid{x}::\coqdocid{l'} \ensuremath{\Rightarrow} \coqdocid{Q} \coqdocid{x} \ensuremath{\land} \coqdocid{listforall} \coqdocid{Q} \coqdocid{l'}\coqdoceol
\noindent
\coqdocid{end}.\coqdoceol

In the first line above, \coqdockw{Fixpoint} starts the recursive definition, \coqdocid{listforall} is the name of the defined function, \coqdocid{Q} is the predicate argument, \coqdocid{l} is the list argument, \{\coqdocid{struct} \coqdocid{l}\} means that the recursion involves \coqdocid{l}, and \coqdocid{Prop} is the type of the output. The \coqdocid{match} performs a case splitting on the structure of \coqdocid{l}. The line thereafter specifies that the function returns \coqdocid{True} for empty list arguments. The last line is the core of the recursion: in order to compute the result for the list, it refers to the result of the computation involving the tail, which is a strictly smaller argument. This ensures termination of the computation, therefore the function is well defined. An example of computation of \coqdocid{listforall} is given below. The symbol $\leadsto$ represents a computation step. 

\medskip
\noindent
\coqdocid{listforall} \coqdocid{Q} (\coqdocid{x}::\coqdocid{y}::\coqdocid{z}::\coqdocid{nil}) $\leadsto$ 
\coqdocid{Q} \coqdocid{x} \ensuremath{\land} \coqdocid{listforall} \coqdocid{Q} (\coqdocid{y}::\coqdocid{z}::\coqdocid{nil}) $\leadsto$\\
\coqdocid{Q} \coqdocid{x} \ensuremath{\land} \coqdocid{Q} \coqdocid{y} \ensuremath{\land} \coqdocid{listforall} \coqdocid{Q} (\coqdocid{z}::\coqdocid{nil}) $\leadsto$
\coqdocid{Q} \coqdocid{x} \ensuremath{\land} \coqdocid{Q} \coqdocid{y} \ensuremath{\land} \coqdocid{Q} \coqdocid{z} \ensuremath{\land} \coqdocid{listforall} \coqdocid{Q} \coqdocid{nil} $\leadsto$\\
\coqdocid{Q} \coqdocid{x} \ensuremath{\land} \coqdocid{Q} \coqdocid{y} \ensuremath{\land} \coqdocid{Q} \coqdocid{z} \ensuremath{\land} \coqdocid{True}
\medskip

Note that \coqdocid{Q} \coqdocid{x} \ensuremath{\land} \coqdocid{Q} \coqdocid{y} \ensuremath{\land} \coqdocid{Q} \coqdocid{z} \ensuremath{\land} \coqdocid{True} is equivalent to \coqdocid{Q} \coqdocid{x} \ensuremath{\land} \coqdocid{Q} \coqdocid{y} \ensuremath{\land} \coqdocid{Q} \coqdocid{z}, which is what the function \coqdocid{listforall} is meant for.

The following four lemmas involve the notion of appending (++), also called concatenation, of two lists. It is defined in the Coq Standard Library. The four lemmas express basic properties of the \coqdocid{listforall} function. They are all proved by induction on the list \coqdocid{l}.

\medskip
\noindent
\coqdockw{Lemma} \coqdocid{listforall\_app} :  \ensuremath{\forall} \coqdocid{Q} \coqdocid{l'} \coqdocid{l}, \coqdoceol
\noindent
\coqdocid{listforall} \coqdocid{Q} \coqdocid{l} \ensuremath{\rightarrow} \coqdocid{listforall} \coqdocid{Q} \coqdocid{l'} \ensuremath{\rightarrow} \coqdocid{listforall} \coqdocid{Q} (\coqdocid{l}++\coqdocid{l'}). \coqdoceol

\medskip
\noindent
\coqdockw{Lemma} \coqdocid{listforall\_appl} :  \ensuremath{\forall} \coqdocid{Q} \coqdocid{l'} \coqdocid{l}, \coqdocid{listforall} \coqdocid{Q} (\coqdocid{l}++\coqdocid{l'}) \ensuremath{\rightarrow} \coqdocid{listforall} \coqdocid{Q} \coqdocid{l}. \coqdoceol

\medskip
\noindent
\coqdockw{Lemma} \coqdocid{listforall\_appr} :  \ensuremath{\forall} \coqdocid{Q} \coqdocid{l'} \coqdocid{l}, \coqdocid{listforall} \coqdocid{Q} (\coqdocid{l}++\coqdocid{l'}) \ensuremath{\rightarrow} \coqdocid{listforall} \coqdocid{Q} \coqdocid{l'}. \coqdoceol

\medskip
\noindent
\coqdockw{Lemma} \coqdocid{listforall\_In} : \ensuremath{\forall} \coqdocid{Q} \coqdocid{x} \coqdocid{l}, \coqdocid{In} \coqdocid{x} \coqdocid{l} \ensuremath{\rightarrow} \coqdocid{listforall} \coqdocid{Q} \coqdocid{l} \ensuremath{\rightarrow} \coqdocid{Q} \coqdocid{x}. \coqdoceol
\medskip

\subsection{Extension of Functions to Lists}

The Coq Standard Library provides a function \coqdocid{map} that, given a list and a function \coqdocid{f}, returns a list with the images by \coqdocid{f} of the elements of the original list. It is defined by recursion.

\medskip
\noindent
\coqdockw{Fixpoint} \coqdocid{map} (\coqdocid{A} \coqdocid{B} : \coqdocid{Set})(\coqdocid{f} : \coqdocid{A} \ensuremath{\rightarrow} \coqdocid{B})(\coqdocid{l} : \coqdocid{list} \coqdocid{A}) : \coqdocid{list} \coqdocid{B} :=\coqdoceol
\coqdocindent{1.00em}
\coqdocid{match} \coqdocid{l} \coqdocid{with}\coqdoceol
\coqdocindent{1.00em}
$\mid$ \coqdocid{nil} \ensuremath{\Rightarrow} \coqdocid{nil}\coqdoceol
\coqdocindent{1.00em}
$\mid$ \coqdocid{a}::\coqdocid{t} \ensuremath{\Rightarrow} (\coqdocid{f} \coqdocid{a})::(\coqdocid{map} \coqdocid{A} \coqdocid{B} \coqdocid{f} \coqdocid{t})\coqdoceol
\coqdocindent{1.00em}
\coqdocid{end}.\coqdoceol
\medskip

Consider the simplified computation example below, where domains and codomains of \coqdocid{f} are omitted for better readability.

\medskip
\noindent
\coqdocid{map} \coqdocid{f} (\coqdocid{x}::\coqdocid{y}::\coqdocid{z}::\coqdocid{nil}) $\leadsto\dots\leadsto$ (\coqdocid{f} \coqdocid{x})::(\coqdocid{f} \coqdocid{y})::(\coqdocid{f} \coqdocid{z})::\coqdocid{nil}\coqdoceol
\medskip

The next two lemmas state \coqdocid{map}'s preserving two functions being inverse and commutativity between \coqdocid{map} and appending. Both lemmas are proved by induction on the list \coqdocid{l}. The second one comes from the Coq Standard Library~\cite{coq}.

\medskip
\noindent
\coqdockw{Lemma} \coqdocid{map\_inverse} : \ensuremath{\forall} (\coqdocid{A} \coqdocid{B} : \coqdocid{Set})(\coqdocid{f} : \coqdocid{A} \ensuremath{\rightarrow} \coqdocid{B}) \coqdocid{g},\coqdoceol
\noindent
(\ensuremath{\forall} \coqdocid{x}, \coqdocid{g} (\coqdocid{f} \coqdocid{x})=\coqdocid{x}) \ensuremath{\rightarrow} \ensuremath{\forall} \coqdocid{l}, \coqdocid{map} \coqdocid{g} (\coqdocid{map} \coqdocid{f} \coqdocid{l})=\coqdocid{l}.\coqdoceol
\medskip

\noindent
\coqdockw{Lemma} \coqdocid{map\_app} : \ensuremath{\forall} (\coqdocid{A} \coqdocid{B} : \coqdocid{Set})\coqdocid{l} \coqdocid{l'} (\coqdocid{f} : \coqdocid{A} \ensuremath{\rightarrow} \coqdocid{B}), \coqdocid{map} \coqdocid{f} (\coqdocid{l}++\coqdocid{l'}) = (\coqdocid{map} \coqdocid{f} \coqdocid{l})++(\coqdocid{map} \coqdocid{f} \coqdocid{l'}).\coqdoceol
\medskip

\subsection{Extension of Binary Relations to Lists}

Let \coqdocid{A} be a \coqdocid{Set}. Given a binary relation \coqdocid{P} : \coqdocid{A} \ensuremath{\rightarrow} \coqdocid{A} \ensuremath{\rightarrow} \coqdocid{Prop}, the function \coqdocid{rel\_vector} expects two lists over \coqdocid{A} and states that they are component-wise related by \coqdocid{P}. Note that if they are component-wise related then their lengths are the same.

\medskip
\noindent
\coqdockw{Fixpoint} \coqdocid{rel\_vector} (\coqdocid{P} : \coqdocid{A} \ensuremath{\rightarrow} \coqdocid{A} \ensuremath{\rightarrow} \coqdocid{Prop})(\coqdocid{l} \coqdocid{l'} : \coqdocid{list} \coqdocid{A}) \{\coqdocid{struct} \coqdocid{l}\}: \coqdocid{Prop} :=\coqdoceol
\noindent
\coqdocid{match} \coqdocid{l} \coqdocid{with}\coqdoceol
\noindent
$\mid$ \coqdocid{nil} \ensuremath{\Rightarrow} \coqdocid{l'}=\coqdocid{nil}\coqdoceol
\noindent
$\mid$ \coqdocid{x}::\coqdocid{l2} \ensuremath{\Rightarrow} \coqdocid{match} \coqdocid{l'} \coqdocid{with}\coqdoceol
\coqdocindent{7.50em}
$\mid$ \coqdocid{nil} \ensuremath{\Rightarrow} \coqdocid{False}\coqdoceol
\coqdocindent{7.50em}
$\mid$ \coqdocid{x'}::\coqdocid{l2'} \ensuremath{\Rightarrow} \coqdocid{P} \coqdocid{x} \coqdocid{x'} \ensuremath{\land} \coqdocid{rel\_vector} \coqdocid{P} \coqdocid{l2} \coqdocid{l2'}\coqdoceol
\coqdocindent{7.50em}
\coqdocid{end}\coqdoceol
\noindent
\coqdocid{end}.\coqdoceol
\medskip

The following examples describe three typical computations of the \coqdocid{rel\_vector} predicate.

\medskip
\noindent
\coqdocid{rel\_vector} \coqdocid{P} (\coqdocid{x}::\coqdocid{y}::\coqdocid{nil}) (\coqdocid{x'}::\coqdocid{y'}::\coqdocid{nil}) $\leadsto$ \coqdocid{P} \coqdocid{x} \coqdocid{x'} \ensuremath{\land} \coqdocid{rel\_vector} \coqdocid{P} (\coqdocid{y}::\coqdocid{nil}) (\coqdocid{y'}::\coqdocid{nil}) $\leadsto$ \coqdocid{P} \coqdocid{x} \coqdocid{x'} \ensuremath{\land} \coqdocid{P} \coqdocid{y} \coqdocid{y'} \ensuremath{\land} \coqdocid{rel\_vector} \coqdocid{P} (\coqdocid{nil}) (\coqdocid{nil}) $\leadsto$ \coqdocid{P} \coqdocid{x} \coqdocid{x'} \ensuremath{\land} \coqdocid{P} \coqdocid{y} \coqdocid{y'} \ensuremath{\land} \coqdocid{nil}=\coqdocid{nil}
\medskip

Note that \coqdocid{P} \coqdocid{x} \coqdocid{x'} \ensuremath{\land} \coqdocid{P} \coqdocid{y} \coqdocid{y'} \ensuremath{\land} \coqdocid{nil}=\coqdocid{nil} is equivalent to \coqdocid{P} \coqdocid{x} \coqdocid{x'} \ensuremath{\land} \coqdocid{P} \coqdocid{y} \coqdocid{y'}.

\medskip
\noindent
\coqdocid{rel\_vector} \coqdocid{P} (\coqdocid{x}::\coqdocid{y}::\coqdocid{nil}) (\coqdocid{x'}::\coqdocid{nil}) $\leadsto$ \coqdocid{P} \coqdocid{x} \coqdocid{x'} \ensuremath{\land} \coqdocid{rel\_vector} \coqdocid{P} (\coqdocid{y}::\coqdocid{nil}) (\coqdocid{nil}) $\leadsto$\\
\coqdocid{P} \coqdocid{x} \coqdocid{x'} \ensuremath{\land} \coqdocid{False} 
\medskip

Note that \coqdocid{P} \coqdocid{x} \coqdocid{x'} \ensuremath{\land} \coqdocid{False} is equivalent to \coqdocid{False}.

\medskip
\noindent
\coqdocid{rel\_vector} \coqdocid{P} (\coqdocid{x}::\coqdocid{nil}) (\coqdocid{x'}::\coqdocid{y'}\coqdocid{nil}) $\leadsto$ \coqdocid{P} \coqdocid{x} \coqdocid{x'} \ensuremath{\land} \coqdocid{rel\_vector} \coqdocid{P} (\coqdocid{nil}) (\coqdocid{y'}::\coqdocid{nil}) $\leadsto$\\
\coqdocid{P} \coqdocid{x} \coqdocid{x'} \ensuremath{\land} \coqdocid{y'}::\coqdocid{nil}=\coqdocid{nil}\coqdoceol
\medskip

Note that \coqdocid{P} \coqdocid{x} \coqdocid{x'} \ensuremath{\land} \coqdocid{y'}::\coqdocid{nil}=\coqdocid{nil} is equivalent to \coqdocid{False} since \coqdocid{y'}::\coqdocid{nil}=\coqdocid{nil} is equivalent to \coqdocid{False}.

The following lemma states that if two lists are component-wise related, then two elements occurring at the same place in each list are also related.

\medskip
\noindent
\coqdockw{Lemma} \coqdocid{rel\_vector\_app\_cons\_same\_length} : \ensuremath{\forall} \coqdocid{P} \coqdocid{a} \coqdocid{a'} \coqdocid{m} \coqdocid{m'} \coqdocid{l} \coqdocid{l'}, \coqdoceol
\noindent
\coqdocid{rel\_vector} \coqdocid{P} (\coqdocid{l}++\coqdocid{a}::\coqdocid{m}) (\coqdocid{l'}++\coqdocid{a'}::\coqdocid{m'}) \ensuremath{\rightarrow} \coqdocid{length} \coqdocid{l}=\coqdocid{length} \coqdocid{l'} \ensuremath{\rightarrow} \coqdocid{P} \coqdocid{a} \coqdocid{a'}. \coqdoceol
\medskip

\begin{proof}
Let \coqdocid{P} be a binary relation over \coqdocid{A}, let \coqdocid{a} and \coqdocid{a'} be in \coqdocid{A}, and let \coqdocid{m} and \coqdocid{m'} be lists over \coqdocid{A}. Prove \ensuremath{\forall} \coqdocid{l} \coqdocid{l'}, \coqdocid{rel\_vector} \coqdocid{P} (\coqdocid{l}++\coqdocid{a}::\coqdocid{m}) (\coqdocid{l'}++\coqdocid{a'}::\coqdocid{m'}) \ensuremath{\rightarrow} \coqdocid{length} \coqdocid{l}=\coqdocid{length} \coqdocid{l'} \ensuremath{\rightarrow} \coqdocid{P} \coqdocid{a} \coqdocid{a'} by induction on \coqdocid{l}. For the inductive case, implying that \coqdocid{l'} is non-empty, apply the induction hypothesis with the tail of \coqdocid{l'}.
\end{proof}

The next result shows that if two lists are component-wise related, then given one element in the second list, one can compute an element of the first list such that both elements are related.

\medskip
\noindent
\coqdockw{Lemma} \coqdocid{rel\_vector\_app\_cons\_exists} : \ensuremath{\forall} \coqdocid{P} \coqdocid{a} \coqdocid{q} \coqdocid{l} \coqdocid{m}, \coqdoceol
\noindent
\coqdocid{rel\_vector} \coqdocid{P} \coqdocid{l} (\coqdocid{m}++\coqdocid{a}::\coqdocid{q}) \ensuremath{\rightarrow} \{\coqdocid{x} : \coqdocid{A} $\mid$ \coqdocid{In} \coqdocid{x} \coqdocid{l} \ensuremath{\land} \coqdocid{P} \coqdocid{x} \coqdocid{a}\}.\coqdoceol
\medskip

\begin{proof}
By induction on \coqdocid{l} and case splitting on \coqdocid{m} being empty. For the inductive case, if \coqdocid{m} is empty then the head of \coqdocid{l} is a witness, if \coqdocid{m} is not empty then use the induction hypothesis with the tail of \coqdocid{m}; the computable element is a witness. 
\end{proof}

The following lemma says that if two lists are component-wise related, then given one element in the first list and one in the second list, but at different places, there is another element in the first list, either before or after the element mentioned first, that is related to the element of the second list.

\medskip
\noindent
\coqdockw{Lemma} \coqdocid{rel\_vector\_app\_cons\_different\_length} : \ensuremath{\forall} \coqdocid{P} \coqdocid{l} \coqdocid{a} \coqdocid{m} \coqdocid{l'} \coqdocid{a'} \coqdocid{m'},\coqdoceol
\noindent
\coqdocid{rel\_vector} \coqdocid{P} (\coqdocid{l}++\coqdocid{a}::\coqdocid{m}) (\coqdocid{l'}++\coqdocid{a'}::\coqdocid{m'}) \ensuremath{\rightarrow} \coqdocid{length} \coqdocid{l}\ensuremath{\not=}\coqdocid{length} \coqdocid{l'} \ensuremath{\rightarrow}  \{\coqdocid{x} : \coqdocid{A} $\mid$ (\coqdocid{In} \coqdocid{x} \coqdocid{l} \ensuremath{\lor} \coqdocid{In} \coqdocid{x} \coqdocid{m}) \ensuremath{\land} \coqdocid{P} \coqdocid{x} \coqdocid{a'}\}.\coqdoceol
\medskip

\begin{proof}
By induction on \coqdocid{l}. For the base case, \coqdocid{l} is empty, if \coqdocid{l'} is empty then it is straightforward, if \coqdocid{l'} is not empty then applying \coqdocid{rel\_vector\_app\_cons\_exists} gives a witness. For the inductive case, \coqdocid{l} is not empty, case split on \coqdocid{l'} being empty and use the induction hypothesis when \coqdocid{l'} is not empty.
\end{proof}

\subsection{No Successor\label{subsect:ns}}

Let \coqdocid{A} be a \coqdocid{Set}. Given a binary relation, the predicate \coqdocid{is\_no\_succ} returns a proposition saying that a given element is the predecessor of no element in a given list. 

\medskip
\noindent
\coqdockw{Definition} \coqdocid{is\_no\_succ} (\coqdocid{P} : \coqdocid{A} \ensuremath{\rightarrow} \coqdocid{A} \ensuremath{\rightarrow} \coqdocid{Prop})(\coqdocid{x} : \coqdocid{A})(\coqdocid{l} : \coqdocid{list} \coqdocid{A}) := \coqdoceol
\noindent
\coqdocid{listforall} (\coqdocid{fun} \coqdocid{y} \ensuremath{\rightarrow} \ensuremath{\lnot}\coqdocid{P} \coqdocid{x} \coqdocid{y}) \coqdocid{l}.\coqdoceol
\medskip

The next two results show a transitivity property and decidability of \coqdocid{is\_no\_succ} when built on a decidable binary relation. Both are proved by induction on the list \coqdocid{l}.

\medskip
\noindent
\coqdockw{Lemma} \coqdocid{is\_no\_succ\_trans} : \ensuremath{\forall} \coqdocid{P} \coqdocid{x} \coqdocid{y} \coqdocid{l},\coqdoceol
\noindent
\coqdocid{transitive} \coqdocid{A} \coqdocid{P} \ensuremath{\rightarrow} \coqdocid{P} \coqdocid{x} \coqdocid{y} \ensuremath{\rightarrow} \coqdocid{is\_no\_succ} \coqdocid{P} \coqdocid{x} \coqdocid{l} \ensuremath{\rightarrow} \coqdocid{is\_no\_succ} \coqdocid{P} \coqdocid{y} \coqdocid{l}. \coqdoceol\medskip

\medskip
\noindent
\coqdockw{Lemma} \coqdocid{is\_no\_succ\_dec} : \ensuremath{\forall} \coqdocid{P}, \coqdocid{rel\_dec} \coqdocid{P} \ensuremath{\rightarrow} \ensuremath{\forall} \coqdocid{x} \coqdocid{l},\coqdoceol
\noindent
\{\coqdocid{is\_no\_succ} \coqdocid{P} \coqdocid{x} \coqdocid{l}\}+\{\ensuremath{\lnot}\coqdocid{is\_no\_succ} \coqdocid{P} \coqdocid{x} \coqdocid{l}\}.\coqdoceol
\medskip

The following lemma helps generalise the notion of ``backward induction''  in section~\ref{sect:ee}. Assume \coqdocid{P} a decidable binary relation over \coqdocid{A}, and \coqdocid{x}::\coqdocid{l} a non-empty list over \coqdocid{A}. The list \coqdocid{x}::\coqdocid{l} can be computabily split into a left list, a chosen element, and a right list such that 1) the chosen element has no \coqdocid{P}-successor in the right list, 2) the chosen element is the first (from left to right) element with such a property, and moreover 3) if \coqdocid{P} is irreflexive and transitive then the chosen element has no \coqdocid{P}-successor in the left list either. The form of the statement has being slightly simplified, as compared to the actual Coq code. The first conjunct corresponds to the splitting of the list \coqdocid{x}::\coqdocid{l}, and the last three conjuncts correspond to the points 1), 2) and 3)as mentioned above.

\medskip
\noindent
\coqdockw{Lemma} \coqdocid{Choose\_and\_split} : \ensuremath{\forall} \coqdocid{P}, \coqdocid{rel\_dec} \coqdocid{P} \ensuremath{\rightarrow} \ensuremath{\forall} (\coqdocid{l} : \coqdocid{list} \coqdocid{A})(\coqdocid{x} : \coqdocid{A}), \coqdoceol
\noindent
\{(\coqdocid{left},\coqdocid{choice},\coqdocid{right}) : \coqdocid{list} \coqdocid{A} \ensuremath{\times} \coqdocid{A} \ensuremath{\times} \coqdocid{list} \coqdocid{A} $\mid$ \coqdoceol
\medskip
\coqdocid{x}::\coqdocid{l}=(\coqdocid{left}++(\coqdocid{choice}::\coqdocid{right}))\coqdoceol
\ensuremath{\land}\coqdoceol
\coqdocid{is\_no\_succ} \coqdocid{P} \coqdocid{choice} \coqdocid{right}\coqdoceol
\ensuremath{\land} \coqdoceol
(\ensuremath{\forall} \coqdocid{left'} \coqdocid{choice'} \coqdocid{right'}, \coqdocid{left}=\coqdocid{left'}++(\coqdocid{choice'}::\coqdocid{right'}) \ensuremath{\rightarrow}\coqdoceol
\ensuremath{\lnot}\coqdocid{is\_no\_succ} \coqdocid{P} \coqdocid{choice'} (\coqdocid{right'}++(\coqdocid{choice}::\coqdocid{right})))\coqdoceol
\ensuremath{\land}\coqdoceol
(\coqdocid{irreflexive} \coqdocid{P} \ensuremath{\rightarrow} \coqdocid{transitive} \coqdocid{P} \ensuremath{\rightarrow} \coqdocid{is\_no\_succ} \coqdocid{P} \coqdocid{choice} \coqdocid{left})\}.  \coqdoceol
\medskip

\begin{proof}
Assume that \coqdocid{P} is decidable and proceed by induction on \coqdocid{l}, the tail of the non-empty list. For the base case where the tail is empty, \coqdocid{nil}, \coqdocid{x}, and \coqdocid{nil} are witnesses for the required left list, choice element, and right list. For the step case, \coqdocid{l}=\coqdocid{a}::\coqdocid{l'}, case split on \coqdocid{x} having a successor in \coqdocid{l}. If not, then \coqdocid{nil}, \coqdocid{x}, and \coqdocid{l} are a witness. If \coqdocid{x} has a successor in \coqdocid{l} then use the induction hypothesis with \coqdocid{a}, which splits the list \coqdocid{l} into a left list, a choice element, and a right list. Put \coqdocid{x} on the top of the left list, here are the three witnesses.
\end{proof}

For example, consider the partial order induced by divisibility of natural numbers by natural numbers. For the list $2::3::9::4::9::6::2::16::nil$, the \coqdocid{choice} is $9$, the left list is $2::3::9::4::nil$, and the right list is $6::2::16::nil$.

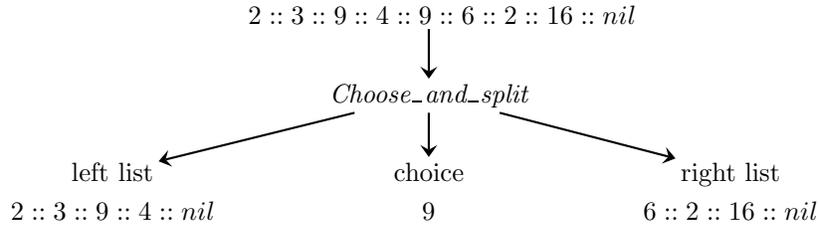
\[\psmatrix
      && [name=n] \quad 2::3::9::4::9::6::2::16::nil \\\\
      &&[name=n0]\coqdocid{Choose\_and\_split}\\\\
      & [name=n1]\mbox{left list} & [name=n2] \mbox{choice} & [name=n3] \mbox{right list}\\ 
      & 2::3::9::4::nil & 9 & 6::2::16::nil
      \ncline[arrows=->]{n}{n0}
      \ncline[arrows=->]{n0}{n1}
      \ncline[arrows=->]{n0}{n2}
      \ncline[arrows=->]{n0}{n3}
      \endpsmatrix\]

Indeed, the first $2$ can divide $4$ written on its right, $3$ can divide $9$ written on its right, the first $9$ can divide the other $9$ written on its right, the $4$ can divide $16$ written on its right, but the second $9$ cannot divide any number written on its right.

\section{Sequential Games\label{sect:sg}}

By abstracting over payoff functions, subsection~\ref{subsect:dsg} generalises the notion of sequential game. In the remainder of this paper, the expression ``sequential game'' refers to the new and abstract sequential games, unless stated otherwise. In addition, subsection~\ref{subsect:dsg} introduces a new formalism for sequential games. In subsection~\ref{subsect:ipsg}, an inductive proof principle is designed for these sequential games. Last, subsection~\ref{subsect:ousg} defines a function related to sequential games.

\subsection{Definition of Sequential Games\label{subsect:dsg}}

This subsection first presents sequential games informally in the way traditional sequential games were presented in section~\ref{sect:tsgt}. Then it describes sequential games in a both inductive and graphical formalism, and makes a link with the traditional formalism. Last, the inductive and graphical formalism is naturally translated into a definition of sequential games in Coq.

\subsubsection{The Traditional Way:}

Informally, consider a collection of outcomes that are the possible end-of-the-game situations, and a collection of agents that are the possible stake-holders and decision makers. Roughly speaking, an abstract sequential game is a traditional sequential game where each real-valued  payoff function (enclosed in a leaf) has been replaced by an outcome. Below, the left-hand picture represents a traditional sequential game, and the right-hand picture represents an abstract sequential game on the same tree structure.

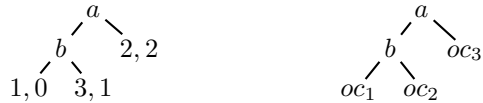
\[\psmatrix
      &&&[name=n1] a&& &&&[name=n1'] a\\
      &&[name=n2] b &&[name=l3]  2,2 &\phantom{aaaaaaaaaaaa}&&[name=n2'] b &&[name=l3'] oc_3 \\
      &[name=l1] 1,0 &&[name=l2] 3,1 && &[name=l1'] oc_1 &&[name=l2'] oc_2
      \ncline{n1}{n2}
      \ncline{n1}{l3}
      \ncline{n2}{l1}
      \ncline{n2}{l2}
      \ncline{n1'}{n2'}
      \ncline{n1'}{l3'}
      \ncline{n2'}{l1'}
      \ncline{n2'}{l2'}
    \endpsmatrix\]

\subsubsection{The Inductive and Graphical Way:}

In what follows, a generic game $g$ will be represented by the left-hand picture below and a generic (possibly empty) list of games by the right-hand picture.

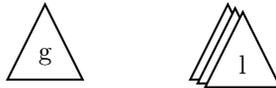
\[\begin{array}{ccc}
\begin{picture}(0,0)
  \rput(0,0){\pspolygon[fillstyle=solid](-.5,-1)(.5,-1)}
  \rput(0,-.7){g}
\end{picture}
  &\phantom{aaaaaaaaaaaa}&
  \begin{picture}(0,0)
    \multiput(0,0)(0.1,-0.05){3}{\pspolygon[fillstyle=solid](-.5,-1)(.5,-1)}
    \rput(0.2,-.8){l}
  \end{picture}
\\\\\\\\
\end{array}\]

Sequential games are inductively defined in two steps. First step: If $oc$ is an outcome then the object below is a game.

\[\begin{array}{c}
  \psframebox[doubleline=true]{oc}
\end{array}\]

Second step: if $a$ is an agent and if the first two objects below are a game and a list of games, then the rightmost object is also a game.

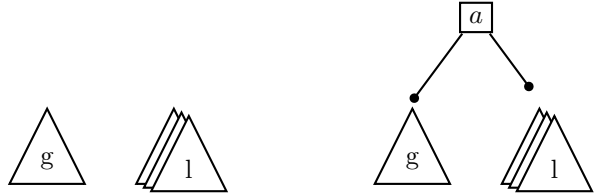
\[
\psmatrix
      &&&&&&[name=n]\psframebox[fillstyle=solid,boxsep=false]{a}\\&
      \begin{picture}(0,0)
	\rput(0,0){\pspolygon[fillstyle=solid](-.5,-1)(.5,-1)}
	\rput(0,-.7){g}
      \end{picture}
      &\phantom{aaaaaaaa}&
      \begin{picture}(0,0)
	\multiput(0,0)(0.1,-0.05){3}{\pspolygon[fillstyle=solid](-.5,-1)(.5,-1)}
	\rput(0.2,-.8){l}
      \end{picture}&\phantom{aaaaaaaaaaaaaaaa}&[name=n1]
      \begin{picture}(0,0)
	\put(0,0){\pspolygon[fillstyle=solid](-.5,-1)(.5,-1)}
	\rput(0,-.7){g}
      \end{picture}
      &\phantom{aaaaaaaa}&
      [name=n2]
      \begin{picture}(0,)
	\multiput(0,0)(0.1,-0.05){3}{\pspolygon[fillstyle=solid](-.5,-1)(.5,-1)}
	\rput(0.2,-.8){l}
      \end{picture}\\\\\\\\      
     \ncline[nodesep=4pt,arrows=-*]{n}{n1}
     \ncline[nodesep=4pt,arrows=-*]{n}{n2}
\endpsmatrix\]

The major difference between the traditional formalism and the new formalism is as follows: In the traditional formalism, an internal node had one or arbitrarily many more children, which are games. In the new formalism instead, an internal node has one left child, which is a game, and one right child, which is a list of games. The next two pictures represent the same game in both formalisms, where for all $i$ between $0$ and $n$, the object $g'_i$ is the translation of $g_i$ from the traditional formalism into the new formalism. 

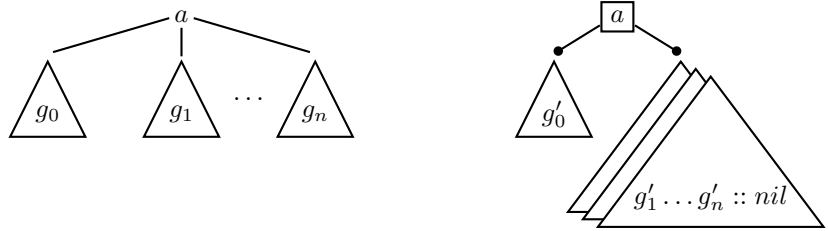
\[\psmatrix
     &&&[name=n] a &&&&& [name=n']\psframebox[fillstyle=solid,boxsep=false]{a}\\&
     [name=l1]\begin{picture}(0,0)
       \put(0,0){\pspolygon[fillstyle=solid](-.5,-1)(.5,-1)}
       \rput(0,-.7){$g_0$}
     \end{picture}&\phantom{aaaaaaaa}&
     [name=l2]\begin{picture}(0,0)
       \put(0,0){\pspolygon[fillstyle=solid](-.5,-1)(.5,-1)}
       \rput(0,-.7){$g_1$}
     \end{picture}&\rput(.7,-.5){\dots}\phantom{aaaaaaaa}&
     [name=l3]\begin{picture}(0,0)
       \put(0,0){\pspolygon[fillstyle=solid](-.5,-1)(.5,-1)}
       \rput(0,-.7){$g_n$}
     \end{picture}&\phantom{aaaaaaaaaaaaaaaa}&
     [name=l'1]\begin{picture}(0,0)
       \put(0,0){\pspolygon[fillstyle=solid](-.5,-1)(.5,-1)}
       \rput(0,-.7){$g'_0$}
     \end{picture}&\phantom{aaaaaaaa}&
     [name=l'2]\begin{picture}(0,0)
	\multiput(0,0)(0.2,-0.1){3}{\pspolygon[fillstyle=solid](-1.5,-2)(1.5,-2)}
	\rput(0.4,-1.8){$g'_1\dots g'_n::nil$}
      \end{picture}\\\\\\\\\\\\
     \ncline{n}{l1}
     \ncline{n}{l2}
     \ncline{n}{l3}
     \ncline[nodesep=4pt,arrows=-*]{n'}{l'1}
     \ncline[nodesep=4pt,arrows=-*]{n'}{l'2}
\endpsmatrix\]

For instance, let $a$, $b$ and $c$ be three agents, and $oc_1$, $oc_2$, $oc_3$ and $oc_4$ be four outcomes. Consider the following game in the traditional formalism.

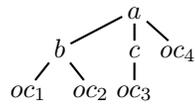
\[\psmatrix
      &&&&[name=n] a\\
      &&[name=n1] b &&[name=n2] c& [name=l4] oc_4 \\
      &[name=l1] oc_1 &&[name=l2] oc_2&[name=l3] oc_3
      \ncline{n}{n1}
      \ncline{n}{n2}
      \ncline{n}{l4}
      \ncline{n1}{l1}
      \ncline{n1}{l2}
      \ncline{n2}{l3}
    \endpsmatrix\]

The game above is represented by the following game in the new formalism.

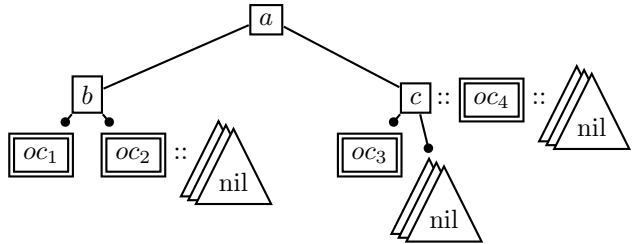
\[\psmatrix
      &&&&&&[name=n]\psframebox[fillstyle=solid,boxsep=false]{a}\\\\ 
      &&[name=n1]\psframebox[fillstyle=solid,boxsep=false]{b}&&&&&&[name=n2]\psframebox[fillstyle=solid,boxsep=false]{c}&&::&\psframebox[doubleline=true]{oc_4}&::&\begin{picture}(0,0)
    \multiput(.3,.5)(0.1,-0.05){3}{\pspolygon[fillstyle=solid](-.5,-1)(.5,-1)}
    \rput(.5,-.3){nil}
  \end{picture}\\
      &[name=l1]\psframebox[doubleline=true]{oc_1}&&[name=l2]\psframebox[doubleline=true]{oc_2}&::&\begin{picture}(0,0)
    \multiput(.3,.5)(0.1,-0.05){3}{\pspolygon[fillstyle=solid](-.5,-1)(.5,-1)}
    \rput(.5,-.3){nil}
  \end{picture} &\phantom{aaaaaaaaa}&[name=l3]\psframebox[doubleline=true]{oc_3}&&[name=l4] \begin{picture}(0,0)
    \multiput(0,0)(0.1,-0.05){3}{\pspolygon[fillstyle=solid](-.5,-1)(.5,-1)}
    \rput(.2,-.8){nil}
  \end{picture}\\
           \\\\\\      
     \ncline[nodesep=4pt,arrows=-]{n}{n1}
     \ncline[nodesep=4pt,arrows=-]{n}{n2}
     \ncline[nodesep=4pt,arrows=-*]{n1}{l1}
     \ncline[nodesep=4pt,arrows=-*]{n1}{l2}
     \ncline[nodesep=4pt,arrows=-*]{n2}{l3}
     \ncline[nodesep=4pt,arrows=-*]{n2}{l4}
\endpsmatrix\]

\subsubsection{The Inductive and Formal Way:}

Formally in Coq, games are defined as follows. First define \coqdocid{Outcome} and \coqdocid{Agent}, two arbitrary objects of type \coqdocid{Set}.

\medskip
\noindent
\coqdockw{Variable} \coqdocid{Outcome} : \coqdocid{Set}.\coqdoceol
\noindent
\coqdockw{Variable} \coqdocid{Agent} : \coqdocid{Set}.\coqdoceol
\medskip

Next, the object \coqdocid{Game} is defined by induction.

\medskip
\noindent
\coqdockw{Inductive} \coqdocid{Game} : \coqdocid{Set} := \coqdoceol
\coqdocindent{1.00em}
$\mid$ \coqdocid{gL} : \coqdocid{Outcome} \ensuremath{\rightarrow} \coqdocid{Game} \coqdoceol
\coqdocindent{1.00em}
$\mid$ \coqdocid{gN} : \coqdocid{Agent} \ensuremath{\rightarrow} \coqdocid{Game} \ensuremath{\rightarrow} \coqdocid{list} \coqdocid{Game} \ensuremath{\rightarrow} \coqdocid{Game}.\coqdoceol
\medskip

\coqdocid{gL} stands for ``game leaf'' and \coqdocid{gN} for ``game node''. If \coqdocid{oc} is an \coqdocid{Outcome}, then \coqdocid{gL} \coqdocid{oc} is a \coqdocid{Game}. If \coqdocid{a} is an \coqdocid{Agent}, \coqdocid{g} is a \coqdocid{Game}, and \coqdocid{l} is a list of objects in \coqdocid{Game}, then \coqdocid{gN} \coqdocid{a} \coqdocid{g} \coqdocid{l} is a \coqdocid{Game}. The interpretation of such an object is as follows: the agent \coqdocid{a} ``owns'' the root of \coqdocid{gN} \coqdocid{a} \coqdocid{g} \coqdocid{l}, and \coqdocid{g} and \coqdocid{l} represent \coqdocid{a}'s options, \textit{i.e.}, the subgames \coqdocid{a} can decide to continue the play in. The structure ensures that any node of the tree meant to be internal has a non-zero number of children. The inductive Coq formalism and the inductive graphical formalism are very close to each other. For instance, compare the game \coqdocid{gN} \coqdocid{a} (\coqdocid{gN} \coqdocid{b} \coqdocid{oc1} \coqdocid{oc2}::\coqdocid{nil}) \coqdocid{oc3}::\coqdocid{nil} with its representation in the inductive graphical formalism below.

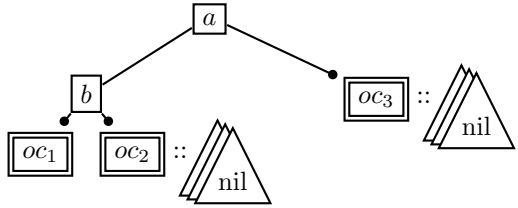
\[\psmatrix
      &&&&&&[name=n]\psframebox[fillstyle=solid,boxsep=false]{a}\\\\ 
      &&[name=n1]\psframebox[fillstyle=solid,boxsep=false]{b}&&&&&\phantom{aaaaaaaa}&[name=l3]\psframebox[doubleline=true]{oc_3}&::&\begin{picture}(0,0)
    \multiput(.3,.5)(0.1,-0.05){3}{\pspolygon[fillstyle=solid](-.5,-1)(.5,-1)}
    \rput(.5,-.3){nil}
  \end{picture}\\
      &[name=l1]\psframebox[doubleline=true]{oc_1}&&[name=l2]\psframebox[doubleline=true]{oc_2}&::&\begin{picture}(0,0)
    \multiput(.3,.5)(0.1,-0.05){3}{\pspolygon[fillstyle=solid](-.5,-1)(.5,-1)}
    \rput(.5,-.3){nil}
  \end{picture} 
           \\\\\\      
     \ncline[nodesep=4pt,arrows=-]{n}{n1}
     \ncline[nodesep=4pt,arrows=-*]{n}{l3}
     \ncline[nodesep=4pt,arrows=-*]{n1}{l1}
     \ncline[nodesep=4pt,arrows=-*]{n1}{l2}
\endpsmatrix\]

Note that the inductive definition of games involves lists, which are already an inductive structure. Therefore, the game structure is inductive ``horizontally and vertically''.

\subsection{Induction Principle for Sequential Games\label{subsect:ipsg}}

This subsection first discusses what would be an inductive proof principle for games in the traditional formalism. The failure of the former leads to an induction principle in the inductive graphical formalism. Then, the induction principle is easily translated in Coq formalism.

\subsubsection{The Traditional Way:}

In the traditional formalism, a so-called induction principle would read as follows. In order to check that a predicate holds for all games, check two properties: First, check that the predicate holds for all games that are leaves (enclosing an outcome). Second, check that if the predicate holds for all the following games,

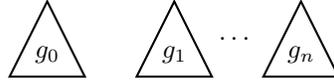
\[\psmatrix
    &
     [name=l1]\begin{picture}(0,0)
       \put(0,0){\pspolygon[fillstyle=solid](-.5,-1)(.5,-1)}
       \rput(0,-.7){$g_0$}
     \end{picture}&\phantom{aaaaaaaa}&
     [name=l2]\begin{picture}(0,0)
       \put(0,0){\pspolygon[fillstyle=solid](-.5,-1)(.5,-1)}
       \rput(0,-.7){$g_1$}
     \end{picture}&\rput(.7,-.5){\dots}\phantom{aaaaaaaa}&
     [name=l3]\begin{picture}(0,0)
       \put(0,0){\pspolygon[fillstyle=solid](-.5,-1)(.5,-1)}
       \rput(0,-.7){$g_n$}
     \end{picture}\\\\\\\\
\endpsmatrix\]

 then, for any agent $a$, it holds for the following game.  

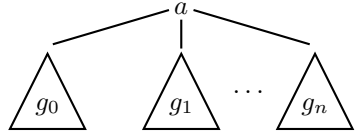
\[\psmatrix
     &&&[name=n] a\\&
     [name=l1]\begin{picture}(0,0)
       \put(0,0){\pspolygon[fillstyle=solid](-.5,-1)(.5,-1)}
       \rput(0,-.7){$g_0$}
     \end{picture}&\phantom{aaaaaaaa}&
     [name=l2]\begin{picture}(0,0)
       \put(0,0){\pspolygon[fillstyle=solid](-.5,-1)(.5,-1)}
       \rput(0,-.7){$g_1$}
     \end{picture}&\rput(.7,-.5){\dots}\phantom{aaaaaaaa}&
     [name=l3]\begin{picture}(0,0)
       \put(0,0){\pspolygon[fillstyle=solid](-.5,-1)(.5,-1)}
       \rput(0,-.7){$g_n$}
     \end{picture}\\\\\\\\
     \ncline{n}{l1}
     \ncline{n}{l2}
     \ncline{n}{l3}
\endpsmatrix\]

However, dots, etc's and so on, very seldom suit formal proofs. While some dots may be easily formalised, some others cannot. In the new formalism proposed for games, they are formalised by lists. So, an induction principle for games must take lists into account.

\subsubsection{The Inductive and Graphical Way:}

In order to prove that the predicate $P$ holds for all games, it suffices to design a predicate $Q$ expecting a game and a list of games, and then to check that all the properties listed below hold.

\begin{itemize}
\item For all outcome $oc$, $P$ ( \psframebox[doubleline=true]{$oc$} ).\\\\

\item For all game \qquad\begin{picture}(0,0)
  \rput(0,.7){\pspolygon[fillstyle=solid](-.5,-1)(.5,-1)}
  \rput(0,0){g} \end{picture}\qquad, if $P$ (\qquad\begin{picture}(0,0)
  \rput(0,.7){\pspolygon[fillstyle=solid](-.5,-1)(.5,-1)}
  \rput(0,0){g} \end{picture}\qquad ) then $Q$ (\qquad\begin{picture}(0,0)
  \rput(0,.7){\pspolygon[fillstyle=solid](-.5,-1)(.5,-1)}
  \rput(0,0){g} \end{picture}\qquad, \qquad\begin{picture}(0,0)
    \multiput(0,.7)(0.1,-0.05){3}{\pspolygon[fillstyle=solid](-.5,-1)(.5,-1)}
    \rput(0.2,-.1){nil}\end{picture}\qquad)\\\\\\

\item For all game \qquad\begin{picture}(0,0)
  \rput(0,.7){\pspolygon[fillstyle=solid](-.5,-1)(.5,-1)}
  \rput(0,0){g} \end{picture}\qquad and \qquad\begin{picture}(0,0)
  \rput(0,.7){\pspolygon[fillstyle=solid](-.5,-1)(.5,-1)}
  \rput(0,0){g'} \end{picture}\qquad and all list of game \qquad\begin{picture}(0,0)
    \multiput(0,.7)(0.1,-0.05){3}{\pspolygon[fillstyle=solid](-.5,-1)(.5,-1)}
    \rput(0.2,-.1){l}\end{picture}\qquad,\\\\\\

 if $P$ (\qquad\begin{picture}(0,0)
  \rput(0,.7){\pspolygon[fillstyle=solid](-.5,-1)(.5,-1)}
  \rput(0,0){g} \end{picture}\qquad ) and $Q$ (\qquad\begin{picture}(0,0)
  \rput(0,.7){\pspolygon[fillstyle=solid](-.5,-1)(.5,-1)}
  \rput(0,0){g'} \end{picture}\qquad, \qquad\begin{picture}(0,0)
    \multiput(0,.7)(0.1,-0.05){3}{\pspolygon[fillstyle=solid](-.5,-1)(.5,-1)}
    \rput(0.2,-.1){l}\end{picture}\qquad) then $Q$ (\qquad\begin{picture}(0,0)
  \rput(0,.7){\pspolygon[fillstyle=solid](-.5,-1)(.5,-1)}
  \rput(0,0){g'} \end{picture}\qquad, \qquad\begin{picture}(0,0)
    \multiput(0,.7)(0.1,-0.05){3}{\pspolygon[fillstyle=solid](-.5,-1)(.5,-1)}
    \rput(0.2,-.1){g::l}\end{picture}\qquad)\\\\\\

\item For all agent $a$, all game \qquad\begin{picture}(0,0)
  \rput(0,.7){\pspolygon[fillstyle=solid](-.5,-1)(.5,-1)}
  \rput(0,0){g} \end{picture}\qquad and all list of game \qquad\begin{picture}(0,0)
    \multiput(0,.7)(0.1,-0.05){3}{\pspolygon[fillstyle=solid](-.5,-1)(.5,-1)}
    \rput(0.2,-.1){l}\end{picture}\qquad,\\\\\\

if $Q$ (\qquad\begin{picture}(0,0)
  \rput(0,.7){\pspolygon[fillstyle=solid](-.5,-1)(.5,-1)}
  \rput(0,0){g} \end{picture}\qquad, \qquad\begin{picture}(0,0)
    \multiput(0,.7)(0.1,-0.05){3}{\pspolygon[fillstyle=solid](-.5,-1)(.5,-1)}
    \rput(0.2,-.1){l}\end{picture}\qquad) \phantom{aaaa}then\phantom{aaaa} $P$ (\phantom{aaa} 
$\psmatrix
      &&[name=n]
     \psframebox[fillstyle=solid,boxsep=false]{a}
     \\
      &[name=n1]
      \begin{picture}(0,0)
	\put(0,0){\pspolygon[fillstyle=solid](-.5,-1)(.5,-1)}
	\rput(0,-.7){g}
      \end{picture}
      &\phantom{aaaaaaa}&
      [name=n2]
      \begin{picture}(0,0)
	\multiput(0,0)(0.1,-0.05){3}{\pspolygon[fillstyle=solid](-.5,-1)(.5,-1)}
	\rput(.2,-.8){l}
      \end{picture}&\phantom{aa}  
     \ncline[nodesep=4pt,arrows=-*]{n}{n1}
     \ncline[nodesep=4pt,arrows=-*]{n}{n2}
\endpsmatrix$
\phantom{aa})\\\\\\

\end{itemize}

\subsubsection{The Inductive and Formal Way:}

The induction principle that Coq automatically associates to the inductively defined structure of games is not efficient. A convenient principle has to be built (and hereby proved) manually, \textit{via} a recursive function and the command \coqdocid{fix}. That leads to the following induction principle, which is therefore a theorem in Coq.

\medskip
\noindent
\coqdocid{Game\_ind2} : \ensuremath{\forall} (\coqdocid{P} : \coqdocid{Game} \ensuremath{\rightarrow} \coqdocid{Prop}) (\coqdocid{Q} : \coqdocid{Game} \ensuremath{\rightarrow} \coqdocid{list} \coqdocid{Game} \ensuremath{\rightarrow} \coqdocid{Prop}),\coqdoceol
\noindent
(\ensuremath{\forall} \coqdocid{oc}, \coqdocid{P} (\coqdocid{gL} \coqdocid{oc})) \ensuremath{\rightarrow} \coqdoceol
\noindent
(\ensuremath{\forall} \coqdocid{g}, \coqdocid{P} \coqdocid{g} \ensuremath{\rightarrow} \coqdocid{Q} \coqdocid{g} \coqdocid{nil}) \ensuremath{\rightarrow} \coqdoceol
\noindent
(\ensuremath{\forall} \coqdocid{g}, \coqdocid{P} \coqdocid{g} \ensuremath{\rightarrow} \ensuremath{\forall} \coqdocid{g'}  \coqdocid{l}, \coqdocid{Q} \coqdocid{g'} \coqdocid{l} \ensuremath{\rightarrow} \coqdocid{Q} \coqdocid{g'} (\coqdocid{g} :: \coqdocid{l})) \ensuremath{\rightarrow}\coqdoceol
\noindent
(\ensuremath{\forall} \coqdocid{g} \coqdocid{l}, \coqdocid{Q} \coqdocid{g} \coqdocid{l} \ensuremath{\rightarrow} \ensuremath{\forall} \coqdocid{a}, \coqdocid{P} (\coqdocid{gN} \coqdocid{a} \coqdocid{g} \coqdocid{l})) \ensuremath{\rightarrow}\coqdoceol
\noindent
\ensuremath{\forall} \coqdocid{g}, \coqdocid{P}\coqdocid{g}
\medskip

Two facts are worth noting: First, this principle corresponds to the one stated above in the inductive graphical formalism. Second, in order to prove a property \ensuremath{\forall} \coqdocid{g} : \coqdocid{Game}, \coqdocid{P}\coqdocid{g} with the induction principle \coqdocid{Game\_ind2}, the user has to imagine a workable predicate \coqdocid{Q}.

\subsection{Outcomes Used in a Sequential Game\label{subsect:ousg}}

This subsection discusses the notion of outcomes used in a sequential game and defines a function that returns the ``left-to-right'' list of all the outcomes involved in a given game. Two lemmas follow the definition of the function in the Coq formalism.

\subsubsection{The Traditional Way:}

The example below explains what function is intended.

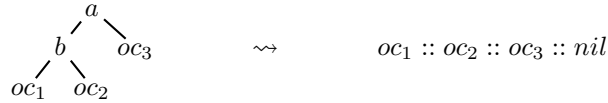
\[\psmatrix
      &&&[name=n1] a\\
      &&[name=n2] b &&[name=l3]  oc_3&\phantom{aaaaaa}&\leadsto&\phantom{aaaaaa}& oc_1::oc_2::oc_3::nil\\
      &[name=l1] oc_1 &&[name=l2] oc_2
      \ncline{n1}{n2}
      \ncline{n1}{l3}
      \ncline{n2}{l1}
      \ncline{n2}{l2}
    \endpsmatrix\]

\subsubsection{The Inductive and Graphical Way:}

To prepare the full formalisation of the idea above, the intended function is defined inductively with the inductive graphical formalism, in two steps along the structure of games. First step: the only outcome used in a leaf is the outcome enclosed in the leaf.

\[\psframebox[doubleline=true]{oc}\qquad\stackrel{\coqdocid{UO}}{\leadsto}\qquad oc::nil\]

Second step: recall that ++ refers to lists appending/concatenation.

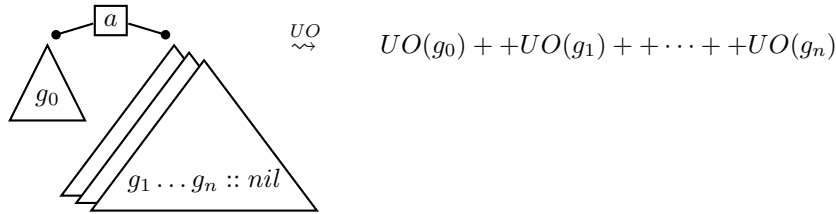
\[\begin{array}{rcl}
\psmatrix
     && [name=n']\psframebox[fillstyle=solid,boxsep=false]{a}\\&
     [name=l'1]\begin{picture}(0,0)
       \put(0,0){\pspolygon[fillstyle=solid](-.5,-1)(.5,-1)}
       \rput(0,-.7){$g_0$}
     \end{picture}&\phantom{aaaaaaaa}&
     [name=l'2]\begin{picture}(0,0)
	\multiput(0,0)(0.2,-0.1){3}{\pspolygon[fillstyle=solid](-1.5,-2)(1.5,-2)}
	\rput(0.4,-1.8){$g_1\dots g_n::nil$}
      \end{picture}\\
     \ncline[nodesep=4pt,arrows=-*]{n'}{l'1}
     \ncline[nodesep=4pt,arrows=-*]{n'}{l'2}
\endpsmatrix
&
\qquad\qquad\stackrel{\coqdocid{UO}}{\leadsto}\qquad
&
UO(g_0)++UO(g_1)++\dots ++UO(g_n)
\\\\\\\\\\\\
\end{array}\]

\subsubsection{The Inductive and Formal Way:} The intended function, called \coqdocid{UsedOutcomes}, is inductively and formally defined in Coq.

\medskip
\noindent
\coqdockw{Fixpoint} \coqdocid{UsedOutcomes} (\coqdocid{g} : \coqdocid{Game}) : \coqdocid{list} \coqdocid{Outcome} := \coqdoceol
\noindent
\coqdocid{match} \coqdocid{g} \coqdocid{with} \coqdoceol
\noindent
$\mid$ \coqdocid{gL} \coqdocid{oc} \ensuremath{\Rightarrow} \coqdocid{oc}::\coqdocid{nil} \coqdoceol
\noindent
$\mid$ \coqdocid{gN} \coqdocid{\_} \coqdocid{g'} \coqdocid{l} \ensuremath{\Rightarrow}  ((\coqdocid{fix} \coqdocid{ListUsedOutcomes} (\coqdocid{l'} : \coqdocid{list} \coqdocid{Game}) : \coqdocid{list} \coqdocid{Outcome} :=\coqdoceol
\coqdocindent{10.50em}
\coqdocid{match} \coqdocid{l'} \coqdocid{with}\coqdoceol
\coqdocindent{10.50em}
$\mid$ \coqdocid{nil} \ensuremath{\Rightarrow} \coqdocid{nil}\coqdoceol
\coqdocindent{10.50em}
$\mid$ \coqdocid{x}::\coqdocid{m} \ensuremath{\Rightarrow} (\coqdocid{UsedOutcomes} \coqdocid{x})++(\coqdocid{ListUsedOutcomes} \coqdocid{m}) \coqdoceol
\coqdocindent{10.50em}
\coqdocid{end}) (\coqdocid{g'}::\coqdocid{l}))\coqdoceol
\noindent
\coqdocid{end}.\coqdoceol
\medskip

The following lemma states that the outcomes used in a game are also used in a structurally bigger game.

\medskip
\noindent
\coqdockw{Lemma} \coqdocid{UsedOutcomes\_gN} : \ensuremath{\forall} \coqdocid{a} \coqdocid{g} \coqdocid{g'} \coqdocid{l},\coqdoceol
\noindent
\coqdocid{In} \coqdocid{g} (\coqdocid{g'}::\coqdocid{l}) \ensuremath{\rightarrow} \coqdocid{incl} (\coqdocid{UsedOutcomes} \coqdocid{g}) (\coqdocid{UsedOutcomes} (\coqdocid{gN} \coqdocid{a} \coqdocid{g'} \coqdocid{l})).\coqdoceol
\medskip

\begin{proof}
By induction on \coqdocid{l}. For the inductive step case split as follows. If \coqdocid{g} equals the head of \coqdocid{l} then invoke \coqdocid{incl\_appr}, \coqdocid{incl\_appl}, and \coqdocid{incl\_refl}. If \coqdocid{g} either equals \coqdocid{g'} or occurs in the tail of \coqdocid{l} then the induction hypothesis says that the outcomes used by \coqdocid{g} are used by \coqdocid{g'} and \coqdocid{l} together. The remainder mainly applies \coqdocid{in\_app\_or} and \coqdocid{in\_or\_app}.
\end{proof}

The next result shows that if the outcomes used in a game all occur in a given list, then so do the outcomes used in any subgame of the original game.

\medskip
\noindent
\coqdockw{Lemma} \coqdocid{UsedOutcomes\_gN\_listforall} : \ensuremath{\forall} \coqdocid{a} \coqdocid{g} \coqdocid{loc} \coqdocid{l},\coqdoceol
\noindent
\coqdocid{incl} (\coqdocid{UsedOutcomes} (\coqdocid{gN} \coqdocid{a} \coqdocid{g} \coqdocid{l})) \coqdocid{loc} \ensuremath{\rightarrow} \coqdocid{listforall} (\coqdocid{fun} \coqdocid{g'} : \coqdocid{Game} \ensuremath{\Rightarrow} \coqdocid{incl} (\coqdocid{UsedOutcomes} \coqdocid{g'}) \coqdocid{loc}) (\coqdocid{g}::\coqdocid{l}).\coqdoceol
\medskip

\begin{proof}
All the following cases invoke \coqdocid{in\_or\_app}. It is straightforward for \coqdocid{g}. For elements of \coqdocid{l} proceed by induction on \coqdocid{l}. For the inductive step, it is straightforward for the head of \coqdocid{l}, and use the induction hypothesis for elements of the tail of \coqdocid{l}.
\end{proof}

\section{Strategy Profiles\label{sect:sp}}

Consistent with subsection~\ref{subsect:dsg}, subsection~\ref{subsect:dsp} generalises the notion of strategy profile by abstracting over payoff functions, and introduces a new formalism for them. In the remainder of this paper, the expression ``strategy profile'' refers to the new and abstract strategy profiles, unless stated otherwise. In subsection~\ref{subsect:ipsp}, an inductive proof principle is designed for these strategy profiles. Subsection~\ref{subsect:ugsp} defines the underlying game of a strategy profile. Last, subsection~\ref{subsect:iosp} defines the induced outcome of a strategy profile.

\subsection{Definition of Strategy Profiles\label{subsect:dsp}}

This subsection first presents strategy profiles informally in the way traditional strategy profiles were presented in section~\ref{sect:tsgt}. Then it describes strategy profiles in a both inductive and graphical formalism consistent with the one used to define sequential games. A correspondence between the traditional formalism and the new one is established. Finally, the inductive and graphical formalism is naturally translated into a definition of strategy profiles in Coq.

\subsubsection{The Traditional Way:}

Roughly speaking, an abstract strategy profile is a traditional strategy profile where each real-valued payoff function (enclosed in a leaf) has been replaced by an outcome. Below, the left-hand picture represents a traditional strategy profile, and the right-hand picture represents an abstract strategy profile on the same tree structure.

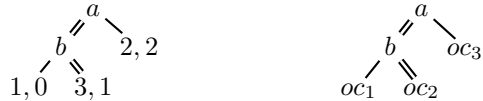
\[\psmatrix
      &&&[name=n1] a&& &&&[name=n1'] a\\
      &&[name=n2] b &&[name=l3]  2,2 &\phantom{aaaaaaaaaaaa}&&[name=n2'] b &&[name=l3'] oc_3 \\
      &[name=l1] 1,0 &&[name=l2] 3,1 && &[name=l1'] oc_1 &&[name=l2'] oc_2
      \choiceline{n1}{n2}
      \ncline{n1}{l3}
      \ncline{n2}{l1}
      \choiceline{n2}{l2}
      \choiceline{n1'}{n2'}
      \ncline{n1'}{l3'}
      \ncline{n2'}{l1'}
      \choiceline{n2'}{l2'}
    \endpsmatrix\]

\subsubsection{The Inductive and Graphical Way:}

A generic strategy profile $s$ will be represented by the left-hand picture below and a generic (possibly empty) list of strategy profiles by the right-hand picture.

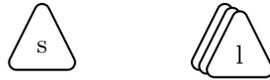
\[\begin{array}{ccc}
\begin{picture}(0,0)
  \rput(0,0){\pspolygon[fillstyle=solid,linearc=.1](-.5,-1)(.5,-1)}
  \rput(0,-.7){s}
\end{picture}
  &\phantom{aaaaaaaaaaaa}&
  \begin{picture}(0,0)
    \multiput(0,0)(0.1,-0.05){3}{\pspolygon[fillstyle=solid,linearc=.1](-.5,-1)(.5,-1)}
    \rput(0.2,-.8){l}
  \end{picture}
\\\\\\\\
\end{array}\]

Strategy profiles are inductively defined in two steps. First step: If $oc$ is an outcome then the object below is a strategy profile.

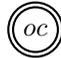
\[\begin{array}{c}
  \pscirclebox[doubleline=true]{oc}
\end{array}\]

Second step: if $a$ is an agent and the first three objects below are a strategy profile and two lists of strategy profiles, then the rightmost object is also a strategy profile.

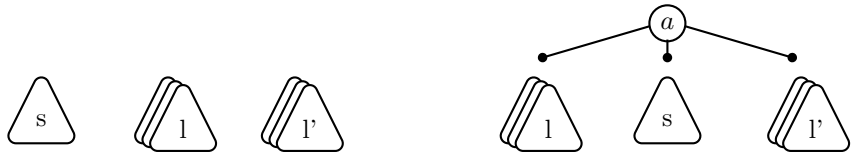
\[
\psmatrix
      &&&&&&&&&[name=n]\pscirclebox[fillstyle=solid,boxsep=false]{a}\\&
      \begin{picture}(0,0)
	\rput(0,0){\pspolygon[fillstyle=solid,linearc=.1](-.5,-1)(.5,-1)}
	\rput(0,-.7){s}
      \end{picture}
      &\phantom{aaaaaaaa}&
      \begin{picture}(0,0)
	\multiput(0,0)(0.1,-0.05){3}{\pspolygon[fillstyle=solid,linearc=.1](-.5,-1)(.5,-1)}
	\rput(0.2,-.8){l}
      \end{picture}&\phantom{aaaaaaaa}&
      \begin{picture}(0,0)
	\multiput(0,0)(0.1,-0.05){3}{\pspolygon[fillstyle=solid,linearc=.1](-.5,-1)(.5,-1)}
	\rput(0.2,-.8){l'}
      \end{picture}&\phantom{aaaaaaaaaaaaaaaa}&[name=n1]
      \begin{picture}(0,0)
	\multiput(0,0)(0.1,-0.05){3}{\pspolygon[fillstyle=solid,linearc=.1](-.5,-1)(.5,-1)}
	\rput(0.2,-.8){l}
      \end{picture}
      &\phantom{aaaaaaaa}&[name=n2]
      \begin{picture}(0,0)
	\put(0,0){\pspolygon[fillstyle=solid,linearc=.1](-.5,-1)(.5,-1)}
	\rput(0,-.7){s}
      \end{picture}
      &\phantom{aaaaaaaa}&
      [name=n3]
      \begin{picture}(0,0)
	\multiput(0,0)(0.1,-0.05){3}{\pspolygon[fillstyle=solid,linearc=.1](-.5,-1)(.5,-1)}
	\rput(0.2,-.8){l'}
      \end{picture}\\\\\\\\      
     \ncline[nodesep=4pt,arrows=-*]{n}{n1}
     \ncline[nodesep=4pt,arrows=-*]{n}{n2}
     \ncline[nodesep=4pt,arrows=-*]{n}{n3}
\endpsmatrix\]

The major difference between the traditional formalism and the new formalism is as follows: In the traditional formalism, an internal node has one or arbitrarily many more children, which are strategy profiles, and in addition an internal node is linked to one and only one of its children by a double line. In the new formalism, an internal node has one left child, which is a list of strategy profiles, one central child, which is a strategy profile corresponding to the double line, and one right child, which is a list of strategy profiles. The next two pictures represent the same strategy profile in both formalisms, where for all $i$ between $0$ and $n$, the object $s'_i$ is the translation of $s_i$ from the traditional formalism into the new formalism. 

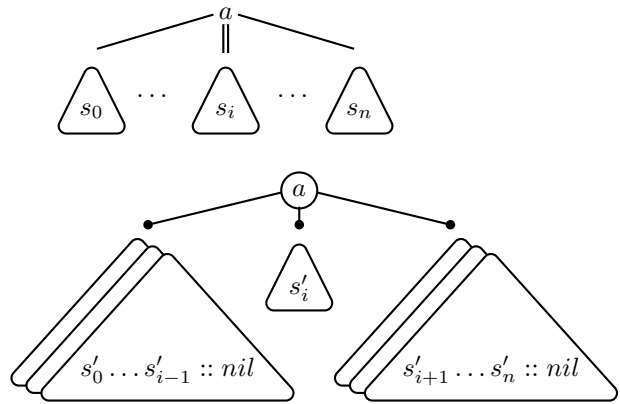
\[\psmatrix
     &&&[name=n] a\\&
     [name=l1]\begin{picture}(0,0)
       \put(0,0){\pspolygon[fillstyle=solid,linearc=.1](-.5,-1)(.5,-1)}
       \rput(0,-.7){$s_0$}
     \end{picture}&\rput(.7,-.5){\dots}\phantom{aaaaaaaa}&
     [name=l2]\begin{picture}(0,0)
       \put(0,0){\pspolygon[fillstyle=solid,linearc=.1](-.5,-1)(.5,-1)}
       \rput(0,-.7){$s_i$}
     \end{picture}&\rput(.7,-.5){\dots}\phantom{aaaaaaaa}&
     [name=l3]\begin{picture}(0,0)
       \put(0,0){\pspolygon[fillstyle=solid,linearc=.1](-.5,-1)(.5,-1)}
       \rput(0,-.7){$s_n$}
     \end{picture}&\phantom{aaaaaaaaaa}\\\\
     \ncline{n}{l1}
     \ncline[doubleline=true]{n}{l2}
     \ncline{n}{l3}
\endpsmatrix\]

\[\psmatrix
     &&& [name=n']\pscirclebox[fillstyle=solid,boxsep=false]{a}\\&
     [name=l'1]\begin{picture}(0,0)
	\multiput(0,0)(0.2,-0.1){3}{\pspolygon[fillstyle=solid,linearc=.1](-1.8,-2)(1.8,-2)}
	\rput(0.4,-1.8){$s'_0\dots s'_{i-1}::nil$}
      \end{picture}
       &\phantom{aaaaaaaaaa}&
     [name=l'2]\begin{picture}(0,0)
       \put(0,0){\pspolygon[fillstyle=solid,linearc=.1](-.5,-1)(.5,-1)}
       \rput(0,-.7){$s'_i$}
     \end{picture}&\phantom{aaaaaaaaaa}&
     [name=l'3]\begin{picture}(0,0)
	\multiput(0,0)(0.2,-0.1){3}{\pspolygon[fillstyle=solid,linearc=.1](-1.8,-2)(1.8,-2)}
	\rput(0.4,-1.8){$s'_{i+1}\dots s'_n::nil$}
      \end{picture}\\\\\\\\\\\\
     \ncline[nodesep=4pt,arrows=-*]{n'}{l'1}
     \ncline[nodesep=4pt,arrows=-*]{n'}{l'2}
     \ncline[nodesep=4pt,arrows=-*]{n'}{l'3}
\endpsmatrix\]

For instance, let $a$ and $b$ be two agents, and $oc_1$, $oc_2$ and $oc_3$ be three outcomes. Consider the following strategy profile in the traditional formalism.

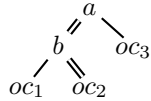
\[\psmatrix
      &&&[name=n1] a\\
      &&[name=n2] b &&[name=l3]  oc_3 \\
      &[name=l1] oc_1 &&[name=l2] oc_2 
      \choiceline{n1}{n2}
      \ncline{n1}{l3}
      \ncline{n2}{l1}
      \choiceline{n2}{l2}
    \endpsmatrix\]

The strategy profile above is represented by the following strategy profile in the new formalism.

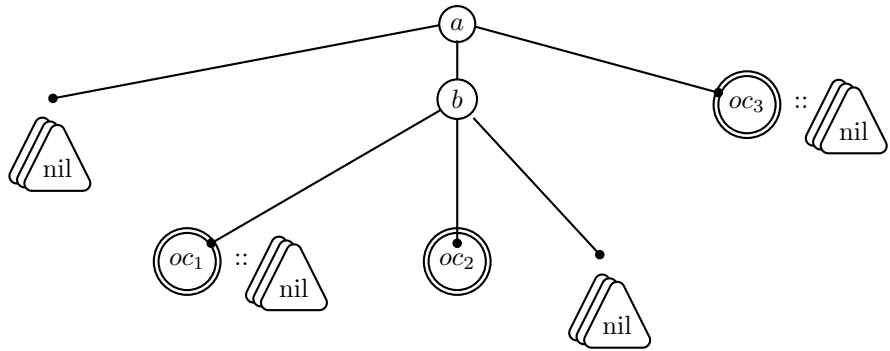
\[\psmatrix
      &&&&&&&[name=n]\pscirclebox[fillstyle=solid,boxsep=false]{a}\\\\ 
      &[name=n1]
      \begin{picture}(0,0)
       \multiput(0,0)(0.1,-0.05){3}{\pspolygon[fillstyle=solid,linearc=.1](-.5,-1)(.5,-1)}
       \rput(0.2,-.8){nil}
     \end{picture}
     &\phantom{aaaaaaaa}&&&&\phantom{aaaaaaaaaaaa}&[name=n2]\pscirclebox[fillstyle=solid,boxsep=false]{b}
     &&\phantom{aaaaaaaaaaaaaaaa}&&[name=n3]\pscirclebox[fillstyle=solid,boxsep=false,doubleline=true]{oc_3}&\phantom{a}::&\begin{picture}(0,0)
       \multiput(.3,.5)(0.1,-0.05){3}{\pspolygon[fillstyle=solid,linearc=.1](-.5,-1)(.5,-1)}
       \rput(0.5,-.3){nil}
     \end{picture}\\\\\\\\
     &&&
     [name=n21]\pscirclebox[fillstyle=solid,boxsep=false,doubleline=true]{oc_1}&\phantom{a}::&\begin{picture}(0,0)
       \multiput(.3,.5)(0.1,-0.05){3}{\pspolygon[fillstyle=solid,linearc=.1](-.5,-1)(.5,-1)}
       \rput(0.5,-.3){nil}
     \end{picture}
     &&
     [name=n22]\pscirclebox[fillstyle=solid,boxsep=false,doubleline=true]{oc_2}
     &&
     [name=n23]
     \begin{picture}(0,0)
       \multiput(0,0)(0.1,-0.05){3}{\pspolygon[fillstyle=solid,linearc=.1](-.5,-1)(.5,-1)}
       \rput(0.2,-.8){nil}
     \end{picture}
              \\\\\\      
     \ncline[nodesep=4pt,arrows=-*]{n}{n1}
     \ncline[nodesep=4pt,arrows=-]{n}{n2}
     \ncline[nodesep=4pt,arrows=-*]{n}{n3}
     \ncline[nodesep=4pt,arrows=-*]{n2}{n21}
     \ncline[nodesep=4pt,arrows=-*]{n2}{n22}
     \ncline[nodesep=4pt,arrows=-*]{n2}{n23}
\endpsmatrix\]

\subsubsection{The Inductive and Formal Way:}

Formally in Coq, strategy profiles are defined as follows.

\medskip
\noindent
\coqdockw{Inductive} \coqdocid{Strat} : \coqdocid{Set} := \coqdoceol
\coqdocindent{1.00em}
$\mid$ \coqdocid{sL} : \coqdocid{Outcome} \ensuremath{\rightarrow} \coqdocid{Strat} \coqdoceol
\coqdocindent{1.00em}
$\mid$ \coqdocid{sN} : \coqdocid{Agent} \ensuremath{\rightarrow} \coqdocid{list} \coqdocid{Strat} \ensuremath{\rightarrow} \coqdocid{Strat} \ensuremath{\rightarrow} \coqdocid{list} \coqdocid{Strat} \ensuremath{\rightarrow} \coqdocid{Strat}.\coqdoceol
\medskip

\coqdocid{sL} stands for ``strategy profile leaf'' and \coqdocid{sN} for ``strategy profile node''. If \coqdocid{oc} is an \coqdocid{Outcome}, then \coqdocid{sL} \coqdocid{oc} is a \coqdocid{Strat}. If \coqdocid{a} is an \coqdocid{Agent}, if \coqdocid{s} is a \coqdocid{Strat}, and if \coqdocid{l} and \coqdocid{l'} are lists of objects in \coqdocid{Strat}, then \coqdocid{sN} \coqdocid{a} \coqdocid{l} \coqdocid{s} \coqdocid{l'} is a \coqdocid{Strat}. The interpretation of such an object is as follows: as for sequential games, the agent \coqdocid{a} ``owns'' the root of \coqdocid{sN} \coqdocid{a} \coqdocid{l} \coqdocid{s} \coqdocid{l'}. Moreover, \coqdocid{s} is the substrategy profile \coqdocid{a} has decided the play shall continue in, and \coqdocid{l} and \coqdocid{l'} represent the options dismissed by \coqdocid{a} on the left and on the right of \coqdocid{s}. The structure ensures that any node of the tree meant to be internal has a non-zero number of children, and that one and only one child has been chosen. Like for sequential games, the inductive graphical formalism and the inductive Coq formalism are very close to each other for strategy profiles.  For instance, compare the strategy profile \coqdocid{sN} \coqdocid{a} \coqdocid{nil} (\coqdocid{sN} \coqdocid{b} \coqdocid{oc1}::\coqdocid{nil} \coqdocid{oc2} \coqdocid{nil}) \coqdocid{oc3}::\coqdocid{nil} with its representation in the inductive graphical formalism above.

\subsection{Induction Principle for Strategy Profiles\label{subsect:ipsp}}

This subsection discusses what would be an inductive proof principle for strategy profiles in the traditional formalism. The failure of the former leads to an induction principle in the inductive Coq formalism. 

\subsubsection{The Traditional Way:}

In the traditional formalism, a so-called induction principle would read as follows. In order to check that a predicate holds for all strategy profiles, check two properties: First, check that the predicate holds for all strategy profiles that are leaves (enclosing an outcome). Second, check that if the predicate holds for all the following strategy profiles,

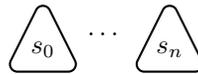
\[\psmatrix
    &
     [name=l1]\begin{picture}(0,0)
       \put(0,0){\pspolygon[fillstyle=solid,linearc=.1](-.5,-1)(.5,-1)}
       \rput(0,-.7){$s_0$}
     \end{picture}&\rput(.7,-.5){\dots}\phantom{aaaaaaaa}&
     [name=l2]\begin{picture}(0,0)
       \put(0,0){\pspolygon[fillstyle=solid,linearc=.1](-.5,-1)(.5,-1)}
       \rput(0,-.7){$s_n$}
     \end{picture}
    \\\\\\
\endpsmatrix\]

 then, for any agent $a$ and any $i$ between $0$ and $n$, it holds for the following strategy profile.  

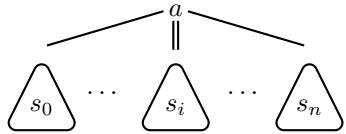
\[\psmatrix
     &&&[name=n] a\\&
     [name=l1]\begin{picture}(0,0)
       \put(0,0){\pspolygon[fillstyle=solid,linearc=.1](-.5,-1)(.5,-1)}
       \rput(0,-.7){$s_0$}
     \end{picture}&\rput(.7,-.5){\dots}\phantom{aaaaaaaa}&
     [name=l2]\begin{picture}(0,0)
       \put(0,0){\pspolygon[fillstyle=solid,linearc=.1](-.5,-1)(.5,-1)}
       \rput(0,-.7){$s_i$}
     \end{picture}&\rput(.7,-.5){\dots}\phantom{aaaaaaaa}&
     [name=l3]\begin{picture}(0,0)
       \put(0,0){\pspolygon[fillstyle=solid,linearc=.1](-.5,-1)(.5,-1)}
       \rput(0,-.7){$s_n$}
     \end{picture}&\phantom{aaaaaaaaaa}\\\\
     \ncline{n}{l1}
     \ncline[doubleline=true]{n}{l2}
     \ncline{n}{l3}
     \ncline[nodesep=4pt,arrows=-*]{n'}{l'1}
     \ncline[nodesep=4pt,arrows=-*]{n'}{l'2}
     \ncline[nodesep=4pt,arrows=-*]{n'}{l'3}
\endpsmatrix\]

However, dots, etc's and so on, very seldom suit formal proofs. While some dots may be easily formalised, some others cannot. In the inductive formalism proposed in this paper, they are formalised by lists. So, an induction principle for games must take lists into account.

\subsubsection{The Inductive and Formal Way:}

The induction principle that Coq automatically associates to the inductively defined structure of strategy profiles is not efficient. A convenient principle has to be built (and hereby proved) manually, \textit{via} a recursive function and the command \coqdocid{fix}. That leads to the following induction principle, which is therefore a theorem in Coq. In order to prove that a predicate $P$ holds for all strategy profiles, it suffices to design a predicate $Q$ expecting a strategy profile and a list of strategy profiles, and then to check that several fixed properties hold, as formally stated below.

\medskip
\noindent
\coqdocid{Strat\_ind2}  : \ensuremath{\forall} (\coqdocid{P} : \coqdocid{Strat} \ensuremath{\rightarrow} \coqdocid{Prop})
         (\coqdocid{Q} : \coqdocid{list} \coqdocid{Strat} \ensuremath{\rightarrow} \coqdocid{Strat} \ensuremath{\rightarrow} \coqdocid{list} \coqdocid{Strat} \ensuremath{\rightarrow} \coqdocid{Prop}),\coqdoceol
\noindent
(\ensuremath{\forall} \coqdocid{oc}, \coqdocid{P} (\coqdocid{sL} \coqdocid{oc})) \ensuremath{\rightarrow}\coqdoceol
\noindent
(\ensuremath{\forall} \coqdocid{sc}, \coqdocid{P} \coqdocid{sc} \ensuremath{\rightarrow} \coqdocid{Q} \coqdocid{nil} \coqdocid{sc} \coqdocid{nil}) \ensuremath{\rightarrow}\coqdoceol
\noindent
(\ensuremath{\forall} \coqdocid{s}, \coqdocid{P} \coqdocid{s} \ensuremath{\rightarrow} \ensuremath{\forall} \coqdocid{sc} \coqdocid{sr}, \coqdocid{Q} \coqdocid{nil} \coqdocid{sc} \coqdocid{sr} \ensuremath{\rightarrow} \coqdocid{Q} \coqdocid{nil} \coqdocid{sc} (\coqdocid{s} :: \coqdocid{sr})) \ensuremath{\rightarrow}\coqdoceol
\noindent
(\ensuremath{\forall} \coqdocid{s}, \coqdocid{P} \coqdocid{s} \ensuremath{\rightarrow} \ensuremath{\forall} \coqdocid{sc} \coqdocid{sl} \coqdocid{sr}, \coqdocid{Q} \coqdocid{sl} \coqdocid{sc} \coqdocid{sr} \ensuremath{\rightarrow} \coqdocid{Q} (\coqdocid{s} :: \coqdocid{sl}) \coqdocid{sc} \coqdocid{sr}) \ensuremath{\rightarrow}\coqdoceol
\noindent
(\ensuremath{\forall} \coqdocid{sc} \coqdocid{sl} \coqdocid{sr}, \coqdocid{Q} \coqdocid{sl} \coqdocid{sc} \coqdocid{sr} \ensuremath{\rightarrow} \ensuremath{\forall} \coqdocid{a}, \coqdocid{P} (\coqdocid{sN} \coqdocid{a} \coqdocid{sl} \coqdocid{sc} \coqdocid{sr})) \ensuremath{\rightarrow}\coqdoceol
\noindent
\ensuremath{\forall} \coqdocid{s}, \coqdocid{P} \coqdocid{s}
\medskip

\subsection{Underlying Game of a Strategy Profile\label{subsect:ugsp}}

This subsection discusses the notion of underlying game of a given strategy profile and defines a function that returns such a game. One lemma follows the definition of the function in the Coq formalism.

\subsubsection{The Traditional Way:}

The example below explains what function is intended. It amounts to forgetting all the choices.

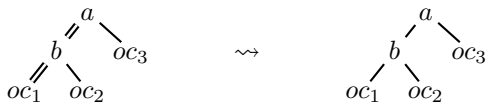
\[\psmatrix
      &&&[name=n1] a&& && &&&[name=n1'] a\\
      &&[name=n2] b &&[name=l3]  oc_3&\phantom{aaaaa}&\leadsto&\phantom{aaaaa}&&[name=n2'] b &&[name=l3']oc_3 \\
      &[name=l1] oc_1 &&[name=l2] oc_2&& &&&[name=l1'] oc_1 &&[name=l2'] oc_2
      \choiceline{n1}{n2}
      \ncline{n1}{l3}
      \choiceline{n2}{l1}
      \ncline{n2}{l2}
      \ncline{n1'}{n2'}
      \ncline{n1'}{l3'}
      \ncline{n2'}{l1'}
      \ncline{n2'}{l2'}
    \endpsmatrix\]

\subsubsection{The Inductive and Graphical Way:}

To prepare the full formalisation of the idea above, the intended function is defined inductively with the inductive graphical formalism, in two steps along the structure of strategy profiles. First step: the underlying game of a strategy profile where no choice has been made, \textit{i.e.}, a leaf strategy profile, is a game where no choice is required, \textit{i.e.}, a leaf game.

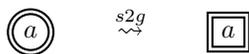
\[\pscirclebox[fillstyle=solid,doubleline=true]{a}\qquad \stackrel{s2g}{\leadsto}\qquad \psframebox[fillstyle=solid,doubleline=true]{a}\]

The second step needs case splitting along the structure of the first list, \textit{i.e.}, whether it is empty or not.

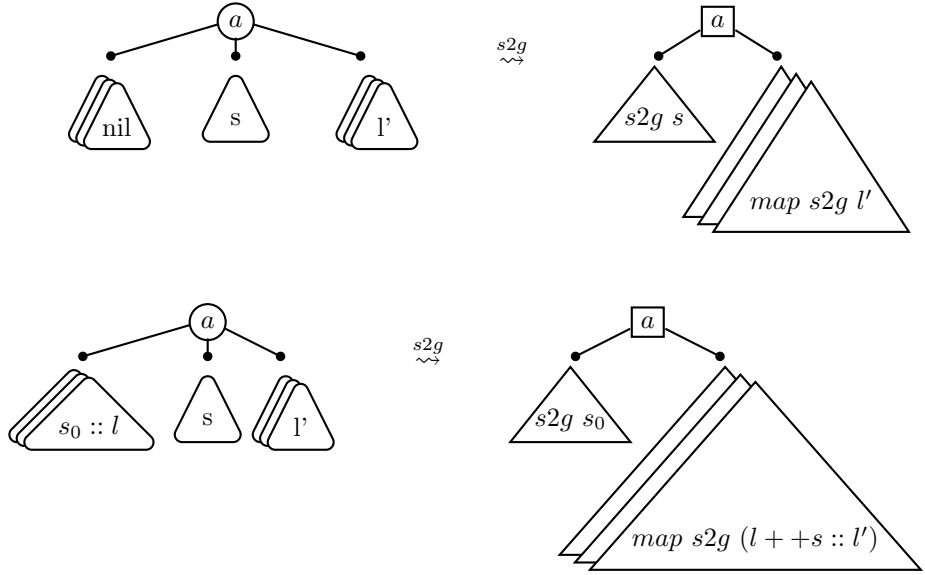
\[\psmatrix
      &&&[name=n]\pscirclebox[fillstyle=solid,boxsep=false]{a} &&&&&[name=n'] \psframebox[fillstyle=solid,boxsep=false]{a}\\
&[name=n1]
      \begin{picture}(0,0)
	\multiput(0,0)(0.1,-0.05){3}{\pspolygon[fillstyle=solid,linearc=.1](-.5,-1)(.5,-1)}
	\rput(0.2,-.8){nil}
      \end{picture}
      &\phantom{aaaaaaaa}&[name=n2]
      \begin{picture}(0,0)
	\put(0,0){\pspolygon[fillstyle=solid,linearc=.1](-.5,-1)(.5,-1)}
	\rput(0,-.7){s}
      \end{picture}
      &\phantom{aaaaaaaa}&
      [name=n3]
      \begin{picture}(0,0)
	\multiput(0,0)(0.1,-0.05){3}{\pspolygon[fillstyle=solid,linearc=.1](-.5,-1)(.5,-1)}
	\rput(0.2,-.8){l'}
      \end{picture}
      &\phantom{aaaaaaaa}\stackrel{s2g}{\leadsto}\phantom{aaaaaaaa}
      &[name=l'1]\begin{picture}(0,0)
	\put(0,0){\pspolygon[fillstyle=solid](-.8,-1)(.8,-1)}
	\rput(0,-.7){$s2g$ $s$}
      \end{picture}&\phantom{aaaaaaaa}
      &[name=l'2]\begin{picture}(0,0)
	\multiput(0,0)(0.2,-0.1){3}{\pspolygon[fillstyle=solid](-1.3,-2)(1.3,-2)}
	\rput(0.4,-1.8){$map$ $s2g$ $l'$}
      \end{picture}&\phantom{aaaa}\\\\\\\\\\ 
      \ncline[nodesep=4pt,arrows=-*]{n}{n1}
      \ncline[nodesep=4pt,arrows=-*]{n}{n2}
      \ncline[nodesep=4pt,arrows=-*]{n}{n3}
      \ncline[nodesep=4pt,arrows=-*]{n'}{l'1}
      \ncline[nodesep=4pt,arrows=-*]{n'}{l'2}  
\endpsmatrix\]

\[\psmatrix
      &&&[name=n]\pscirclebox[fillstyle=solid,boxsep=false]{a} &&&&&[name=n'] \psframebox[fillstyle=solid,boxsep=false]{a}\\
&[name=n1]
      \begin{picture}(0,0)
	\multiput(0,0)(0.1,-0.05){3}{\pspolygon[fillstyle=solid,linearc=.1](-1,-1)(1,-1)}
	\rput(0.2,-.8){$s_0::l$}
      \end{picture}
      &\phantom{aaaaaaaa}&[name=n2]
      \begin{picture}(0,0)
	\put(0,0){\pspolygon[fillstyle=solid,linearc=.1](-.5,-1)(.5,-1)}
	\rput(0,-.7){s}
      \end{picture}
      &\phantom{aaaa}&
      [name=n3]
      \begin{picture}(0,0)
	\multiput(0,0)(0.1,-0.05){3}{\pspolygon[fillstyle=solid,linearc=.1](-.5,-1)(.5,-1)}
	\rput(0.2,-.8){l'}
      \end{picture}
      &\phantom{aaaaaaaa}\stackrel{s2g}{\leadsto}\phantom{aaaaaaaa}
      &[name=l'1]\begin{picture}(0,0)
	\put(0,0){\pspolygon[fillstyle=solid](-.8,-1)(.8,-1)}
	\rput(0,-.7){$s2g$ $s_0$}
      \end{picture}&\phantom{aaaaaaaaaa}
      &[name=l'2]\begin{picture}(0,0)
	\multiput(0,0)(0.2,-0.1){3}{\pspolygon[fillstyle=solid](-2.2,-2.5)(2.2,-2.5)}
	\rput(0.4,-2.3){$map$ $s2g$ $(l++s::l')$}
      \end{picture}&\phantom{aaaaaaaaaa}\\\\\\\\\\ 
      \ncline[nodesep=4pt,arrows=-*]{n}{n1}
      \ncline[nodesep=4pt,arrows=-*]{n}{n2}
      \ncline[nodesep=4pt,arrows=-*]{n}{n3}
      \ncline[nodesep=4pt,arrows=-*]{n'}{l'1}
      \ncline[nodesep=4pt,arrows=-*]{n'}{l'2}  
\endpsmatrix\]

\subsubsection{The Inductive and Formal Way:}

The intended function is inductively and formally defined in Coq, and called \coqdocid{s2g}, which stands for ``strategy profile to game''.

\medskip
\noindent
\coqdockw{Fixpoint} \coqdocid{s2g} (\coqdocid{s} : \coqdocid{Strat}) : \coqdocid{Game} := \coqdoceol
\noindent
\coqdocid{match} \coqdocid{s} \coqdocid{with} \coqdoceol
\noindent
$\mid$ \coqdocid{sL} \coqdocid{oc} \ensuremath{\Rightarrow} \coqdocid{gL} \coqdocid{oc} \coqdoceol
\noindent
$\mid$ \coqdocid{sN} \coqdocid{a} \coqdocid{sl} \coqdocid{sc} \coqdocid{sr} \ensuremath{\Rightarrow}\coqdoceol
\coqdocindent{5em}
  \coqdocid{match} \coqdocid{sl} \coqdocid{with}\coqdoceol
\coqdocindent{5em}
$\mid$ \coqdocid{nil} \ensuremath{\Rightarrow} \coqdocid{gN} \coqdocid{a} (\coqdocid{s2g} \coqdocid{sc}) (\coqdocid{map} \coqdocid{s2g} \coqdocid{sr})\coqdoceol
\coqdocindent{5em}
$\mid$ \coqdocid{s'}::\coqdocid{sl'} \ensuremath{\Rightarrow} \coqdocid{gN} \coqdocid{a} (\coqdocid{s2g} \coqdocid{s'}) ((\coqdocid{map} \coqdocid{s2g} \coqdocid{sl'})++(\coqdocid{s2g} \coqdocid{sc})::(\coqdocid{map} \coqdocid{s2g} \coqdocid{sr}))\coqdoceol
\coqdocindent{5em}
\coqdocid{end}\coqdoceol
\noindent
\coqdocid{end}.\coqdoceol
\medskip

The next result states that if the two lists of substrategy profiles of two strategy profiles have, component-wise, the same underlying games, then the two original strategy profiles also have the same underlying game.

\medskip
\noindent
\coqdockw{Lemma} \coqdocid{map\_s2g\_sN\_s2g} : \ensuremath{\forall} \coqdocid{a} \coqdocid{sl} \coqdocid{sc} \coqdocid{sr} \coqdocid{sl'} \coqdocid{sc'} \coqdocid{sr'},\coqdoceol
\noindent
\coqdocid{map} \coqdocid{s2g} (\coqdocid{sl} ++ \coqdocid{sc} :: \coqdocid{sr}) = \coqdocid{map} \coqdocid{s2g} (\coqdocid{sl'} ++ \coqdocid{sc'} :: \coqdocid{sr'}) \ensuremath{\rightarrow}\coqdoceol
\noindent
\coqdocid{s2g} (\coqdocid{sN} \coqdocid{a} \coqdocid{sl} \coqdocid{sc} \coqdocid{sr}) = \coqdocid{s2g} (\coqdocid{sN} \coqdocid{a} \coqdocid{sl'} \coqdocid{sc'} \coqdocid{sr'}).\coqdoceol
\medskip

\begin{proof}
Double case split on \coqdocid{sl} and \coqdocid{sl'} being empty and use \coqdocid{map\_app}. 
\end{proof}

\subsection{Induced Outcome of a Strategy Profile\label{subsect:iosp}}

This subsection discusses the notion of outcome induced by a strategy profile, and defines a function that computes such an outcome. Three lemmas follow the definition of the function in the Coq formalism.

\subsubsection{The Traditional Way:}

Starting at the root of a strategy profile and following the agents' consecutive choices leads to a leaf, and hereby to an outcome. The following example explains what function is intended. 

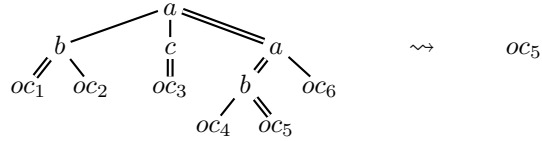
\[\psmatrix
   &&&&&[name=n] a\\
   &&[name=n1] b &&&[name=n2] c &&&[name=n3] a&&\phantom{aaaa}\leadsto\phantom{aaaa}&oc_5\\
   &[name=l11] oc_1 &&[name=l12] oc_2 &\phantom{aa}&[name=l21] oc_3 &\phantom{aa}&[name=n31] b &&[name=l32] oc_6\\
   &&&&&&[name=l311] oc_4&&[name=l312] oc_5 
   \ncline{n}{n1}
   \ncline{n}{n2}
   \choiceline{n}{n3}
   \choiceline{n1}{l11}
   \ncline{n1}{l12}
   \choiceline{n2}{l21}
   \choiceline{n3}{n31}
   \ncline{n3}{l32}
   \ncline{n31}{l311}
   \choiceline{n31}{l312}
\endpsmatrix\]

\subsubsection{The Inductive and Graphical Way:}

To prepare the full formalisation of the idea above, the intended function is defined inductively with the inductive graphical formalism, in two steps along the structure of strategy profiles. First step: the outcome induced by a leaf strategy profile is the enclosed outcome. 

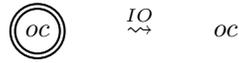
\[\pscirclebox[doubleline=true]{oc}\qquad\stackrel{IO}{\leadsto}\qquad oc\]\\

Second step: follow the choice at the root and consider the chosen substrategy profile.

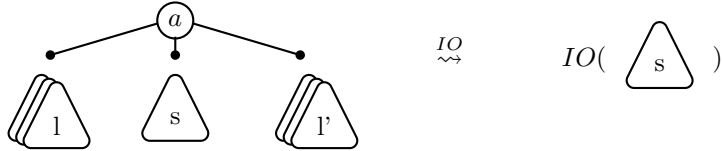
\[\psmatrix
      &&&[name=n]\pscirclebox[fillstyle=solid,boxsep=false]{a}\\&[name=n1]
      \begin{picture}(0,0)
	\multiput(0,0)(0.1,-0.05){3}{\pspolygon[fillstyle=solid,linearc=.1](-.5,-1)(.5,-1)}
	\rput(0.2,-.8){l}
      \end{picture}
      &\phantom{aaaaaaaa}&[name=n2]
      \begin{picture}(0,0)
	\put(0,0){\pspolygon[fillstyle=solid,linearc=.1](-.5,-1)(.5,-1)}
	\rput(0,-.7){s}
      \end{picture}
      &\phantom{aaaaaaaa}&
      [name=n3]
      \begin{picture}(0,0)
	\multiput(0,0)(0.1,-0.05){3}{\pspolygon[fillstyle=solid,linearc=.1](-.5,-1)(.5,-1)}
	\rput(0.2,-.8){l'}
      \end{picture}&\phantom{aaaaaaaa}\stackrel{IO}{\leadsto}\phantom{aaaaaa} &
IO(\qquad\begin{picture}(0,0)
	\put(0,.7){\pspolygon[fillstyle=solid,linearc=.1](-.5,-1)(.5,-1)}
	\rput(0,0){s}
      \end{picture}\qquad)\\\\\\\\      
     \ncline[nodesep=4pt,arrows=-*]{n}{n1}
     \ncline[nodesep=4pt,arrows=-*]{n}{n2}
     \ncline[nodesep=4pt,arrows=-*]{n}{n3}
\endpsmatrix\]

\subsubsection{The Inductive and Formal Way:}

The intended function, called \coqdocid{InducedOutcome}, is inductively and formally defined in Coq.

\medskip
\noindent
\coqdockw{Fixpoint} \coqdocid{InducedOutcome} (\coqdocid{s} : \coqdocid{Strat}) : \coqdocid{Outcome} := \coqdoceol
\noindent
\coqdocid{match} \coqdocid{s} \coqdocid{with} \coqdoceol
\noindent
$\mid$ \coqdocid{sL} \coqdocid{oc} \ensuremath{\Rightarrow} \coqdocid{oc} \coqdoceol
\noindent
$\mid$ \coqdocid{sN} \coqdocid{a} \coqdocid{sl} \coqdocid{sc} \coqdocid{sr} \ensuremath{\Rightarrow} \coqdocid{InducedOutcome} \coqdocid{sc} \coqdoceol
\noindent
\coqdocid{end}. \coqdoceol
\medskip

The following lemma, proved by induction on \coqdocid{sl}, states that the outcomes used by the underlying game of a strategy profile are all used by the underlying game of a bigger strategy profile whose chosen substrategy profile is the original strategy profile. 

\medskip
\noindent
\coqdockw{Lemma} \coqdocid{UsedOutcomes\_sN\_incl} : \ensuremath{\forall} \coqdocid{a} \coqdocid{sl} \coqdocid{sc} \coqdocid{sr}, \coqdoceol
\noindent
\coqdocid{incl} (\coqdocid{UsedOutcomes} (\coqdocid{s2g} \coqdocid{sc})) (\coqdocid{UsedOutcomes} (\coqdocid{s2g} (\coqdocid{sN} \coqdocid{a} \coqdocid{sl} \coqdocid{sc} \coqdocid{sr}))).\coqdoceol
\medskip

The next result says that the outcome induced by a strategy profile is also an outcome used in the underlying game.

\medskip
\noindent
\coqdockw{Lemma} \coqdocid{Used\_Induced\_Outcomes} : \ensuremath{\forall} \coqdocid{s} : \coqdocid{Strat},\coqdoceol
\noindent
\coqdocid{In} (\coqdocid{InducedOutcome} \coqdocid{s}) (\coqdocid{UsedOutcomes} (\coqdocid{s2g} \coqdocid{s})). \coqdoceol
\medskip

\begin{proof}
Write the claim as a predicate on \coqdocid{s} and apply the induction principle \coqdocid{Strat\_ind2} where \coqdocid{Q} \coqdocid{l} \coqdocid{s} \coqdocid{l'} is \coqdocid{In} (\coqdocid{InducedOutcome} \coqdocid{s}) (\coqdocid{UsedOutcomes} (\coqdocid{s2g} \coqdocid{s})). For the last induction case invoke \coqdocid{UsedOutcomes\_sN\_incl}.
\end{proof}

Note that if the underlying game of a strategy profile is a leaf game enclosing a given outcome, then the strategy profile is a leaf strategy profile enclosing the same outcome, so that the outcome induced by the strategy profile is also the same outcome: if \coqdocid{s2g} \coqdocid{s}=\coqdocid{gL} \coqdocid{oc} then \coqdocid{InducedOutcome} \coqdocid{s}=\coqdocid{oc}. The following lemma is the list version of the remark above. It is prove by induction on \coqdocid{ls}.

\medskip
\noindent
\coqdockw{Lemma} \coqdocid{map\_s2g\_gL\_InducedOutcome} : \ensuremath{\forall} \coqdocid{ls} \coqdocid{loc},\coqdoceol
\noindent
\coqdocid{map} \coqdocid{s2g} \coqdocid{ls} =\coqdocid{map} (\coqdocid{fun} \coqdocid{y} \ensuremath{\Rightarrow} \coqdocid{gL} \coqdocid{y}) \coqdocid{loc} \ensuremath{\rightarrow} \coqdocid{map} \coqdocid{InducedOutcome} \coqdocid{ls} = \coqdocid{loc}.\coqdoceol
\medskip

\section{Convertibility\label{sect:c}}

The first subsection defines convertibility, which will be used to formally define the notion of Nash equilibrium in section~\ref{sect:ce}. The second subsection designs an inductive proof principle for convertibility.

\subsection{Definition of Convertibility}

This subsection discusses the notion of convertibility, which is the ability of an agent to convert a strategy profile into another strategy profile, and defines a predicate accounting for it. Four lemmas follow its definition in Coq.

\subsubsection{The Traditional Way:}

An agent is (usually implicitly) granted the ability to change his choices at all nodes he owns, as shown in the following example. Agent $b$ can change his choices at the nodes where $\bold{b}$ is displayed in bold font, and only at those nodes.

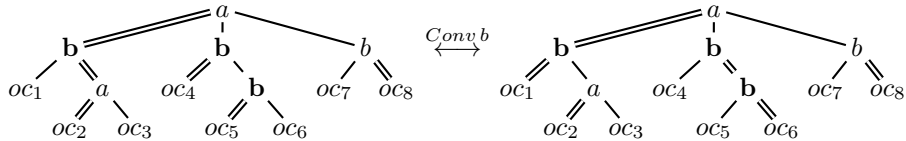
\[\psmatrix
      &&&&&&[name=n0] a&&&&&&   &&&&&&[name=n0'] a \\
      &&[name=n1] \bold{b} &&&&[name=n2] \bold{b}&&&&[name=n3] b &&\stackrel{Conv\,b}{\longleftrightarrow} &&[name=n1'] \bold{b} &&&&[name=n2'] \bold{b} &&&&[name=n3'] b \\
      &[name=l11] oc_1 &&[name=n12] a &&[name=l21] oc_4 &&[name=n22] \bold{b}  &&[name=l31] oc_7 &&[name=l32] oc_8
      &\phantom{aaaaa}& [name=l11'] oc_1 &&[name=n12'] a &&[name=l21'] oc_4 &&[name=n22'] \bold{b} &&[name=l31'] oc_7 &&[name=l32'] oc_8\\
      &&[name=l121] oc_2&&[name=l122] oc_3&&[name=l221] oc_5&&[name=l222] oc_6 &&&&&&[name=l121'] oc_2&&[name=l122'] oc_3&&[name=l221'] oc_5&&[name=l222'] oc_6
      \choiceline{n0}{n1}
      \ncline{n0}{n2}
      \ncline{n0}{n3}
      \ncline{n1}{l11}
      \choiceline{n1}{n12}
      \choiceline{n2}{l21}
      \ncline{n2}{n22}
      \ncline{n3}{l31}
      \choiceline{n3}{l32}
      \choiceline{n12}{l121}
      \ncline{n12}{l122}
      \choiceline{n22}{l221}
      \ncline{n22}{l222}
      \choiceline{n0'}{n1'}
      \ncline{n0'}{n2'}
      \ncline{n0'}{n3'}
      \choiceline{n1'}{l11'}
      \ncline{n1'}{n12'}
      \ncline{n2'}{l21'}
      \choiceline{n2'}{n22'}
      \ncline{n3'}{l31'}
      \choiceline{n3'}{l32'}
      \choiceline{n12'}{l121'}
      \ncline{n12'}{l122'}
      \ncline{n22'}{l221'}
      \choiceline{n22'}{l222'}
      \endpsmatrix\]

\subsubsection{The Inductive and Graphical Way:}

To prepare the full formalisation of the idea above, the intended function is defined inductively with the inductive graphical formalism, in two steps along the structure of strategy profiles. Let $b$ be an agent. First step: Let $oc$ be an outcome. Agent $b$ can convert the leaf strategy profile enclosing $oc$ into itself by changing some, actually none of his choices since none has been made.

\[\pscirclebox[fillstyle=solid,doubleline=true]{oc} \stackrel{Conv\; b}{\longleftrightarrow} \pscirclebox[fillstyle=solid,doubleline=true]{oc}\]\\\\\\\\

 Second step: let \begin{picture}(0,0)
	\multiput(1.5,1.5)(0.2,-0.1){3}{\pspolygon[fillstyle=solid,linearc=.1](-1.5,-2)(1.5,-2)}
	\rput(1.9,-.3){$s_0 \dots s_n::nil$}
      \end{picture}\phantom{aaaaaaaaaaaaaaaaaa} and \begin{picture}(0,0)
	\multiput(1.5,1.5)(0.2,-0.1){3}{\pspolygon[fillstyle=solid,linearc=.1](-1.5,-2)(1.5,-2)}
	\rput(1.9,-.3){$s'_0 \dots s'_n::nil$}
      \end{picture}\phantom{aaaaaaaaaaaaaaaaaa} be two lists\\\\\\

of strategy profiles such that for all $i$ between $0$ and $n$, \begin{picture}(0,0)
	\put(.5,.7){\pspolygon[fillstyle=solid,linearc=.1](-.5,-1)(.5,-1)}
	\rput(.5,0){$s_i$}
      \end{picture} \phantom{aaaa} $\stackrel{Conv\, b}{\longleftrightarrow}$ \begin{picture}(0,0)
	\put(.5,.7){\pspolygon[fillstyle=solid,linearc=.1](-.5,-1)(.5,-1)}
	\rput(.5,0){$s'_i$}
      \end{picture}\\

The second step involves compound strategy profiles, and needs case splitting on the ``converting agent'' owning the roots or not. Also let $a$ be another agent. For all $i$ between $0$ and $n$, agent $b$ can perform the following conversion by combining at once all his conversion abilities in all the substrategy profiles.

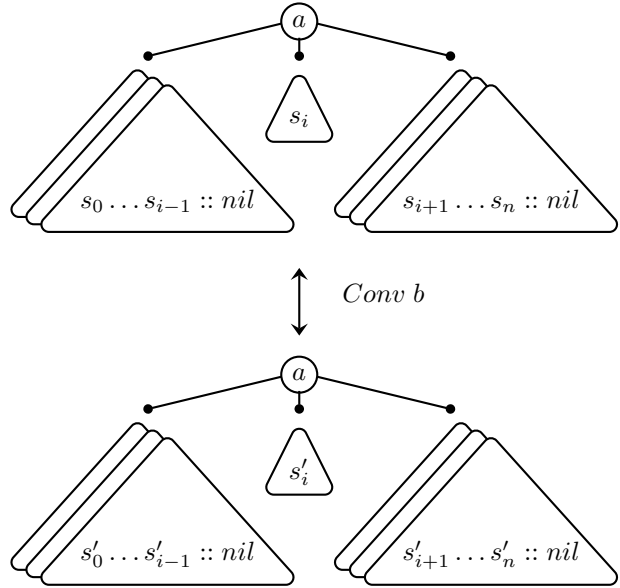
\[\psmatrix
 &&& [name=n]\pscirclebox[fillstyle=solid,boxsep=false]{a}\\&
     [name=l1]\begin{picture}(0,0)
	\multiput(0,0)(0.2,-0.1){3}{\pspolygon[fillstyle=solid,linearc=.1](-1.8,-2)(1.8,-2)}
	\rput(0.4,-1.8){$s_0\dots s_{i-1}::nil$}
      \end{picture}
       &\phantom{aaaaaaaaaa}&
     [name=l2]\begin{picture}(0,0)
       \put(0,0){\pspolygon[fillstyle=solid,linearc=.1](-.5,-1)(.5,-1)}
       \rput(0,-.7){$s_i$}
     \end{picture}&\phantom{aaaaaaaaaa}&
     [name=l3]\begin{picture}(0,0)
	\multiput(0,0)(0.2,-0.1){3}{\pspolygon[fillstyle=solid,linearc=.1](-1.8,-2)(1.8,-2)}
	\rput(0.4,-1.8){$s_{i+1}\dots s_n::nil$}
      \end{picture}\\\\\\\\\\
     \ncline[nodesep=4pt,arrows=-*]{n}{l1}
     \ncline[nodesep=4pt,arrows=-*]{n}{l2}
     \ncline[nodesep=4pt,arrows=-*]{n}{l3}
     &&&[name=t]\\
     &&&& Conv\; b\\
     &&&[name=b]
     \ncline[arrows=<->]{t}{b}\\
     &&& [name=n']\pscirclebox[fillstyle=solid,boxsep=false]{a}\\&
     [name=l'1]\begin{picture}(0,0)
	\multiput(0,0)(0.2,-0.1){3}{\pspolygon[fillstyle=solid,linearc=.1](-1.8,-2)(1.8,-2)}
	\rput(0.4,-1.8){$s'_0\dots s'_{i-1}::nil$}
      \end{picture}
     &\phantom{aaaaaaaaaa}&
     [name=l'2]\begin{picture}(0,0)
       \put(0,0){\pspolygon[fillstyle=solid,linearc=.1](-.5,-1)(.5,-1)}
       \rput(0,-.7){$s'_i$}
     \end{picture}&\phantom{aaaaaaaaaa}&
     [name=l'3]\begin{picture}(0,0)
	\multiput(0,0)(0.2,-0.1){3}{\pspolygon[fillstyle=solid,linearc=.1](-1.8,-2)(1.8,-2)}
	\rput(0.4,-1.8){$s'_{i+1}\dots s'_n::nil$}
      \end{picture}\\\\\\\\\\\\
     \ncline[nodesep=4pt,arrows=-*]{n'}{l'1}
     \ncline[nodesep=4pt,arrows=-*]{n'}{l'2}
     \ncline[nodesep=4pt,arrows=-*]{n'}{l'3}
\endpsmatrix\]

When $a$ equals $b$, the agent $b$ can also change his choice at the root owned by $a$. In that case, the agent $b$ can perform the following conversion for all $i$ and $k$ between $0$ and $n$.

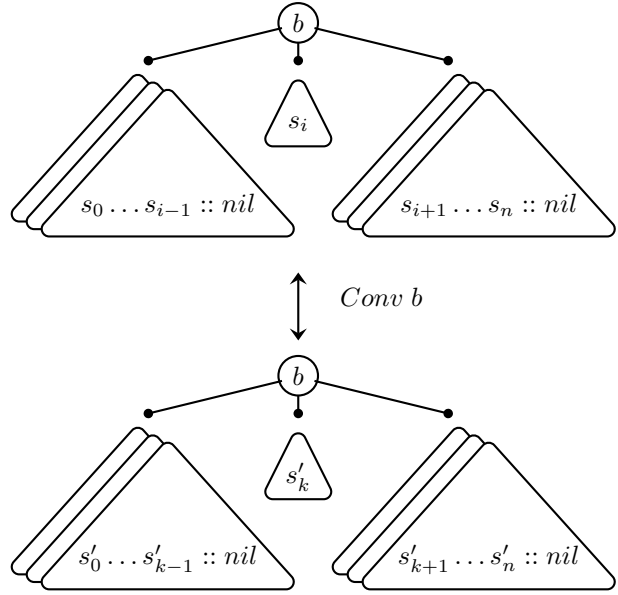
\[\psmatrix
 &&& [name=n]\pscirclebox[fillstyle=solid,boxsep=false]{b}\\&
     [name=l1]\begin{picture}(0,0)
	\multiput(0,0)(0.2,-0.1){3}{\pspolygon[fillstyle=solid,linearc=.1](-1.8,-2)(1.8,-2)}
	\rput(0.4,-1.8){$s_0\dots s_{i-1}::nil$}
      \end{picture}
       &\phantom{aaaaaaaaaa}&
     [name=l2]\begin{picture}(0,0)
       \put(0,0){\pspolygon[fillstyle=solid,linearc=.1](-.5,-1)(.5,-1)}
       \rput(0,-.7){$s_i$}
     \end{picture}&\phantom{aaaaaaaaaa}&
     [name=l3]\begin{picture}(0,0)
	\multiput(0,0)(0.2,-0.1){3}{\pspolygon[fillstyle=solid,linearc=.1](-1.8,-2)(1.8,-2)}
	\rput(0.4,-1.8){$s_{i+1}\dots s_n::nil$}
      \end{picture}\\\\\\\\\\
     \ncline[nodesep=4pt,arrows=-*]{n}{l1}
     \ncline[nodesep=4pt,arrows=-*]{n}{l2}
     \ncline[nodesep=4pt,arrows=-*]{n}{l3}
     &&&[name=t]\\
     &&&& Conv\; b\\
     &&&[name=b]
     \ncline[arrows=<->]{t}{b}\\
     &&& [name=n']\pscirclebox[fillstyle=solid,boxsep=false]{b}\\&
     [name=l'1]\begin{picture}(0,0)
	\multiput(0,0)(0.2,-0.1){3}{\pspolygon[fillstyle=solid,linearc=.1](-1.8,-2)(1.8,-2)}
	\rput(0.4,-1.8){$s'_0\dots s'_{k-1}::nil$}
      \end{picture}
     &\phantom{aaaaaaaaaa}&
     [name=l'2]\begin{picture}(0,0)
       \put(0,0){\pspolygon[fillstyle=solid,linearc=.1](-.5,-1)(.5,-1)}
       \rput(0,-.7){$s'_k$}
     \end{picture}&\phantom{aaaaaaaaaa}&
     [name=l'3]\begin{picture}(0,0)
	\multiput(0,0)(0.2,-0.1){3}{\pspolygon[fillstyle=solid,linearc=.1](-1.8,-2)(1.8,-2)}
	\rput(0.4,-1.8){$s'_{k+1}\dots s'_n::nil$}
      \end{picture}\\\\\\\\\\\\
     \ncline[nodesep=4pt,arrows=-*]{n'}{l'1}
     \ncline[nodesep=4pt,arrows=-*]{n'}{l'2}
     \ncline[nodesep=4pt,arrows=-*]{n'}{l'3}
\endpsmatrix\]

\subsubsection{The Inductive and Formal Way:}

The following definition accounts for agents being able to unilaterally change part of their choices. \coqdocid{Conv}, which means strategy profile convertibility, and \coqdocid{ListConv}, which means component-wise convertibility of list of strategy profiles, are defined by mutual induction {\it via} the word \coqdocid{with}, within the same definition.

\medskip
\noindent
\coqdockw{Inductive} \coqdocid{Conv} : \coqdocid{Agent} \ensuremath{\rightarrow} \coqdocid{Strat} \ensuremath{\rightarrow} \coqdocid{Strat} \ensuremath{\rightarrow} \coqdocid{Prop} :=\coqdoceol
\noindent
$\mid$ \coqdocid{convLeaf} : \ensuremath{\forall} \coqdocid{b} \coqdocid{oc}, \coqdocid{Conv} \coqdocid{b} (\coqdocid{sL} \coqdocid{oc})(\coqdocid{sL} \coqdocid{oc})\coqdoceol
\noindent
$\mid$ \coqdocid{convNode} : \ensuremath{\forall} \coqdocid{b} \coqdocid{a} \coqdocid{sl} \coqdocid{sl'} \coqdocid{sc} \coqdocid{sc'} \coqdocid{sr} \coqdocid{sr'},  (\coqdocid{length} \coqdocid{sl}=\coqdocid{length} \coqdocid{sl'} \ensuremath{\lor} \coqdocid{a} = \coqdocid{b}) \ensuremath{\rightarrow}\coqdoceol
\noindent
\coqdocid{ListConv} \coqdocid{b} (\coqdocid{sl}++(\coqdocid{sc}::\coqdocid{sr})) (\coqdocid{sl'}++(\coqdocid{sc'}::\coqdocid{sr'})) \ensuremath{\rightarrow}\coqdoceol
\noindent
\coqdocid{Conv} \coqdocid{b} (\coqdocid{sN} \coqdocid{a} \coqdocid{sl} \coqdocid{sc} \coqdocid{sr})(\coqdocid{sN} \coqdocid{a} \coqdocid{sl'} \coqdocid{sc'} \coqdocid{sr'})\coqdoceol
\medskip
\noindent
\coqdocid{with}\coqdoceol
\medskip
\noindent
\coqdocid{ListConv} : \coqdocid{Agent} \ensuremath{\rightarrow} \coqdocid{list} \coqdocid{Strat} \ensuremath{\rightarrow} \coqdocid{list} \coqdocid{Strat} \ensuremath{\rightarrow} \coqdocid{Prop} :=\coqdoceol
\noindent
$\mid$ \coqdocid{lconvnil} : \ensuremath{\forall} \coqdocid{b}, \coqdocid{ListConv} \coqdocid{b} \coqdocid{nil} \coqdocid{nil}\coqdoceol
\noindent
$\mid$ \coqdocid{lconvcons} : \ensuremath{\forall} \coqdocid{b} \coqdocid{s} \coqdocid{s'} \coqdocid{tl} \coqdocid{tl'}, \coqdocid{Conv} \coqdocid{b} \coqdocid{s} \coqdocid{s'} \ensuremath{\rightarrow} \coqdocid{ListConv} \coqdocid{b} \coqdocid{tl} \coqdocid{tl'} \ensuremath{\rightarrow}\coqdoceol
\noindent
\coqdocid{ListConv} \coqdocid{b} (\coqdocid{s}::\coqdocid{tl})(\coqdocid{s'}::\coqdocid{tl'}).\coqdoceol
\medskip

The formula  \coqdocid{length} \coqdocid{sl}=\coqdocid{length} \coqdocid{sl'} \ensuremath{\lor} \coqdocid{a} = \coqdocid{b} ensures that only the owner of a node can change his choice at that node. It corresponds to the case splitting in the inductive and graphical definition above. \coqdocid{ListConv} \coqdocid{b} (\coqdocid{sl}++(\coqdocid{sc}::\coqdocid{sr})) (\coqdocid{sl'}++(\coqdocid{sc'}::\coqdocid{sr'})) guarantees that this property holds also in the substrategy profiles. \coqdocid{ListConv} corresponds to the dots in the inductive and graphical definition above. One may think of avoiding a mutual definition for \coqdocid{Conv} by using the already defined \coqdocid{rel\_vector} instead of \coqdocid{ListConv}. However, the Coq versions 8.0 and 8.1 do not permit it, presumably because it would require to pass the whole \coqdocid{Conv} object as an argument (to \coqdocid{rel\_vector}) while not yet fully defined.

The following two lemmas establish the equivalence between \coqdocid{ListConv} on the one hand and \coqdocid{rel\_vector} and \coqdocid{Conv} on the other hand. They are both proved by induction on the list \coqdocid{l} and by a case splitting on \coqdocid{l'}.

\medskip
\noindent
\coqdockw{Lemma} \coqdocid{ListConv\_rel\_vector} : \ensuremath{\forall} \coqdocid{a} \coqdocid{l} \coqdocid{l'},\coqdoceol
\noindent
\coqdocid{ListConv} \coqdocid{a} \coqdocid{l} \coqdocid{l'} \ensuremath{\rightarrow} \coqdocid{rel\_vector} (\coqdocid{Conv} \coqdocid{a}) \coqdocid{l} \coqdocid{l'}.\coqdoceol
\medskip

\medskip
\noindent
\coqdockw{Lemma} \coqdocid{rel\_vector\_ListConv} : \ensuremath{\forall} \coqdocid{a} \coqdocid{l} \coqdocid{l'},\coqdoceol
\noindent
\coqdocid{rel\_vector} (\coqdocid{Conv} \coqdocid{a}) \coqdocid{l} \coqdocid{l'} \ensuremath{\rightarrow} \coqdocid{ListConv} \coqdocid{a} \coqdocid{l} \coqdocid{l'} .\coqdoceol
\medskip

The next two results state reflexivity property of \coqdocid{Conv} and \coqdocid{ListConv}.

\medskip
\noindent
\coqdockw{Lemma}  \coqdocid{Conv\_refl} : \ensuremath{\forall} \coqdocid{a} \coqdocid{s} , \coqdocid{Conv} \coqdocid{a} \coqdocid{s} \coqdocid{s}.\coqdoceol
\medskip

\begin{proof}
Let \coqdocid{a} be an agent and \coqdocid{s} a strategy profile. It suffices to prove that \coqdocid{P} \coqdocid{s} where \coqdocid{P} \coqdocid{s'} is \coqdocid{Conv} \coqdocid{a} \coqdocid{s'} \coqdocid{s'} by definition. Apply the induction principle \coqdocid{Strat\_ind2} where \coqdocid{Q} \coqdocid{sl} \coqdocid{sc} \coqdocid{sr} is \coqdocid{ListConv} \coqdocid{a} (\coqdocid{sl}++\coqdocid{sc}::\coqdocid{sr}) (\coqdocid{sl}++\coqdocid{sc}::\coqdocid{sr}) by definition. Checking all cases is straightforward.
\end{proof}

\medskip
\noindent
\coqdockw{Lemma} \coqdocid{ListConv\_refl} : \ensuremath{\forall} \coqdocid{a} \coqdocid{l}, \coqdocid{ListConv} \coqdocid{a} \coqdocid{l} \coqdocid{l}.\coqdoceol
\medskip

\begin{proof}
By \coqdocid{rel\_vector\_ListConv}, induction on the list \coqdocid{l}, and \coqdocid{Conv\_refl}.
\end{proof}

\subsection{Induction Principle for Convertibility}

\subsubsection{The Inductive and Formal Way:}

 A suitable induction principle for convertibility is automatically generated in Coq with the command \coqdocid{Scheme}, and hereby is a theorem in Coq. The principle states that under four conditions, \coqdocid{Conv} \coqdocid{a} is a subrelation of the binary relation \coqdocid{P} \coqdocid{a} for all agents \coqdocid{a}. Note that in order to prove such a property, the user has to imagine a workable predicate \coqdocid{P0}. 

\medskip
\noindent
\coqdocid{conv\_lconv\_ind}
     : \ensuremath{\forall} (\coqdocid{P} : \coqdocid{Agent} \ensuremath{\rightarrow} \coqdocid{Strat} \ensuremath{\rightarrow} \coqdocid{Strat} \ensuremath{\rightarrow} \coqdocid{Prop})\coqdoceol
\noindent
(\coqdocid{P0} : \coqdocid{Agent} \ensuremath{\rightarrow} \coqdocid{list} \coqdocid{Strat} \ensuremath{\rightarrow} \coqdocid{list} \coqdocid{Strat} \ensuremath{\rightarrow} \coqdocid{Prop}),\coqdoceol
\medskip
\noindent
{\bf(}\ensuremath{\forall} \coqdocid{b} \coqdocid{oc},
\coqdocid{P} \coqdocid{b} (\coqdocid{sL} \coqdocid{oc}) (\coqdocid{sL} \coqdocid{oc}){\bf)} \ensuremath{\rightarrow}\coqdoceol
\medskip
\noindent
{\bf(}\ensuremath{\forall} \coqdocid{b} \coqdocid{a} \coqdocid{sl} \coqdocid{sl'} \coqdocid{sc} \coqdocid{sc'} \coqdocid{sr} \coqdocid{sr'},
        \coqdocid{length} \coqdocid{sl} = \coqdocid{length} \coqdocid{sl'} \ensuremath{\lor} \coqdocid{a} = \coqdocid{b} \ensuremath{\rightarrow}
        \coqdocid{ListConv} \coqdocid{b} (\coqdocid{sl} ++ \coqdocid{sc} :: \coqdocid{sr}) (\coqdocid{sl'} ++ \coqdocid{sc'} :: \coqdocid{sr'}) \ensuremath{\rightarrow}
        \coqdocid{P0} \coqdocid{b} (\coqdocid{sl} ++ \coqdocid{sc} :: \coqdocid{sr}) (\coqdocid{sl'} ++ \coqdocid{sc'} :: \coqdocid{sr'}) \ensuremath{\rightarrow}
        \coqdocid{P} \coqdocid{b} (\coqdocid{sN} \coqdocid{a} \coqdocid{sl} \coqdocid{sc} \coqdocid{sr}) (\coqdocid{sN} \coqdocid{a} \coqdocid{sl'} \coqdocid{sc'} \coqdocid{sr'}){\bf)} \ensuremath{\rightarrow}\coqdoceol
\medskip
\noindent
{\bf(}\ensuremath{\forall} \coqdocid{b}, \coqdocid{P0} \coqdocid{b} \coqdocid{nil} \coqdocid{nil}{\bf)} \ensuremath{\rightarrow}\coqdoceol
\medskip
\noindent
{\bf(}\ensuremath{\forall} \coqdocid{b} \coqdocid{s} \coqdocid{s'} \coqdocid{tl} \coqdocid{tl'},
        \coqdocid{Conv} \coqdocid{b} \coqdocid{s} \coqdocid{s'} \ensuremath{\rightarrow}
        \coqdocid{P} \coqdocid{b} \coqdocid{s} \coqdocid{s'} \ensuremath{\rightarrow}
        \coqdocid{ListConv} \coqdocid{b} \coqdocid{tl} \coqdocid{tl'} \ensuremath{\rightarrow}
        \coqdocid{P0} \coqdocid{b} \coqdocid{tl} \coqdocid{tl'} \ensuremath{\rightarrow} \coqdocid{P0} \coqdocid{b} (\coqdocid{s} :: \coqdocid{tl}) (\coqdocid{s'} :: \coqdocid{tl'}){\bf)} \ensuremath{\rightarrow}\coqdoceol
\medskip
\noindent
\ensuremath{\forall} \coqdocid{a} \coqdocid{s} \coqdocid{s0}, \coqdocid{Conv} \coqdocid{a} \coqdocid{s} \coqdocid{s0} \ensuremath{\rightarrow} \coqdocid{P} \coqdocid{a} \coqdocid{s} \coqdocid{s0}
\noindent
\medskip

This induction principle is used below to prove that two convertible strategy profiles have the same underlying game.

\medskip
\noindent
\coqdockw{Lemma} \coqdocid{Conv\_s2g} : \ensuremath{\forall} (\coqdocid{a} : \coqdocid{Agent})(\coqdocid{s} \coqdocid{s'} : \coqdocid{Strat}), \coqdocid{Conv} \coqdocid{a} \coqdocid{s} \coqdocid{s'} \ensuremath{\rightarrow} \coqdocid{s2g} \coqdocid{s} = \coqdocid{s2g} \coqdocid{s'}. \coqdoceol
\medskip

\begin{proof}
Assume \coqdocid{s} and \coqdocid{s'} convertible by \coqdocid{a}. Write  \coqdocid{s2g} \coqdocid{s} = \coqdocid{s2g} \coqdocid{s'} as \coqdocid{P} \coqdocid{a} \coqdocid{s} \coqdocid{s'} for some \coqdocid{P} and proceed by the induction principle \coqdocid{conv\_lconv\_ind} where \coqdocid{P0} \coqdocid{b} \coqdocid{l} \coqdocid{l'} is (\coqdocid{ListConv} \coqdocid{b} \coqdocid{l} \coqdocid{l'} \ensuremath{\rightarrow} \coqdocid{map} \coqdocid{s2g} \coqdocid{l}=\coqdocid{map} \coqdocid{s2g} \coqdocid{l'}) by definition. The remainder invokes \coqdocid{map\_s2g\_sN\_s2g}.
\end{proof}

\section{Concepts of Equilibrium\label{sect:ce}}

This section defines the notions of preference, happiness, Nash equilibrium, and subgame perfect equilibrium in the new and abstract formalism. Two lemmas follow these definitions.

In traditional game theory, the agents implicitly prefer strictly greater payoffs, and thus also prefer payoff functions granting them strictly greater payoffs. In the abstract formalism, the agents' preferences for outcomes are explicitly represented by binary relations over the outcomes, one relation per agent. Below, \coqdocid{OcPref} \coqdocid{a} is the preference of agent \coqdocid{a}.

\medskip
\noindent
\coqdockw{Variable} \coqdocid{OcPref} : \coqdocid{Agent} \ensuremath{\rightarrow} \coqdocid{Outcome} \ensuremath{\rightarrow} \coqdocid{Outcome} \ensuremath{\rightarrow} \coqdocid{Prop}. \coqdoceol
\medskip

Since every strategy profile induces an outcome, the preferences over outcomes yield preferences over strategy profiles. For instance, if agent $a$ prefers $oc_1$ to $oc_3$, then he prefers the following right-hand strategy profile to the left-hand one. As to agent $b$, it could be either way since no specific property is assumed about preferences.

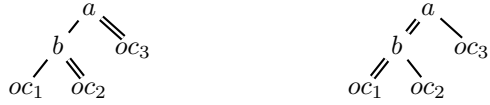
\[\psmatrix
      &&&[name=n1] a&& &&&[name=n1'] a \\
      &&[name=n2] b &&[name=l3]  oc_3& &&[name=n2'] b &&[name=l3']  oc_3\\
      &[name=l1] oc_1 &&[name=l2] oc_2 && \phantom{aaaaaaaaaaaaa}&[name=l1'] oc_1 &&[name=l2'] oc_2
      \choiceline{n1}{l3}
      \ncline{n1}{n2}
      \ncline{n2}{l1}
      \choiceline{n2}{l2}
      \ncline{n1'}{l3'}
      \choiceline{n1'}{n2'}
      \choiceline{n2'}{l1'}
      \ncline{n2'}{l2'}
    \endpsmatrix\]

Formally in Coq, preference over strategy profiles is defined as follows.

\medskip
\noindent
\coqdockw{Definition} \coqdocid{StratPref} (\coqdocid{a} : \coqdocid{Agent})(\coqdocid{s} \coqdocid{s'} : \coqdocid{Strat}) : \coqdocid{Prop} := \coqdoceol
\coqdocindent{0.50em}
\coqdocid{OcPref} \coqdocid{a} (\coqdocid{InducedOutcome} \coqdocid{s})(\coqdocid{InducedOutcome} \coqdocid{s'}).\coqdoceol
\medskip

If an agent cannot convert a given strategy profile to any preferred one, then he is said to be happy with respect to the given strategy profile. For instance the following strategy profile makes agent $a$ happy \emph{iff} agent $a$ does not prefer $oc_2$ to $oc_3$, whatever his other preferences are.

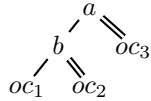
\[\psmatrix
      &&&[name=n1] a\\
      &&[name=n2] b &&[name=l3]  oc_3\\
      &[name=l1] oc_1 &&[name=l2] oc_2 
      \choiceline{n1}{l3}
      \ncline{n1}{n2}
      \ncline{n2}{l1}
      \choiceline{n2}{l2}
    \endpsmatrix\]

Formally in Coq, happiness of an agent is defined as follows.

\medskip
\noindent
\coqdockw{Definition} \coqdocid{Happy} (\coqdocid{s} : \coqdocid{Strat})(\coqdocid{a} : \coqdocid{Agent}) : \coqdocid{Prop} := \ensuremath{\forall} \coqdocid{s'},\coqdoceol
\noindent
\coqdocid{Conv} \coqdocid{a} \coqdocid{s} \coqdocid{s'} \ensuremath{\rightarrow} \ensuremath{\lnot}\coqdocid{StratPref} \coqdocid{a} \coqdocid{s} \coqdocid{s'}.\coqdoceol
\medskip

A strategy profile that makes every agent happy is called a Nash equilibrium.

\medskip
\noindent
\coqdockw{Definition} \coqdocid{Eq} (\coqdocid{s} : \coqdocid{Strat}) : \coqdocid{Prop} := \ensuremath{\forall} \coqdocid{a}, \coqdocid{Happy} \coqdocid{s} \coqdocid{a}. \coqdoceol
\medskip

A subgame perfect equilibrium is a Nash equilibrium such that all of its substrategy profiles are subgame perfect equilibria. Compare this definition with the informal and more complicated one in section~\ref{sect:tsgt}.

\medskip
\noindent
\coqdockw{Fixpoint} \coqdocid{SPE} (\coqdocid{s} : \coqdocid{Strat}) : \coqdocid{Prop} := \coqdocid{Eq} \coqdocid{s} \ensuremath{\land} \coqdoceol
\noindent
\coqdocid{match} \coqdocid{s} \coqdocid{with} \coqdoceol
\noindent
$\mid$ \coqdocid{sL} \coqdocid{oc} \ensuremath{\Rightarrow}  \coqdocid{True} \coqdoceol
\noindent
$\mid$ \coqdocid{sN} \coqdocid{a} \coqdocid{sl} \coqdocid{sc} \coqdocid{sr} \ensuremath{\Rightarrow} (\coqdocid{listforall} \coqdocid{SPE} \coqdocid{sl}) \ensuremath{\land} \coqdocid{SPE} \coqdocid{sc} \ensuremath{\land} (\coqdocid{listforall} \coqdocid{SPE} \coqdocid{sr}) \coqdoceol
\noindent
\coqdocid{end}.\coqdoceol
\medskip

Therefore, a subgame perfect equilibrium is a Nash equilibrium. 

\medskip
\noindent
\coqdockw{Lemma} \coqdocid{SPE\_is\_Eq} : \ensuremath{\forall} \coqdocid{s} : \coqdocid{Strat}, \coqdocid{SPE} \coqdocid{s} \ensuremath{\rightarrow} \coqdocid{Eq} \coqdocid{s}. \coqdoceol
\medskip

The following provides a sufficient condition for a strategy profile to be a Nash equilibrium: at the root of a compound strategy profile, if the chosen substrategy profile is a Nash equilibrium, and if the owner of the root cannot convert any of his other options into a substrategy profile that he prefers to his current choice, then the compound strategy profile is also a Nash equilibrium. This is stated in Coq below.

\medskip
\noindent
\coqdockw{Lemma} \coqdocid{Eq\_subEq\_choice} : \ensuremath{\forall} \coqdocid{a} \coqdocid{sl} \coqdocid{sc} \coqdocid{sr},\coqdoceol
\noindent
(\ensuremath{\forall} \coqdocid{s} \coqdocid{s'}, \coqdocid{In} \coqdocid{s} \coqdocid{sl} \ensuremath{\lor} \coqdocid{In} \coqdocid{s} \coqdocid{sr} \ensuremath{\rightarrow} \coqdocid{Conv} \coqdocid{a} \coqdocid{s} \coqdocid{s'} \ensuremath{\rightarrow} \ensuremath{\lnot}\coqdocid{StratPref} \coqdocid{a} \coqdocid{sc} \coqdocid{s'}) \ensuremath{\rightarrow}\coqdoceol
\noindent
\coqdocid{Eq} \coqdocid{sc} \ensuremath{\rightarrow} \coqdocid{Eq} (\coqdocid{sN} \coqdocid{a} \coqdocid{sl} \coqdocid{sc} \coqdocid{sr}).\coqdoceol
\medskip

\begin{proof}
Let be \coqdocid{a}, \coqdocid{sl}, \coqdocid{sc}, \coqdocid{sr} and assume the two premises. Also assume that an agent \coqdocid{a'} can convert \coqdocid{sN} \coqdocid{a} \coqdocid{sl} \coqdocid{sc} \coqdocid{sr} to \coqdocid{s'} that he prefers. Now try to derive a contradiction from the hypothesis. For this, note that \coqdocid{s'} equals \coqdocid{sN} \coqdocid{a} \coqdocid{sl'} \coqdocid{sc'} \coqdocid{sr'} for some \coqdocid{sl'}, \coqdocid{sc'}, and \coqdocid{sr'}. Case split on \coqdocid{sl} and \coqdocid{sl'} having or not the same length. If they have the same length then use the equilibrium assumption together with \coqdocid{rel\_vector\_app\_cons\_same\_length} and \coqdocid{ListConv\_rel\_vector}. If \coqdocid{sl} and \coqdocid{sl'} have different lengths then \coqdocid{a'} equals \coqdocid{a}. The remainder invokes \coqdocid{rel\_vector\_app\_cons\_different\_length} and \coqdocid{ListConv\_rel\_vector}.
\end{proof}

The converse of this lemma also holds, but it is omitted in this paper.

\section{Existence of Equilibria\label{sect:ee}}

This section generalises the notion of ``backward induction'' for abstract sequential games, and shows that it yields a subgame perfect equilibrium when preferences are totally ordered. But it also shows that it may not always yield a subgame perfect equilibrium for arbitrary preferences. However, this section eventually proves that acyclicity of decidable preferences is a necessary and sufficient condition for guaranteeing computable existence of Nash equilibrium/subgame perfect equilibrium.

\subsection{``Backward Induction''\label{subsect:bi}}

This subsection starts with an informal discussion, and continues with definitions in the inductive graphical formalism. Eventually, a ``backward induction'' function is defined in Coq, and one lemma follows.

\subsubsection{The Traditional Way:}

Informally, the idea is to perform ``backward induction'' on all subgames first, and then to let the owner of the root choose one strategy profile that suits him best among the newly built strategy profiles. When preferences are partially ordered, the agent can choose in order to maximise his preference; when they are not partially ordered, a procedure slightly more general may be needed. Lemma \coqdocid{Choose\_and\_split} defined in subsection~\ref{subsect:ns} relates to such a procedure. (More specifically, the proof of the lemma is such a procedure.) For instance, let $oc_1\dots oc_6$ be six outcomes such that $a$ prefers $oc_5$ to $oc_2$, and $b$ prefers $oc_2$ to $oc_1$ and $oc_1$ to $oc_2$, and nothing else. A  ``backward induction'' process is detailed below. On the left-hand picture, $b$ chooses by \coqdocid{Choose\_and\_split}. On the right-hand picture, agent $a$ being ``aware'' of $b$'s choice procedure chooses accordingly, also by \coqdocid{Choose\_and\_split}.

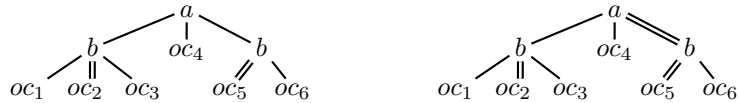
\[\psmatrix
     &&&&[name=n] a&&&&  &&&&[name=n'] a\\
     &&[name=n1] b && [name=l4] oc_4 && [name=n2] b&& &&[name=n'1] b && [name=l'4] oc_4 && [name=n'2] b\\
     &[name=l1] oc_1 \phantom{a}&[name=l2] oc_2 \phantom{a}&[name=l3] oc_3 && [name=l5] oc_5 &&[name=l6] oc_6&\phantom{aaaaaaaa}&[name=l'1] oc_1 \phantom{a}&[name=l'2] oc_2 \phantom{a}&[name=l'3] oc_3 && [name=l'5] oc_5 &&[name=l'6] oc_6
     \ncline{n}{n1}
     \ncline{n}{l4}
     \ncline{n}{n2}
     \ncline{n1}{l1}
     \choiceline{n1}{l2}
     \ncline{n1}{l3}
     \choiceline{n2}{l5}
     \ncline{n2}{l6}
     \ncline{n'}{n'1}
     \ncline{n'}{l'4}
     \choiceline{n'}{n'2}
     \ncline{n'1}{l'1}
     \choiceline{n'1}{l'2}
     \ncline{n'1}{l'3}
     \choiceline{n'2}{l'5}
     \ncline{n'2}{l'6}
\endpsmatrix\]

\subsubsection{The Inductive and Graphical Way:}

To prepare the full formalisation of the idea above, the intended function is defined inductively with the inductive graphical formalism, in two steps along the structure of strategy profiles. First step: performing ``backward induction'' on a leaf game that encloses an outcome yields a leaf strategy profile that encloses the same outcome.

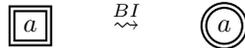
\[\psframebox[fillstyle=solid,doubleline=true]{a}\qquad \stackrel{BI}{\leadsto}\qquad \pscirclebox[fillstyle=solid,doubleline=true]{a}\]

Second step: Assume that ``backward induction'' is defined for the games $g_0,\dots ,g_n$. An agent $a$ can choose one strategy profile among $BI\,g_0,\dots ,BI\,g_n$ by \coqdocid{Choose\_and\_split} and his own preference, as below. 

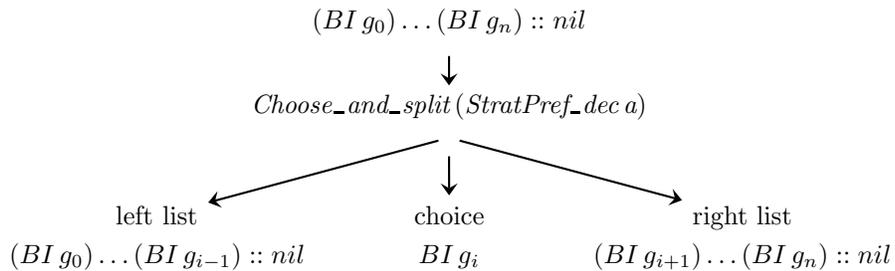
\[\psmatrix
    && [name=n] (BI\, g_0)\dots(BI\, g_n)::nil\\\\
    &&[name=n0']\rput(0,0){\coqdocid{Choose\_and\_split}\,(\coqdocid{StratPref\_dec}\,\coqdocid{a})}\\
    &&[name=n0]\\\\
    &[name=n1]\mbox{left list} & [name=n2] \mbox{choice} & [name=n3] \mbox{right list}\\ 
    & (BI\,g_0)\dots (BI\,g_{i-1})::nil & BI\,g_i & (BI\, g_{i+1})\dots (BI\,g_n)::nil\\
    \ncline[nodesep=8pt,arrows=->]{n}{n0'}
    \ncline[nodesep=4pt,arrows=->]{n0}{n1}
    \ncline[nodesep=4pt,arrows=->]{n0}{n2}
    \ncline[nodesep=4pt,arrows=->]{n0}{n3}
\endpsmatrix\]

Then, ``backward induction'' can be defined on the following compound game.\\

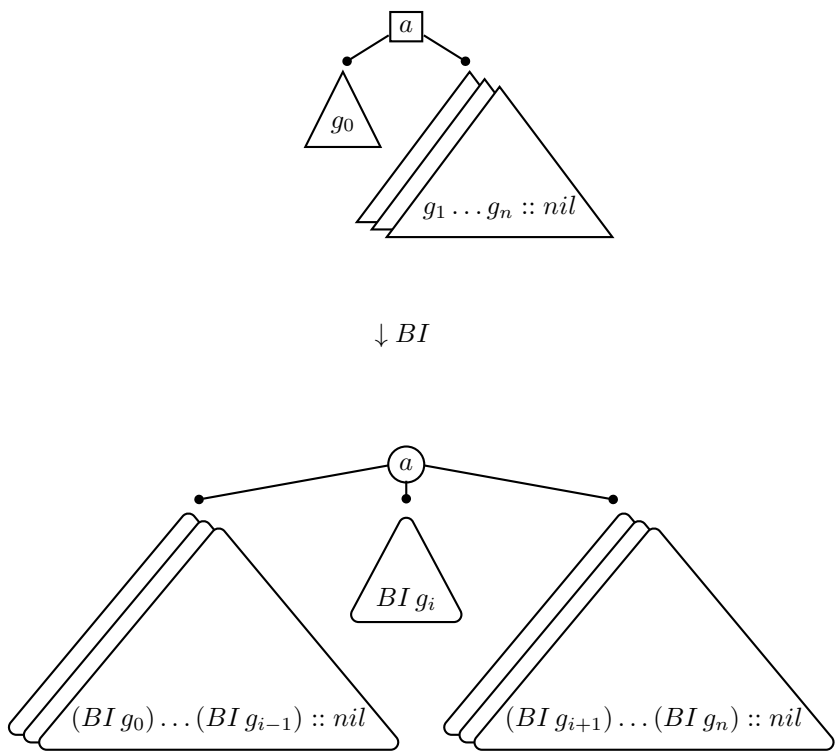
\[\psmatrix
      &&[name=n]\psframebox[fillstyle=solid,boxsep=false]{a}\\
      &[name=n1]
      \begin{picture}(0,0)
	\put(0,0){\pspolygon[fillstyle=solid](-.5,-1)(.5,-1)}
	\rput(0,-.7){$g_0$}
      \end{picture}
      &\phantom{aaaaaaaa}&
      [name=n2]
      \begin{picture}(0,0)
	\multiput(0,0)(0.2,-0.1){3}{\pspolygon[fillstyle=solid](-1.5,-2)(1.5,-2)}
	\rput(0.4,-1.8){$g_1\dots g_n::nil$}
      \end{picture}
      \\\\\\\\\\ 
      \ncline[nodesep=4pt,arrows=-*]{n}{n1}
      \ncline[nodesep=4pt,arrows=-*]{n}{n2}
\endpsmatrix\]

\[\downarrow BI\]\\

\[\psmatrix
 &&& [name=n]\pscirclebox[fillstyle=solid,boxsep=false]{a}\\&
     [name=l1]\begin{picture}(0,0)
	\multiput(0,0)(0.2,-0.1){3}{\pspolygon[fillstyle=solid,linearc=.1](-2.5,-3)(2.5,-3)}
	\rput(0.4,-2.8){$(BI\,g_0)\dots (BI\,g_{i-1})::nil$}
      \end{picture}
       &\phantom{aaaaaaaaaaaaaa}&
     [name=l2]\begin{picture}(0,0)
       \put(0,0){\pspolygon[fillstyle=solid,linearc=.1](-.8,-1.5)(.8,-1.5)}
       \rput(0,-1.2){$BI\,g_i$}
     \end{picture}&\phantom{aaaaaaaaaaaaaa}&
     [name=l3]\begin{picture}(0,0)
	\multiput(0,0)(0.2,-0.1){3}{\pspolygon[fillstyle=solid,linearc=.1](-2.5,-3)(2.5,-3)}
	\rput(0.4,-2.8){$(BI\,g_{i+1})\dots (BI\,g_n)::nil$}
      \end{picture}\\\\\\\\\\\\
     \ncline[nodesep=4pt,arrows=-*]{n}{l1}
     \ncline[nodesep=4pt,arrows=-*]{n}{l2}
     \ncline[nodesep=4pt,arrows=-*]{n}{l3}
\endpsmatrix\]

\subsubsection{The Inductive and Formal Way:}

The definition using the inductive graphical formalism above is translated into Coq formalism. First, assume that preferences over outcomes are decidable.

\medskip
\noindent
\coqdockw{Hypothesis} \coqdocid{OcPref\_dec} : \ensuremath{\forall} (\coqdocid{a} : \coqdocid{Agent}), \coqdocid{rel\_dec} (\coqdocid{OcPref} \coqdocid{a}). \coqdoceol
\medskip

 Subsequently, preferences over strategy profiles are also decidable.

\medskip
\noindent
\coqdockw{Lemma} \coqdocid{StratPref\_dec} : \ensuremath{\forall} (\coqdocid{a} : \coqdocid{Agent}), \coqdocid{rel\_dec} (\coqdocid{StratPref} \coqdocid{a}).\coqdoceol
\medskip

Next, the generalisation of ``backward induction'', with respect to the preferences above, is defined by recursion. For the sake of readability, the definition displayed below is a slight simplification of the actual Coq code. 

\medskip
\noindent
\coqdockw{Fixpoint} \coqdocid{BI} (\coqdocid{g} : \coqdocid{Game}) : \coqdocid{Strat} := \coqdoceol
\noindent
\coqdocid{match} \coqdocid{g} \coqdocid{with} \coqdoceol
\noindent
$\mid$ \coqdocid{gL} \coqdocid{oc} \ensuremath{\Rightarrow} \coqdocid{sL} \coqdocid{oc} \coqdoceol
\noindent
$\mid$ \coqdocid{gN} \coqdocid{a} \coqdocid{g} \coqdocid{l} \ensuremath{\Rightarrow} \coqdocid{let} (\coqdocid{sl},\coqdocid{sc},\coqdocid{sr}):=\coqdoceol
\noindent
\coqdocindent{6.00em} \coqdocid{Choose\_and\_split} (\coqdocid{StratPref\_dec} \coqdocid{a}) (\coqdocid{map} \coqdocid{BI} \coqdocid{l}) (\coqdocid{BI} \coqdocid{g}) \coqdocid{in}\coqdoceol
\coqdocindent{6.00em}
\coqdocid{sN} \coqdocid{a} \coqdocid{sl} \coqdocid{sc} \coqdocid{sr}\coqdoceol
\noindent
\coqdocid{end}.\coqdoceol
\medskip

As stated below, the underlying game of the image by \coqdocid{BI} of a given game is the same game. 

\medskip
\noindent
\coqdockw{Lemma} \coqdocid{BI\_s2g} : \ensuremath{\forall} \coqdocid{g} : \coqdocid{Game}, \coqdocid{s2g} (\coqdocid{BI} \coqdocid{g})=\coqdocid{g}. \coqdoceol
\medskip

\begin{proof}
Rewrite the claim as \ensuremath{\forall} \coqdocid{g}, \coqdocid{P} \coqdocid{g} for some \coqdocid{P}. Apply the induction principle \coqdocid{Game\_ind2} where \coqdocid{Q} \coqdocid{g} \coqdocid{l} is \coqdocid{s2g} (\coqdocid{BI} \coqdocid{g})=\coqdocid{g} \ensuremath{\land} \coqdocid{map} \coqdocid{s2g} (\coqdocid{map} \coqdocid{BI}) \coqdocid{l}=\coqdocid{l}. For the last induction case, invoke \coqdocid{Choose\_and\_split} and \coqdocid{map\_app}. 
\end{proof}

\subsection{The Total Order Case}

In this subsection only, assume transitivity and irreflexivity of preferences over outcomes. Subsequently, those propertiesalso hold for preferences over strategy profiles.

\medskip
\noindent
\coqdockw{Hypothesis} \coqdocid{OcPref\_irrefl} : \ensuremath{\forall} (\coqdocid{a} : \coqdocid{Agent}), \coqdocid{irreflexive} (\coqdocid{OcPref} \coqdocid{a}). \coqdoceol
\medskip
\noindent
\coqdockw{Hypothesis} \coqdocid{OcPref\_trans} : \ensuremath{\forall} (\coqdocid{a} : \coqdocid{Agent}), \coqdocid{transitive} (\coqdocid{OcPref} \coqdocid{a}).\coqdoceol
\medskip
\noindent
\coqdockw{Lemma} \coqdocid{StratPref\_irrefl} : \ensuremath{\forall} (\coqdocid{a} : \coqdocid{Agent}), \coqdocid{irreflexive} (\coqdocid{StratPref} \coqdocid{a}).\coqdoceol
\medskip
\noindent
\coqdockw{Lemma} \coqdocid{StratPref\_trans} : \ensuremath{\forall} (\coqdocid{a} : \coqdocid{Agent}), \coqdocid{transitive} (\coqdocid{StratPref} \coqdocid{a}).\coqdoceol
\medskip

Irreflexivity of preferences guarantees that leaf strategy profiles are Nash equilibria.

\medskip
\noindent
\coqdockw{Lemma} \coqdocid{Leaf\_Eq} : \ensuremath{\forall} \coqdocid{oc} : \coqdocid{Outcome}, \coqdocid{Eq} (\coqdocid{sL} \coqdocid{oc}). \coqdoceol
\medskip

If preferences are total over a given list of outcomes, then for any sequential game using only outcomes from the list, ``backward Induction'' yields subgame perfect equilibrium. This is the translation of Kuhn's result into abstract sequential game formalism.

\medskip
\noindent
\coqdockw{Lemma} \coqdocid{BI\_SPE} :  \ensuremath{\forall} \coqdocid{loc}, (\ensuremath{\forall} (\coqdocid{a} : \coqdocid{Agent}), \coqdocid{total} (\coqdocid{OcPref} \coqdocid{a}) \coqdocid{loc}) \ensuremath{\rightarrow}\coqdoceol
\noindent
\ensuremath{\forall} \coqdocid{g} : \coqdocid{Game}, \coqdocid{incl} (\coqdocid{UsedOutcomes} \coqdocid{g}) \coqdocid{loc} \ensuremath{\rightarrow} \coqdocid{SPE} (\coqdocid{BI} \coqdocid{g}). \coqdoceol
\medskip

\begin{proof}
Assume \coqdocid{loc}, a list of outcomes, and the totality property. Write \coqdocid{incl} (\coqdocid{UsedOutcomes} \coqdocid{g}) \coqdocid{loc} \ensuremath{\rightarrow} \coqdocid{SPE} (\coqdocid{BI} \coqdocid{g}) as \coqdocid{P} \coqdocid{g} and proceed by the induction principle \coqdocid{Game\_ind2} where \coqdocid{Q} \coqdocid{g} \coqdocid{l} is \coqdocid{listforall}  (\coqdocid{fun} \coqdocid{g'} \ensuremath{\Rightarrow} \coqdocid{incl} (\coqdocid{UsedOutcomes} \coqdocid{g'}) \coqdocid{loc}) (\coqdocid{g}::\coqdocid{l}) \ensuremath{\rightarrow} \coqdocid{SPE} (\coqdocid{BI} \coqdocid{g}) \ensuremath{\land} \coqdocid{listforall} \coqdocid{SPE} (\coqdocid{map} \coqdocid{BI} \coqdocid{l}). The first three cases are straightforward. For the fourth and last case assume \coqdocid{g}, \coqdocid{l}, and the other premises, such as an agent \coqdocid{a}. In order to prove that the ``backward induction'' of the compound game \coqdocid{gN} \coqdocid{a} \coqdocid{g} \coqdocid{l} is a subgame perfect equilibrium, first note that all substrategy profiles of this ``backward induction'' are subgame perfect equilibria, by  the induction hypotheses, \coqdocid{listforall\_appr} and  \coqdocid{listforall\_appr}. Then, the main difficulty is to prove that it is a Nash equilibrium. For this, invoke \coqdocid{Eq\_subEq\_choice} after proving its two premises: The induction hypothesis and lemma \coqdocid{SPE\_is\_Eq} shows that the substrategy profile chosen by \coqdocid{a} is a Nash equilibrium. For the other premise required for invoking \coqdocid{Eq\_subEq\_choice}, assume a substrategy profile not chosen by \coqdocid{a}. The induction hypothesis and lemma \coqdocid{SPE\_is\_Eq} show that it is a Nash equilibrium. Next, assume that this Nash equilibrium is convertible by \coqdocid{a} into another strategy profile, and show that $a$ does not prefer this new strategy profile to his current choice. For this, invoke decidability, irreflexivity, transitivity, and totality of preferences, as well as lemmas \coqdocid{listforall\_In}, \coqdocid{UsedOutcomes\_gN}, \coqdocid{map\_inverse}, \coqdocid{Used\_Induced\_Outcomes} and \coqdocid{Conv\_s2g}.
\end{proof}

\subsection{Limitation\label{subsect:l}}

Now, no property is assumed about the preferences. Consider the three outcomes $oc_1$, $oc_2$ and $oc_3$, and an agent $a$ that prefers $oc_2$ to $oc_3$, and nothing else. The strategy profile below is obtained by the ``backward induction'' function define in subsection~\ref{subsect:bi}. However, it is not a Nash equilibrium since the current induced outcome is $oc_3$, but it can be converted by $a$ into the preferred $oc_2$.

 \[\psmatrix
      &&&[name=n1] a\\
      &&[name=n2] a &&[name=l3] oc_3 \\
      &[name=l1] oc_1 &&[name=l2] oc_2 
      \ncline{n1}{n2}
      \choiceline{n1}{l3}
      \choiceline{n2}{l1}
      \ncline{n2}{l2}
    \endpsmatrix\]

Informally, when dealing with totally ordered preferences, the notions of ``backward induction'' and subgame perfect equilibrium coincide although their definitions differ; they are the same in extension, but are different in intension. This difference in intension is critical when dealing with partially ordered preferences: as shown by the example above, ``backward induction'' no longer yields Nash equilibrium, let alone subgame perfect equilibrium.

\subsection{General Case}

Until this subsection, equilibrium concepts and related notions have been defined with respect to given preferences: a preference binary relation was associated with an agent once for all. In the following, equilibrium concepts and related notions are abstracted over preferences. It means that preferences become a parameter of the definitions and lemmas. For instance,  instead of writing \coqdocid{Eq} \coqdocid{s} to say that \coqdocid{s} is a Nash equilibrium, one shall write \coqdocid{Eq} \coqdocid{OcPref} \coqdocid{s} to say that \coqdocid{s} is a Nash equilibrium with respect to the family of preferences defined by \coqdocid{OcPref}.

As formally stated in the two lemmas below, given two families of preferences and given a strategy profile, if for every agent the restriction of his first preference to the outcomes used by the strategy profile is a subrelation of his second preference, and if the strategy profile is a Nash equilibrium/subgame perfect equilibrium with respect to the second preferences then it is also a Nash equilibrium/subgame perfect equilibrium with respect to the first preferences. Informally, the less arcs an agent's preference has, the more strategy profiles make the agent happy.

\medskip
\noindent
\coqdockw{Lemma} \coqdocid{Eq\_order\_inclusion} :  \ensuremath{\forall} \coqdocid{OcPref} \coqdocid{OcPref'} \coqdocid{s}, \coqdoceol
\noindent
(\ensuremath{\forall} \coqdocid{a} : \coqdocid{Agent},\coqdoceol
\noindent
 \coqdocid{sub\_rel} (\coqdocid{restriction} (\coqdocid{OcPref} \coqdocid{a}) (\coqdocid{UsedOutcomes} (\coqdocid{s2g} \coqdocid{s}))) (\coqdocid{OcPref'} \coqdocid{a})) \ensuremath{\rightarrow}\coqdoceol
\noindent
\coqdocid{Eq} \coqdocid{OcPref'} \coqdocid{s} \ensuremath{\rightarrow} \coqdocid{Eq} \coqdocid{OcPref} \coqdocid{s}.\coqdoceol
\medskip

\begin{proof}
Invoke \coqdocid{Used\_Induced\_Outcomes} and \coqdocid{Conv\_s2g}. 
\end{proof}

\medskip
\noindent
\coqdockw{Lemma} \coqdocid{SPE\_order\_inclusion} : \ensuremath{\forall} \coqdocid{OcPref} \coqdocid{OcPref'} \coqdocid{s}, \coqdoceol
\noindent
(\ensuremath{\forall} \coqdocid{a} : \coqdocid{Agent},\coqdoceol
\noindent
\coqdocid{sub\_rel} (\coqdocid{restriction} (\coqdocid{OcPref} \coqdocid{a}) (\coqdocid{UsedOutcomes} (\coqdocid{s2g} \coqdocid{s}))) (\coqdocid{OcPref'} \coqdocid{a})) \ensuremath{\rightarrow}\coqdoceol
\noindent
\coqdocid{SPE} \coqdocid{OcPref'} \coqdocid{s} \ensuremath{\rightarrow} \coqdocid{SPE} \coqdocid{OcPref} \coqdocid{s}. \coqdoceol
\medskip

\begin{proof}
Assume two families of preferences and rewrite the claim as \ensuremath{\forall} \coqdocid{s}, \coqdocid{P} \coqdocid{s} for some \coqdocid{P}. Then apply \coqdocid{Strat\_ind2} where \coqdocid{Q} \coqdocid{sl} \coqdocid{sc} \coqdocid{sr} is  (\ensuremath{\forall} \coqdocid{a}, \coqdocid{sub\_rel} (\coqdocid{restriction} (\coqdocid{OcPref} \coqdocid{a}) (\coqdocid{UsedOutcomes} (\coqdocid{s2g} (\coqdocid{sN} \coqdocid{a} \coqdocid{sl} \coqdocid{sc} \coqdocid{sr})))) (\coqdocid{OcPref'} \coqdocid{a})) \ensuremath{\rightarrow} \coqdocid{listforall} (\coqdocid{SPE} \coqdocid{OcPref'}) \coqdocid{sl} \ensuremath{\rightarrow} \coqdocid{SPE} \coqdocid{OcPref'} \coqdocid{sc} \ensuremath{\rightarrow} \coqdocid{listforall} (\coqdocid{SPE} \coqdocid{OcPref'}) \coqdocid{sr} \ensuremath{\rightarrow} \coqdocid{listforall} (\coqdocid{SPE} \coqdocid{OcPref}) \coqdocid{sl} \ensuremath{\land} \coqdocid{SPE} \coqdocid{OcPref} \coqdocid{sc} \ensuremath{\land} \coqdocid{listforall} (\coqdocid{SPE} \coqdocid{OcPref}) \coqdocid{sr}. For the first and fifth induction steps, apply \coqdocid{Eq\_order\_inclusion}. For the third and fourth induction step, invoke lemmas \coqdocid{transitive\_sub\_rel}, \coqdocid{sub\_rel\_restriction\_incl}, \coqdocid{incl\_appr}, \coqdocid{incl\_appl}, and \coqdocid{incl\_refl}.
\end{proof}

The following lemma generalises Kuhn's result to acyclic preferences (instead of totally ordered). It invokes a result related to topological sorting proved in~\cite{SLR2007-14}.

\medskip
\noindent
\coqdockw{Theorem} \coqdocid{acyclic\_SPE} : \ensuremath{\forall} \coqdocid{OcPref},\coqdoceol
\noindent
(\ensuremath{\forall} \coqdocid{a}, \coqdocid{rel\_dec} (\coqdocid{OcPref} \coqdocid{a})) \ensuremath{\rightarrow}\coqdoceol
\noindent
(\ensuremath{\forall} \coqdocid{a}, \coqdocid{irreflexive} (\coqdocid{clos\_trans} \coqdocid{Outcome} (\coqdocid{OcPref} \coqdocid{a}))) \ensuremath{\rightarrow} \coqdoceol
\coqdocindent{1.00em}
\ensuremath{\forall} \coqdocid{g}, \{\coqdocid{s} : \coqdocid{Strat} $\mid$ \coqdocid{s2g} \coqdocid{s}=\coqdocid{g} \ensuremath{\land} \coqdocid{SPE} \coqdocid{OcPref} \coqdocid{s}\}. \coqdoceol
\medskip

\begin{proof}
Assume a family of decidable acyclic preferences. Let \coqdocid{g} be a game. According to~\cite{SLR2007-14}, by topological sorting there exists a family of decidable acyclic preferences that are strict total orders including the original preferences on the outcomes used by \coqdocid{g}. A subgame perfect equilibrium can be computed by \coqdocid{BI\_SPE}. Conclude by \coqdocid{SPE\_order\_inclusion}.
\end{proof}

The following property relates to \coqdocid{SPE\_is\_Eq}.
 
\medskip
\noindent
\coqdockw{Theorem} \coqdocid{SPE\_Eq} : \ensuremath{\forall} \coqdocid{OcPref},\coqdoceol
\noindent
(\ensuremath{\forall} \coqdocid{a}, \coqdocid{rel\_dec} (\coqdocid{OcPref} \coqdocid{a})) \ensuremath{\rightarrow}\coqdoceol
\noindent
(\ensuremath{\forall} \coqdocid{g} : \coqdocid{Game}, \{\coqdocid{s} : \coqdocid{Strat} $\mid$ \coqdocid{s2g} \coqdocid{s}=\coqdocid{g} \ensuremath{\land} \coqdocid{SPE} \coqdocid{OcPref} \coqdocid{s}\}) \ensuremath{\rightarrow} \coqdoceol
\noindent
\ensuremath{\forall} \coqdocid{g} : \coqdocid{Game}, \{\coqdocid{s} : \coqdocid{Strat} $\mid$ \coqdocid{s2g} \coqdocid{s}=\coqdocid{g} \ensuremath{\land} \coqdocid{Eq} \coqdocid{OcPref} \coqdocid{s}\}.  \coqdoceol
\medskip

The next result says that if all games have Nash equilibria with respect to a given family of preferences, then these preferences are acyclic. 

\subsubsection{The Traditional Way}

Informally, let an agent $a$ prefer $x_1$ to $x_0$, $x_2$ to $x_1$, and so on, and $x_0$ to $x_n$. The game displayed below has no Nash equilibrium, as suggested graphically. The symbol \coqdocid{s} $\longrightarrow_a$  \coqdocid{s'} means that agent $a$ both prefers \coqdocid{s'} to \coqdocid{s} and can convert \coqdocid{s} to \coqdocid{s'}. So, the formula \coqdocid{s} $\longrightarrow_a$ \coqdocid{s'} witnesses the agent's non-happiness.

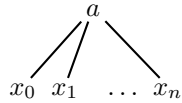
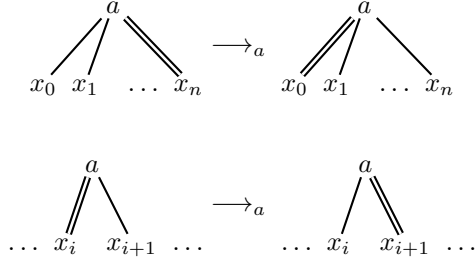
\[\psmatrix
      &&&&[name=n] a \\\\
      &[name=l1] x_0 &&[name=l2] x_1 && [name=l3]\dots && [name=l4] x_n 
      \ncline{n}{l1}
      \ncline{n}{l2}
      \ncline{n}{l4}
 \endpsmatrix\]

 \[\psmatrix
      &&&&[name=n] a &&&&& &&&[name=n'] a\\
      &&&&&&&& \longrightarrow_a \\
      &[name=l1] x_0 &&[name=l2] x_1 && [name=l3]\dots && [name=l4] x_n & \phantom{aaa}
      &[name=l1'] x_0 &&[name=l2'] x_1 && [name=l3'] \dots && [name=l4'] x_n  
      \ncline{n}{l1}
      \ncline{n}{l2}
      \choiceline{n}{l4}
      \choiceline{n'}{l1'}
      \ncline{n'}{l2'}
      \ncline{n'}{l4'}
\endpsmatrix\]

 \[\psmatrix
      &&&&[name=n] a &&&&& &&&[name=n'] a\\
      &&&&&&&& \longrightarrow_a \\
      &[name=l1] \dots &&[name=l2] x_i && [name=l3] x_{i+1} && [name=l4] \dots & \phantom{aaa}
      &[name=l1'] \dots &&[name=l2'] x_i && [name=l3'] x_{i+1} && [name=l4'] \dots 
      \choiceline{n}{l2}
      \ncline{n}{l3}
      \ncline{n'}{l2'}
      \choiceline{n'}{l3'}
 \endpsmatrix\]

\subsubsection{The Formal Way}
Now, the corresponding formal statement and its proof.

\medskip
\noindent
\coqdockw{Theorem} \coqdocid{Eq\_acyclic} : \ensuremath{\forall} \coqdocid{OcPref},\coqdoceol
\noindent
(\ensuremath{\forall} \coqdocid{a}, \coqdocid{rel\_dec} (\coqdocid{OcPref} \coqdocid{a})) \ensuremath{\rightarrow}\coqdoceol
\noindent
(\ensuremath{\forall} \coqdocid{g} : \coqdocid{Game}, \{\coqdocid{s} : \coqdocid{Strat} $\mid$ \coqdocid{s2g} \coqdocid{s}=\coqdocid{g} \ensuremath{\land} \coqdocid{Eq} \coqdocid{OcPref} \coqdocid{s}\}) \ensuremath{\rightarrow} \coqdoceol
\noindent
\ensuremath{\forall} \coqdocid{a} : \coqdocid{Agent}, \coqdocid{irreflexive} (\coqdocid{clos\_trans} \coqdocid{Outcome} (\coqdocid{OcPref} \coqdocid{a})). \coqdoceol
\medskip

\begin{proof}
Assume all premises. In particular, let \coqdocid{a} be an agent and \coqdocid{oc} be an outcome related to itself by the transitive closure of the agent's preference. Prove a contradiction by building a game such that every strategy profile for the game can be improved upon, as follows. By lemma \coqdocid{clos\_trans\_path}, get an actual path \coqdocid{loc} from \coqdocid{oc} to itself with respect to the preference. If \coqdocid{loc} is empty then invoke lemma \coqdocid{Conv\_refl}. If \coqdocid{loc} is not empty then, thanks to the assumption, compute a Nash equilibrium for the game with root owned by agent \coqdocid{a} and the children being leaves enclosing \coqdocid{oc} for the first and the elements of \coqdocid{loc} for the others. Case split on the right-hand substrategy profile list of the Nash equilibrium being empty. If it is empty, and the left-hand substrategy profile list of the Nash equilibrium as well, then the root of the game has two or more children and the strategy profile has one only, hence a contradiction. If the left-hand substrategy profile list is not empty then \coqdocid{ListConv\_refl}, \coqdocid{map\_s2g\_gL\_InducedOutcome}, \coqdocid{path\_app\_elim\_right}, and \coqdocid{map\_app} may be needed. If the right-hand substrategy profile list of the Nash equilibrium is not empty then invoke associativity of list appending as well as the lemmas mentioned just above, and a case splitting on the left-hand substrategy profile list again.
\end{proof}

\subsection{Examples\label{subsect:ex}}

\subsubsection{Partial Order and Subgame Perfect Equilibrium}

As in section~\ref{subsect:l}, consider three outcomes $oc_1$, $oc_2$ and $oc_3$, and an agent $a$ that only prefers $oc_2$ to $oc_3$. Each of the three possible linear extensions of $a$'s preference, prior to ``backward induction'', leads to a subgame perfect equilibrium as shown below, where the symbol $<$ represents the different linear extensions. Compare with the ``backward induction'' without prior linear extension of subsection~\ref{subsect:l}. 

 \[\psmatrix
      &oc_2& <&oc_3&< &oc_1&& oc_3&<&oc_2&< &oc_1&& oc_3& <&oc_1&< &oc_2\\  
      \\
      &&&[name=n1] a &&& &&&[name=n1'] a &&& &&&[name=n1''] a\\
      &&[name=n2] a &&[name=l3] oc_3  && \phantom{aaaaaaaaa} &&[name=n2'] a &&[name=l3'] oc_3  && \phantom{aaaaaaaaa}
      &&[name=n2''] a &&[name=l3''] oc_3\\
      &[name=l1] oc_1 &&[name=l2] oc_2 &&& &[name=l1'] oc_1 &&[name=l2'] oc_2 &&& &[name=l1''] oc_1 &&[name=l2''] oc_2
      \ncline{n1}{n2}
      \choiceline{n1}{n2}
      \choiceline{n2}{l1}
      \ncline{n2}{l2}
      \ncline{n1}{l3}
      \ncline{n1'}{n2'}
      \choiceline{n1'}{n2'}
      \choiceline{n2'}{l1'}
      \ncline{n2'}{l2'}
      \ncline{n1'}{l3'}
      \ncline{n1''}{n2''}
      \choiceline{n1''}{n2''}
      \ncline{n2''}{l1''}
      \choiceline{n2''}{l2''}
      \ncline{n1''}{l3''}
    \endpsmatrix\]

In the same way, the results obtained in this section show that subgame perfect equilibrium exists for games with preferences discussed in section~\ref{sect:wto}. 

\subsubsection{Benevolent and Malevolent Selfishness}

In subsection~\ref{subsect:self} were defined benevolent and malevolent selfishnesses. It is not surprising that, sometimes, benevolent selfishness yields better payoffs for all agents than malevolent selfishness. For instance compare the induced payoff functions below. The lefthand ``backward induction'' corresponds to benevolent selfishness, and the righthand one to malevolent selfishness.

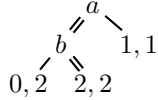
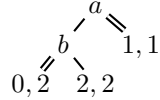
\[\psmatrix
   &&&[name=n1]a & &\phantom{aaaaaaaa}& &&[name=n1']a \\
   &&[name=n2] b&&[name=l1]1,1 && &[name=n2'] b&&[name=l1']1,1 \\
   &[name=l3] 0,2&&[name=l2]2,2& && [name=l3'] 0,2&&[name=l2']2,2
   \choiceline{n1}{n2}
   \ncline{n1}{l1}
   \choiceline{n2}{l2}
   \ncline{n2}{l3}
   \ncline{n1'}{n2'}
   \choiceline{n1'}{l1'}
   \choiceline{n2'}{l3'}
   \ncline{n2'}{l2'}
\endpsmatrix\]

On the contrary, malevolent selfishness may yield better payoffs for all agents than benevolent selfishness, as shown below. The lefthand ``backward induction'' corresponds to benevolent selfishness, and the righthand one to malevolent selfishness.

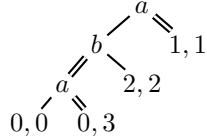
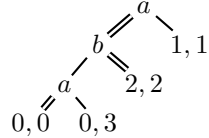
\[\psmatrix
   &&&&[name=n1]a & &\phantom{aaaaaaaa}& &&&[name=n1']a \\
   &&&[name=n2] b&&[name=l1]1,1 && &&[name=n2'] b&&[name=l1']1,1 \\
   &&[name=n3] a&&[name=l2]2,2& && &[name=n3'] a&&[name=l2']2,2\\
   &[name=l3] 0,0&&[name=l4]0,3&& && [name=l3'] 0,0&&[name=l4']0,3
   \ncline{n1}{n2}
   \choiceline{n1}{l1}
   \choiceline{n2}{n3}
   \ncline{n2}{l2}
   \ncline{n3}{l3}
   \choiceline{n3}{l4}
   \choiceline{n1'}{n2'}
   \ncline{n1'}{l1'}
   \ncline{n2'}{n3'}
   \choiceline{n2'}{l2'}
   \choiceline{n3'}{l3'}
   \ncline{n3'}{l4'}
\endpsmatrix\]

\section{Conclusion}

This paper introduces a new inductive formalism for sequential games. it also replaces real-valued payoff functions with atomic objects called outcomes, and the usual total order over the reals with arbitrary preferences. This way, it also defines an abstract version of sequential games, with similar tree structure and notion of convertibility as for traditional sequential games. The notions of Nash equilibrium, subgame perfect equilibrium, and ``backward induction'' are translated into the new formalism. When preferences are totally ordered, ``backward induction''guarantees existence of subgame perfect equilibrium for all sequential games, thus translating Kuhn's result in the new formalism. However, an example shows that ``backward induction'' fails to provide equilibrium for non-totally ordered preferences, {\it e.g.}, partial orders. But when preferences are acyclic, it is still possible to perform ``backward induction'' on a game whose preferences have been linearly extended. This yields a subgame perfect equilibrium of the game with respect to the acyclic preferences, because removing arcs from a preference relation amounts to removing reasons for being unhappy. So, given a collection of outcomes, the following three propositions are equivalent:

\begin{itemize}
\item The agents' preferences over the given outcomes are acyclic.
\item Every game built over the given outcomes has a Nash equilibrium.
\item Every game built over the given outcomes has a subgame perfect equilibrium.
\end{itemize}

The formalism introduced in this paper is suitable for proofs in Coq, which is a (highly reliable) constructive proof assistant. This way, the above-mentioned equivalence is fully computer-certified using Coq. Beside the additional guarantee of correctness provided by the Coq proof, the activity of formalisation also helps clearly identify the useful definitions and the main articulations of the proof.

Informally, a result due to Aumann states that, when dealing with traditional sequential games, ``common knowledge of rationality among agents'' is equivalent to ``backward induction'' (where ``rationality'' means playing in order to maximise one's payoff). This is arguable in abstract sequential games if one expects ``common knowledge of rationality among agents'' to imply Nash equilibrium. Indeed, ``backward induction'' may not imply Nash equilibrium, as seen in this paper. Therefore, ``backward induction'' may not imply ``common knowledge of rationality among agents''. Instead, one may wonder whether ``common knowledge of rationality among agents'' is equivalent to subgame perfect equilibrium, whatever it may mean, when preferences are acyclic, {\it i.e.} rational in some sense. In this case, the difference between ``backward induction'' and subgame perfect equilibrium seems critical again.

\section{Acknowledgement}

I thank Pierre Lescanne for his helpful comments and Jingdi Zeng for proof reading my English.

\bibliographystyle{plain}
\bibliography{slr_inria_RR2007-18}

\end{document}